\documentclass[10pt,journal,compsoc]{sty/IEEEtran}

\usepackage{subcaption}
\usepackage{url}
\ifCLASSOPTIONcompsoc
  \usepackage[nocompress]{cite}
\else
  \usepackage{cite}
\fi
\usepackage{mathtools}
\usepackage{backnaur}
\usepackage{amssymb}
\usepackage{amsthm}
\usepackage[breaklinks,hidelinks]{hyperref}
\usepackage{multirow, makecell, multicol}
\usepackage{booktabs}
\usepackage{xspace}
\usepackage{fancyvrb}
\usepackage{siunitx}
\usepackage[normal]{threeparttable}
\usepackage{pifont}
\usepackage{wasysym}
\usepackage{tikz}
\usetikzlibrary{shapes,arrows}
\usepackage{comment}
\usepackage{enumitem}

\usepackage{scalerel}
\usepackage{xcolor, colortbl}
\usepackage[figuresright]{rotating}

\usepackage{shortcuts}
\usepackage{xstring}
\newcommand{\PP}[1]{
\noindent{\bf \IfEndWith{#1}{.}{#1}{#1.}}
}

\newcommand{\PI}[1]{
{\bf \IfEndWith{#1}{.}{#1}{#1.}}
}

\newcommand{\ie}{\textit{i.e.}\xspace}

\usepackage{expl3}
\ExplSyntaxOn
\cs_new_protected:Npn \tc #1
{
  \tl_set:Nn \l_tmpa_tl {#1}
  \regex_replace_all:nnN { . } { \c{string} \0 } \l_tmpa_tl
  \tl_set:Nx \l_tmpb_tl { \l_tmpa_tl }
  \texttt{\l_tmpb_tl}
}
\ExplSyntaxOff

\theoremstyle{definition}%

\def\Snospace~{\S{}}

\newcommand\vfrac[2]{\ThisStyle{%
  \setbox0=\hbox{$\SavedStyle#1#2$}%
  \setbox2=\hbox{$\SavedStyle X$}%
  \ifdim\ht0>\ht2\setlength{\ht0}{\ht2}\fi%
  #1\mathord{\stretchto{\raisebox{2.3\LMpt}{$\SavedStyle/$}}{\ht0}}#2}}

\usepackage{tikz}

\newcommand*\CC[1]{%
\begin{tikzpicture}[baseline=(C.base)]
\node[draw,circle,inner sep=0.2pt](C) {#1};
\end{tikzpicture}}

\begin{document}

\title{Revisiting Binary Code Similarity Analysis using Interpretable Feature
Engineering and Lessons Learned}

\author{
  Dongkwan~Kim, 
  Eunsoo~Kim, 
  Sang~Kil~Cha, 
  Sooel~Son, 
  Yongdae~Kim
  \IEEEcompsocitemizethanks{
    \IEEEcompsocthanksitem D. Kim, E. Kim, S. K. Cha, S. Son, and Y. Kim are
    with KAIST.\protect\\
    E-mail: \{ dkay, hahah, sangkilc, sl.son, yongdaek \}@kaist.ac.kr
  }%
  \thanks{Corresponding author: Sang Kil Cha.}%
  \thanks{Manuscript received: November 21, 2020.}%
  \thanks{Manuscript revised (major): July 16, 2021.}%
  \thanks{Manuscript revised (minor): Feb 22, 2022.}%
  \thanks{Manuscript accepted: June 26, 2022.}}
\markboth{IEEE TRANSACTIONS ON SOFTWARE ENGINEERING, November~2020}
{Kim \MakeLowercase{\textit{et al.}}: Revisiting Binary Code Similarity Analysis with Interpretable Feature Engineering}

\IEEEtitleabstractindextext{
\begin{abstract}
  Binary code similarity analysis (BCSA) is widely used for diverse security
  applications, including plagiarism detection, software license violation
  detection, and vulnerability discovery.
  Despite the surging research interest in BCSA, it is significantly challenging
  to perform new research in this field for several reasons.
  First, most existing approaches focus only on the end results, namely,
  increasing the success rate of BCSA, by adopting uninterpretable machine
  learning.
  Moreover, they utilize their own benchmark, sharing neither the source code nor
  the entire dataset.
  %
  %
  %
  Finally, researchers often use different terminologies or even use the same
  technique without citing the previous literature properly, which makes it
  difficult to reproduce or extend previous work.
  %
  %
  %
  To address these problems, we take a step back from the mainstream and
  contemplate fundamental research questions for BCSA. Why does a certain
  technique or a certain feature show better results than the others?
  %
  Specifically, we conduct the first systematic study on the basic features used
  in BCSA by leveraging interpretable feature engineering on a large-scale
  benchmark.
  Our study reveals various useful insights on BCSA. For example, we show that a
  simple interpretable model with a few basic features can achieve a comparable
  result to that of recent deep learning-based approaches.
  Furthermore, we show that the way we compile binaries or the correctness of
  underlying binary analysis tools can significantly affect the performance of
  BCSA.
  Lastly, we make all our source code and benchmark public and suggest future
  directions in this field to help further research.

\end{abstract}


}

\maketitle

\IEEEdisplaynontitleabstractindextext
\IEEEpeerreviewmaketitle

\IEEEraisesectionheading{\section{Introduction}\label{s:intro}}
\IEEEPARstart{P}{rogrammers} reuse existing code to build new software.
It is common practice for them to find the source code from another project and
repurpose that code for their own needs~\cite{reiss:2009}. Inexperienced
developers even copy and paste code samples from the Internet to ease the
development process.

This trend has deep implications for software security and privacy. When a
programmer takes a copy of a buggy function from an existing project, the bug
will remain intact even after the original developer has fixed it. Furthermore,
if a developer in a commercial software company inadvertently uses library
code from an open-source project, the company can be accused of violating an
open-source license such as the GNU General Public License
(GPL)~\cite{gplviolation}.

Unfortunately, detecting such problems from binary code using a
similarity analysis is \emph{not} straightforward, particularly when the source
code is not available. This is because binary code lacks high-level
abstractions, such as data types and functions. For example, it is not obvious
from binary code to determine whether a memory cell represents an integer,
a string, or another data type. Moreover,
identifying precise function boundaries is
radically challenging in the first place~\cite{shin2015recognizing, bao2014byteweight}.

Therefore, measuring the similarity between binaries has been an essential
research topic in many areas, such as malware
detection~\cite{comparetti2010identifying, jang2011bitshred}, plagiarism
detection~\cite{luo2014semantics, zhang2014program}, authorship
identification~\cite{meng:esorics:2017}, and vulnerability
discovery~\cite{\paperVuln}.
%

However, despite the surging research interest in binary code similarity
analysis (BCSA), we found that it is still significantly challenging to conduct
new research on this field for several reasons.

First, most of the methods focus only on the end results without considering the
precise reasoning behind their approaches.
For instance, during our literature study in the field, we observed that there
is a prominent research trend in applying BCSA techniques to cross-architecture
and cross-compiler binaries of the same program~\cite{liu2018alphadiff,
xu2017neural, feng2017extracting, chandramohan2016bingo, eschweiler2016discovre,
pewny2015cross, gao2018vulseeker}.
%
%
Those approaches aim to measure the similarity between two or more seemingly
distinct binaries generated from different compilers targeting different
instruction sets.
To achieve this, multiple approaches have devised complex analyses based on
machine learning to extract the semantics of the binaries, assuming that their
semantics should not change across compilers nor target architectures.
However, none of the existing approaches clearly justifies the necessity of such
complex semantics-based analyses. One may imagine that a compiler may generate
structurally similar binaries for different architectures, even though they are
syntactically different.
Do compilers and architectures really matter for BCSA in this regard?
Unfortunately, it is difficult to answer this question because most of the existing
approaches leverage \textbf{uninterpretable} machine learning
techniques~\cite{\paperML}. Further, it is not even clear why a BCSA algorithm
works only on some benchmarks and not on others.

Second, every existing paper on BCSA that we studied utilizes its own
benchmark to evaluate the proposed technique, which makes it difficult to
compare the approaches with one another. Moreover, reproducing the previous
results is often infeasible because most researchers reveal neither their source
code nor their dataset. Only 10 of the \numAnalyzedPapers~papers that we studied
fully released their source code, and \emph{only two} of them opened their
entire dataset.

Finally, researchers in this field do not use unified terminologies and often
miss out on critical citations that have appeared in top-tier venues of other fields.
Some of them even mistakenly use the same technique without citing the previous
literature properly.
These observations motivate one of our research goals, which is to summarize and
review widely adopted techniques in this field, particularly in terms of
generating features.

%
%

To address these problems, we take a step back from the mainstream and
contemplate fundamental research questions for BCSA.
%
As the first step, we precisely define the terminologies and categorize the
features used in the previous literature to unify terminologies and build
knowledge bases for BCSA.
We then construct a comprehensive and reproducible benchmark for BCSA to help
researchers extend and evaluate their approaches easily.
Lastly, we design an interpretable feature engineering model and conduct a
series of experiments to investigate the influence of
compilers, their options, and their target architectures
on the syntactic and structural features of the resulting binaries.

%
Our benchmark, which we refer to as \sys, encompasses various existing
benchmarks.
It is generated by using major compiler options and targets, which include
\numArch architectures, \numComp different compilers, \numOpti optimization
levels, as well as various other compiler flags.
%
%
\sys contains \numBin distinct binaries and \numFunc functions built for \numOpt
different combinations of compiler options, on \numPack real-world software
packages.
We also provide an automated script that helps extend \sys to handle different
architectures or compiler versions.
We believe this is critical because it is not easy to modify or extend previous
benchmarks, despite us having their source codes. Cross-compiling software
packages using various compiler options is challenging because of numerous
environmental issues.
To the best of our knowledge, \sys is the first \emph{reproducible} and
\emph{extensible} benchmark for BCSA.
%

With our benchmark, we perform a series of rigorous studies on how the way of
compilation can affect the resulting binaries in terms of their syntactic and
structural shapes.
To this end, we design a simple \emph{interpretable} BCSA model, which
essentially computes relative differences between BCSA feature values.
We then build a BCSA tool that we call \techname, which employs our
interpretable model.
With \techname, we found several misconceptions in the field of BCSA as well as
novel insights for future research as follows.

First, the current research trend in BCSA is founded on a rather exaggerated
assumption: binaries are radically different across architectures, compiler
types, or compiler versions. However, our study shows that this is not
necessarily the case.
For example, we demonstrate that simple numeric features, such as the number of
incoming/outgoing calls in a function, are largely similar across binaries
compiled for different architectures.
%
%
We also present other elementary features that are robust
across compiler types, compiler versions, and even intra-procedural obfuscation.
%
With these findings, we show that \techname with those simple features can
achieve comparable accuracy to that of the state-of-the-art BCSA tools, such
as VulSeeker, which relies on a complex deep learning-based model.

%

Second, most researchers focus on vectorizing features from binaries, but not on
recovering lost information during the compilation, such as variable
types. However, our experimental results suggest that focusing on the latter can
be highly effective for BCSA.
Specifically, we show that \techname with recovered type information achieves an
accuracy of over 99\% on all our benchmarks, which was indeed the best result
compared to all the existing tools we studied.
%
%
This result highlights that recovering type information from binaries can be as
critical as developing a novel machine learning algorithm for BCSA.

Finally, the interpretability of the model helps advance the field by deeply
understanding BCSA results.
%
For example, we present several practical issues in the underlying binary
analysis tool, \ie, IDA Pro, which is used by \techname,
and discuss how such errors can affect the performance of BCSA.
Since our benchmark has the ground truth and our tool employs an interpretable
model, we were able to easily pinpoint those fundamental issues, which will
eventually benefit binary analysis tools and the entire field of binary
analysis.

\PP{Contribution.}
In summary, our contributions are as follows:
%
\begin{itemize}[topsep=-1pt,itemsep=0pt,parsep=0pt,partopsep=0pt,leftmargin=8pt]


\item We study the features and benchmarks used in the past literature regarding
  BCSA and clarify less-explored research questions in this field.

\item We propose \sys\footnote{\binkiturl}, the first reproducible and
  expandable BCSA benchmark. It contains \numBin binaries and \numFunc
  functions compiled for \numOpt distinct combinations of compilers,
  compiler options, and target architectures.

\item We develop a BCSA tool, \techname\footnote{\tikniburl}, which employs a
  simple interpretable model. We demonstrate that \techname can achieve an
  accuracy comparable to that of a state-of-the-art deep learning-based tool. We
  believe this will serve as a baseline to evaluate future research in this
  field.

\item We investigate the efficacy of basic BCSA features with \techname on
  our benchmark and unveil several misconceptions and novel insights.

\item We make our source code, benchmark, and experimental data publicly available
  to support open science.

\end{itemize}

\section{Binary Code Similarity Analysis}
\label{sec:back}


Binary Code Similarity Analysis (BCSA) is the process of identifying whether two
given code snippets have similar semantics. Typically, it takes in two code
snippets as input and returns a similarity score ranging from 0 to 1, where 0
indicates the two snippets are completely different, and 1 means that they are
equivalent.
The input code snippet can be a function~\cite{\paperFunc}, or even an entire
binary image~\cite{\paperBinary}.
Additionally, the actual comparison can be based on functions, even if the
inputs are entire binary images~\cite{feng2016scalable, chandramohan2016bingo,
hu2016cross, xu2017neural, huang2017binsequence, hu2017binary,
gao2018vulseeker}.


At a high level, BCSA performs four major steps as described below:

\smallskip
\PP{(S1) Syntactic Analysis}
Given a binary code snippet, one parses the code to obtain a disassembly or an
Abstract Syntax Tree (AST) of the code, which is often referred to as an
Intermediate Representation (IR)~\cite{kim2017testing}. This step corresponds to
the syntax analysis in traditional compiler theory, where source code is parsed
down to an AST.
If the input code is an entire binary file, we first parse it based on its file
format and split it into sections.

%

\smallskip
\PP{(S2) Structural Analysis}
This step analyzes and recovers the control structures inherent in the given
binary code, which are not readily available from the syntactic analysis phase
(S1).
In particular, this step involves recovering the control-flow graphs (CFGs) and
call graphs (CGs) in the binary code~\cite{schwartz:usec:2013,
  yakdan:ndss:2015}.
Once the control-structural information is obtained, one can use any attribute
of these control structures as a feature.
We distinguish this step from semantic analysis (S3) because binary analysis
frameworks typically provide CFGs and CGs for free; the analysts do not have to
write a complex semantic analyzer.

\smallskip
\PP{(S3) Semantic Analysis}
Using the control-structural information obtained from S2, one can perform
traditional program analyses, such as data-flow analysis and symbolic analysis,
on the binary to figure out the underlying semantics.
In this step, one can generate features that represent sophisticated program
semantics, such as how register values flow into various program points.
One can also enhance the features gathered from S1--S2 along with the semantic
information.

\smallskip
\PP{(S4) Vectorization and Comparison}
The final step is to vectorize all the information gathered from S1--S3 to
compute the similarity between the binaries. This step essentially results in a
similarity score between 0 and 1.
%

\begin{figure}[t]
  \centering
  \includegraphics[width=\linewidth]{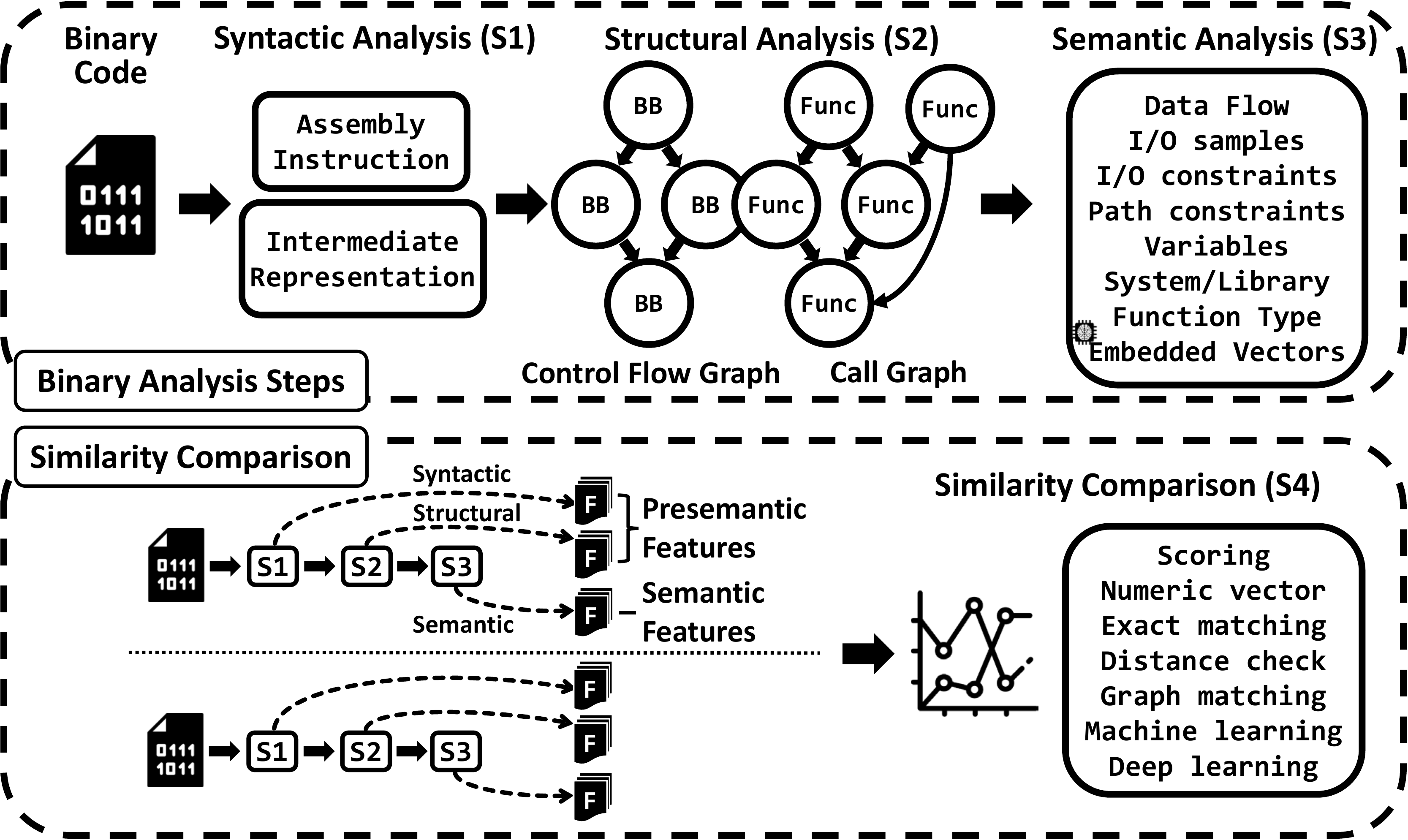}
  \vspace{-0.2in}
  \caption{Typical workflow of binary analysis (upper)
  and similarity comparison (lower)
  in binary code similarity analysis. Tools may skip some of the steps.}

  \label{fig:analysisoverview}
  \vspace{-0.2in}
\end{figure}

\smallskip
\autoref{fig:analysisoverview} depicts the four-step process. The first three
steps determine the inputs to the comparison step (S4), which are often referred
to as \emph{features}.
Some of the first three steps can be skipped depending on the underlying
features being used.
The actual comparison methodology in S4 can also vary depending on the BCSA
technique. For example, one may compute the Jaccard
distance~\cite{real1996probabilistic} between feature sets, calculate the graph
edit distance~\cite{bunke1997relation} between CFGs, or even leverage deep
learning algorithms~\cite{bromley1994signature, geurts2006extremely}.
%
%
However, \textit{as the success of any comparison algorithm significantly
  depends on the chosen features, this paper focuses on features used in
  previous studies rather than the comparison methodologies.}

In this section, we first describe the features used in the previous papers and
their underlying assumptions (\autoref{ss:prevfeatures}). We then discuss the
benchmarks used in those papers and point out their problems
(\autoref{ss:prevbench}).
Lastly, we present several research questions identified during our study
(\autoref{ss:rq}).

\smallskip \PP{Scope.}
Our study focuses on 43 recent BCSA papers (from 2014 to 2020)
that appeared in 27 top-tier venues
of different computer science areas,
such as computer security, software engineering, programming languages,
and machine learning.
%
There are, of course, plentiful research papers in this field, all of which are
invaluable.  Nevertheless, \emph{our focus here is not to conduct a complete
  survey on them but to introduce a prominent trend and the underlying research
  questions in this field, as well as to answer those questions.}
We particularly focus on
features and datasets used in those studies,
which lead us to four underexplored research questions
that we will discuss in~\autoref{ss:rq};
our goal is to investigating these research questions
by conducting a series of rigorous experiments.
Because of the space limit, we excluded papers~\cite{\paperBack} that were
published before 2014 and those not regarding top-tier venues,
or binary diffing tools~\cite{bindiff, diaphora, oh2015darungrim} used in the industry.
%
%
%
Additionally, we excluded papers that aimed to address a specific research problem
such as malware detection, library function identification, or patch
identification.
%
%
Although our study focuses only on recent papers, we found that the features we
studied in this paper are indeed general enough; they cover most of the features
used in the older papers.

\subsection{Features Used in Prior Works}
\label{ss:prevfeatures}

We categorize features into two groups
based on when they are generated during BCSA.
Particularly, we refer to features obtained before and after the semantic analysis
step (S3) as \emph{presemantic features} and \emph{semantic features},
respectively. Presemantic features can be derived from either S1 or S2, and
semantic features can be derived from S3. We summarize both features used in
the recent literature in~\autoref{tab:prevfeatures}.

\subsubsection{Presemantic Features}
\label{ss:presemantic}

Presemantic features denote
direct or indirect outcomes of
the syntactic (S1) and structural (S2) analyses.
Therefore,
we refer to any attribute of binary code,
which can be derived without a semantic analysis,
as a presemantic feature.
We can further categorize presemantic features
used in previous literature based on
whether the feature represents a number or not.
We refer to features representing a number
as \emph{numeric presemantic features},
and others as \emph{non-numeric presemantic features}.
The first half of~\autoref{tab:prevfeatures} summarizes them.

\begin{table*}[t]
    \caption{Summary of the features used in previous studies.}

  \vspace{-0.1in}
  \label{tab:prevfeatures}
  \centering
  \scriptsize
  \setlength\tabcolsep{0.1pt}
  \def\arraystretch{0.50}
  \begin{threeparttable}
    \begin{tabular}{
      l@{\hspace{0.05cm}}
      l
      c@{\hspace{0.00cm}}
    cccccc
      c@{\hspace{0.05cm}}
    c
      c@{\hspace{0.05cm}}
    ccccccc
      c@{\hspace{0.05cm}}
    cc
    ccccccc
      c@{\hspace{0.05cm}}
    ccccccc
    ccc
      c@{\hspace{0.05cm}}
    cccccc
      c@{\hspace{0.05cm}}
    ccccc
    @{\hspace{0.01cm}}
    c
  }
    \toprule
    &
    &
    & \multicolumn{6}{c}{\textbf{\scriptsize{2014}}}
    &
    & \multicolumn{1}{c}{\textbf{\scriptsize{2015}}}
    &
    & \multicolumn{7}{c}{\textbf{\scriptsize{2016}}}
    &
    & \multicolumn{9}{c}{\textbf{\scriptsize{2017}}}
    &
    & \multicolumn{10}{c}{\textbf{\scriptsize{2018}}}
    &
    & \multicolumn{6}{c}{\textbf{\scriptsize{2019}}}
    &
    & \multicolumn{5}{c}{\textbf{\scriptsize{2020}}}
    \\
    \cmidrule(lr){4-9}
    \cmidrule(lr){11-11}
    \cmidrule(lr){13-19}
    \cmidrule(lr){21-29}
    \cmidrule(lr){31-40}
    \cmidrule(lr){42-47}
    \cmidrule(lr){49-53}

    &
    &
    & \multicolumn{1}{c}{\rotatebox[origin=l]{90}{TEDEM}} 
    & \multicolumn{1}{c}{\rotatebox[origin=l]{90}{Tracy}} 
    & \multicolumn{1}{c}{\rotatebox[origin=l]{90}{CoP}} 
    & \multicolumn{1}{c}{\rotatebox[origin=l]{90}{LoPD}} 
    & \multicolumn{1}{c}{\rotatebox[origin=l]{90}{BLEX}} 
    & \multicolumn{1}{c}{\rotatebox[origin=l]{90}{BinClone}} 

    &
    & \multicolumn{1}{c}{\rotatebox[origin=l]{90}{Multi-k-MH}} 

    &
    & \multicolumn{1}{c}{\rotatebox[origin=l]{90}{discovRE}} 
    & \multicolumn{1}{c}{\rotatebox[origin=l]{90}{Genius}} 
    & \multicolumn{1}{c}{\rotatebox[origin=l]{90}{Esh}} 
    & \multicolumn{1}{c}{\rotatebox[origin=l]{90}{BinGo}} 
    & \multicolumn{1}{c}{\rotatebox[origin=l]{90}{MockingBird}} 
    & \multicolumn{1}{c}{\rotatebox[origin=l]{90}{Kam1n0}} 
    & \multicolumn{1}{c}{\rotatebox[origin=l]{90}{BinDNN}} 

    &
    & \multicolumn{1}{c}{\rotatebox[origin=l]{90}{BinSign}} 
    & \multicolumn{1}{c}{\rotatebox[origin=l]{90}{Xmatch}} 
    & \multicolumn{1}{c}{\rotatebox[origin=l]{90}{Gemini}} 
    & \multicolumn{1}{c}{\rotatebox[origin=l]{90}{GitZ}} 
    & \multicolumn{1}{c}{\rotatebox[origin=l]{90}{BinSim}} 
    & \multicolumn{1}{c}{\rotatebox[origin=l]{90}{BinSequence}} 
    & \multicolumn{1}{c}{\rotatebox[origin=l]{90}{IMF-sim}} 
    & \multicolumn{1}{c}{\rotatebox[origin=l]{90}{CACompare}} 
    & \multicolumn{1}{c}{\rotatebox[origin=l]{90}{ASE17}} 

    &
    & \multicolumn{1}{c}{\rotatebox[origin=l]{90}{BinArm}} 
    & \multicolumn{1}{c}{\rotatebox[origin=l]{90}{SANER18}} 
    & \multicolumn{1}{c}{\rotatebox[origin=l]{90}{BinGo-E}} 
    & \multicolumn{1}{c}{\rotatebox[origin=l]{90}{WSB}} 
    & \multicolumn{1}{c}{\rotatebox[origin=l]{90}{BinMatch}} 
    & \multicolumn{1}{c}{\rotatebox[origin=l]{90}{MASES18}} 
    & \multicolumn{1}{c}{\rotatebox[origin=l]{90}{Zeek}} 

    & \multicolumn{1}{c}{\rotatebox[origin=l]{90}{FirmUp}} 
    & \multicolumn{1}{c}{\rotatebox[origin=l]{90}{$\alpha$Diff}} 
    & \multicolumn{1}{c}{\rotatebox[origin=l]{90}{VulSeeker}} 

    &
    & \multicolumn{1}{c}{\rotatebox[origin=l]{90}{InnerEye}} 
    & \multicolumn{1}{c}{\rotatebox[origin=l]{90}{Asm2Vec}} 
    & \multicolumn{1}{c}{\rotatebox[origin=l]{90}{SAFE}} 
    & \multicolumn{1}{c}{\rotatebox[origin=l]{90}{BAR19i}} 
    & \multicolumn{1}{c}{\rotatebox[origin=l]{90}{BAR19ii}} 
    & \multicolumn{1}{c}{\rotatebox[origin=l]{90}{FuncNet}} 

    &
    & \multicolumn{1}{c}{\rotatebox[origin=l]{90}{DeepBinDiff}} 
    & \multicolumn{1}{c}{\rotatebox[origin=l]{90}{ImOpt}} 
    & \multicolumn{1}{c}{\rotatebox[origin=l]{90}{ACCESS20}} 
    & \multicolumn{1}{c}{\rotatebox[origin=l]{90}{Patchecko}} %
    & \multicolumn{1}{c}{\rotatebox[origin=l]{90}{\sys}} 
    \\

    &
    &
    & \tiny{\cite{pewny2014leveraging}} 
    & \tiny{\cite{david2014tracelet}} 
    & \tiny{\cite{luo2014semantics}} 
    & \tiny{\cite{zhang2014program}} 
    & \tiny{\cite{egele2014blanket}} 
    & \tiny{\cite{farhadi:2014}} 

    &
    & \tiny{\cite{pewny2015cross}} 

    &
    & \tiny{\cite{eschweiler2016discovre}} 
    & \tiny{\cite{feng2016scalable}} 
    & \tiny{\cite{david2016statistical}} 
    & \tiny{\cite{chandramohan2016bingo}} 
    & \tiny{\cite{hu2016cross}} 
    & \tiny{\cite{ding2016kam1n0}} 
    & \tiny{\cite{lageman2016bindnn}} 

    &
    & \tiny{\cite{nouh2017binsign}} 

    & \tiny{\cite{feng2017extracting}} 
    & \tiny{\cite{xu2017neural}} 
    & \tiny{\cite{david2017similarity}} 
    & \tiny{\cite{ming2017binsim}} 
    & \tiny{\cite{huang2017binsequence}} 
    & \tiny{\cite{wang2017memory}} 
    & \tiny{\cite{hu2017binary}} 
    & \tiny{\cite{kargen2017towards}} 

    &
    & \tiny{\cite{shirani2018binarm}} 
    & \tiny{\cite{karamitas2018efficient}} 
    & \tiny{\cite{xue2018accurate}} 
    & \tiny{\cite{yuan2018new}} 
    & \tiny{\cite{hu2018binmatch}} 
    & \tiny{\cite{marastoni2018deep}} 
    & \tiny{\cite{shalev2018binary}} 

    & \tiny{\cite{david2018firmup}} 
    & \tiny{\cite{liu2018alphadiff}} 
    & \tiny{\cite{gao2018vulseeker}} 

    &
    & \tiny{\cite{zuo2019neural}} 
    & \tiny{\cite{ding2019asm2vec}} 
    & \tiny{\cite{massarelli2019safe}} 
    & \tiny{\cite{redmond2019cross}} 
    & \tiny{\cite{massarelli2019investigating}} 
    & \tiny{\cite{luo2019funcnet}} 

    &
    & \tiny{\cite{duan2020deepbindiff}} 
    & \tiny{\cite{jiang2020similarity}} 
    & \tiny{\cite{guo2020lightweight}} 
    & \tiny{\cite{sun2020hybrid}} 
    & $\bigstar$
    \\

    \midrule

    \multirow{6}[1]{*}{\rotatebox[origin=c]{90}{\textbf{Presemantic}}}
    & BB-level Numbers
    &
    & $\cdot$ & $\cdot$ & $\cdot$ & $\cdot$ & $\cdot$ & $\cdot$

    &
    & $\cdot$

    &
    & \Circle & \RIGHTcircle & $\cdot$ & $\cdot$ & $\cdot$ & $\cdot$ & $\cdot$

    &
    & $\cdot$ & $\cdot$ & \RIGHTcircle & $\cdot$ & $\cdot$ & $\cdot$ & $\cdot$ & $\cdot$
    & $\cdot$

    &
    & \Circle & $\cdot$ & $\cdot$ & $\cdot$ & $\cdot$
    & $\cdot$ & $\cdot$
    & $\cdot$ & $\cdot$ & \RIGHTcircle

    &
    & $\cdot$
    & $\cdot$
    & $\cdot$
    & $\cdot$
    & $\cdot$
    & $\RIGHTcircle$

    &
    & $\cdot$
    & $\cdot$
    & $\cdot$
    & $\RIGHTcircle$
    & $\Circle$

    \\

%
%
%
%
%

%
%
%
%
%

%
%
%
%
%
%

    & CFG-level Numbers
    &
    & $\cdot$ & $\cdot$ & $\cdot$ & $\cdot$ & $\cdot$ & \Circle

    &
    & $\cdot$

    &
    & \Circle & \RIGHTcircle & $\cdot$ & $\cdot$ & $\cdot$ & $\cdot$ & $\cdot$

    &
    & \Circle & $\cdot$ & \RIGHTcircle & $\cdot$ & $\cdot$ & \Circle & $\cdot$ & $\cdot$
    & $\cdot$

    &
    & \Circle & \Circle & \Circle & $\cdot$ & $\cdot$
    & $\cdot$ & $\cdot$
    & $\cdot$ & $\cdot$ & $\RIGHTcircle$

    &
    & $\cdot$
    & $\cdot$
    & $\cdot$
    & $\cdot$
    & $\cdot$
    & $\RIGHTcircle$

    &
    & $\cdot$
    & $\cdot$
    & $\cdot$
    & $\RIGHTcircle$
    & $\Circle$
    \\

    & CG-level Numbers
    &
    & $\cdot$ & $\cdot$ & $\cdot$ & $\cdot$ & $\cdot$ & $\cdot$

    &
    & $\cdot$

    &
    & \Circle & $\cdot$ & $\cdot$ & \Circle & $\cdot$ & $\cdot$ & $\cdot$

    &
    & \Circle & $\cdot$ & $\cdot$ & $\cdot$ & $\cdot$ & $\cdot$ & $\cdot$ & $\cdot$
    & $\cdot$

    &
    & \Circle & \Circle & \Circle & $\cdot$ & $\cdot$
    & $\cdot$ & $\cdot$
    & $\cdot$ & \Circle & $\cdot$

    &
    & $\cdot$
    & \RIGHTcircle
    & $\cdot$
    & $\cdot$
    & $\cdot$
    & $\cdot$

    &
    & $\cdot$
    & $\cdot$
    & $\cdot$
    & $\RIGHTcircle$
    & $\Circle$
    \\
    \cmidrule(){2-53}

    & Raw Bytes
    &
    & $\cdot$ & $\cdot$ & $\cdot$ & $\cdot$ & $\cdot$ & $\cdot$

    &
    & $\cdot$

    &
    & $\cdot$ & $\cdot$ & $\cdot$ & $\cdot$ & $\cdot$ & $\cdot$ & $\cdot$

    &
    & $\cdot$ & $\cdot$ & $\cdot$ & $\cdot$ & $\cdot$ & $\cdot$ & $\cdot$ & $\cdot$
    & $\cdot$

    &
    & $\cdot$ & $\cdot$ & $\cdot$ & $\cdot$ & $\cdot$
    & \RIGHTcircle & $\cdot$
    & $\cdot$ & \RIGHTcircle & $\cdot$

    &
    & $\cdot$
    & $\cdot$
    & $\cdot$
    & $\cdot$
    & $\cdot$
    & $\cdot$

    &
    & $\cdot$
    & $\cdot$
    & \Circle
    & $\cdot$
    & $\cdot$
    \\

    & Instructions
    &
    & \Circle & \Circle & $\cdot$ & $\cdot$ & $\cdot$ & \Circle

    &
    & $\cdot$

    &
    & $\cdot$ & $\cdot$ & $\cdot$ & \Circle & $\cdot$ & \Circle & \Circle

    &
    & $\cdot$ & $\cdot$ & $\cdot$ & $\cdot$ & $\cdot$ & \Circle & $\cdot$ & $\cdot$
    & $\cdot$

    &
    & $\cdot$ & $\cdot$ & \Circle & $\cdot$ & $\cdot$
    & $\cdot$ & $\cdot$
    & $\cdot$ & $\cdot$ & $\cdot$

    &
    & \RIGHTcircle
    & \RIGHTcircle
    & \RIGHTcircle
    & \RIGHTcircle
    & \RIGHTcircle
    & $\cdot$

    &
    & \RIGHTcircle
    & \RIGHTcircle
    & $\cdot$
    & $\cdot$
    & $\cdot$
    \\

    & Functions
    &
    & $\cdot$ & $\cdot$ & $\cdot$ & $\cdot$ & $\cdot$ & $\cdot$

    &
    & $\cdot$

    &
    & $\cdot$ & $\cdot$ & $\cdot$ & $\cdot$ & $\cdot$ & $\cdot$ & $\cdot$

    &
    & \Circle & $\cdot$ & $\cdot$ & $\cdot$ & $\cdot$ & $\cdot$ & $\cdot$ & $\cdot$
    & $\cdot$

    &
    & $\cdot$ & $\cdot$ & $\cdot$ & $\cdot$ & $\cdot$
    & $\cdot$ & $\cdot$
    & $\cdot$ & \Circle & $\cdot$

    &
    & $\cdot$
    & $\cdot$
    & $\cdot$
    & $\cdot$
    & $\cdot$
    & $\cdot$

    &
    & $\cdot$
    & $\cdot$
    & $\cdot$
    & $\cdot$
    & $\cdot$
    \\

    \midrule

    \multirow{7}[4]{*}{\rotatebox[origin=c]{90}{\textbf{Semantic}}}

    & Symbolic Constraints
    &
    & $\cdot$ & $\cdot$ & \Circle & \Circle & $\cdot$ & $\cdot$

    &
    & $\cdot$

    &
    & $\cdot$ & $\cdot$ & \Circle & $\cdot$ & $\cdot$ & $\cdot$ & $\cdot$

    &
    & $\cdot$ & \Circle & $\cdot$ & $\cdot$ & \Circle & $\cdot$ & $\cdot$ & $\cdot$
    & $\cdot$

    &
    & $\cdot$ & $\cdot$ & $\cdot$ & $\cdot$ & $\cdot$
    & $\cdot$ & $\cdot$
    & $\cdot$ & $\cdot$ & $\cdot$

    &
    & $\cdot$
    & $\cdot$
    & $\cdot$
    & $\cdot$
    & $\cdot$
    & $\cdot$

    &
    & $\cdot$
    & $\cdot$
    & $\cdot$
    & $\cdot$
    & $\cdot$
    \\

    & I/O Samples
    &
    & $\cdot$ & $\cdot$ & $\cdot$ & \Circle & $\cdot$ & $\cdot$

    &
    & \Circle

    &
    & $\cdot$ & $\cdot$ & $\cdot$ & \Circle & $\cdot$ & $\cdot$ & $\cdot$

    &
    & $\cdot$ & $\cdot$ & $\cdot$ & $\cdot$ & $\cdot$ & $\cdot$ & $\cdot$ & $\cdot$
    & $\cdot$

    &
    & $\cdot$ & $\cdot$ & \Circle & $\cdot$ & $\cdot$
    & $\cdot$ & $\cdot$
    & $\cdot$ & $\cdot$ & $\cdot$

    &
    & $\cdot$
    & $\cdot$
    & $\cdot$
    & $\cdot$
    & $\cdot$
    & $\cdot$

    &
    & $\cdot$
    & $\cdot$
    & $\cdot$
    & $\cdot$
    & $\cdot$
    \\

    & Runtime Behavior
    &
    & $\cdot$ & $\cdot$ & $\cdot$ & $\cdot$ & \Circle & $\cdot$

    &
    & $\cdot$

    &
    & $\cdot$ & $\cdot$ & $\cdot$ & $\cdot$ & \Circle & $\cdot$ & $\cdot$

    &
    & $\cdot$ & $\cdot$ & $\cdot$ & $\cdot$ & \Circle & $\cdot$ & \Circle & \Circle
    & \Circle

    &
    & $\cdot$ & $\cdot$ & \Circle & \Circle & \Circle
    & $\cdot$ & $\cdot$
    & $\cdot$ & $\cdot$ & $\cdot$

    &
    & $\cdot$
    & $\cdot$
    & $\cdot$
    & $\cdot$
    & $\cdot$
    & $\cdot$

    &
    & $\cdot$
    & $\cdot$
    & $\cdot$
    & $\RIGHTcircle$
    & $\cdot$
    \\

    & Manual Annotation
    &
    & $\cdot$ & $\cdot$ & $\cdot$ & $\cdot$ & $\cdot$ & $\cdot$

    &
    & $\cdot$

    &
    & $\cdot$ & $\cdot$ & $\cdot$ & \Circle & $\cdot$ & $\cdot$ & $\cdot$

    &
    & \Circle & $\cdot$ & $\cdot$ & $\cdot$ & \Circle & $\cdot$ & $\cdot$ & $\cdot$
    & $\cdot$

    &
    & $\cdot$ & $\cdot$ & \Circle & $\cdot$ & $\cdot$
    & $\cdot$ & $\cdot$
    & $\cdot$ & $\cdot$ & $\cdot$

    &
    & $\cdot$
    & $\cdot$
    & $\cdot$
    & $\cdot$
    & $\cdot$
    & $\cdot$

    &
    & $\cdot$
    & $\cdot$
    & $\cdot$
    & $\cdot$
    & $\cdot$
    \\

    & Program Slices, PDG
    &
    & $\cdot$ & $\cdot$ & $\cdot$ & $\cdot$ & $\cdot$ & $\cdot$

    &
    & $\cdot$

    &
    & $\cdot$ & $\cdot$ & $\cdot$ & $\cdot$ & $\cdot$ & $\cdot$ & $\cdot$

    &
    & $\cdot$ & $\cdot$ & $\cdot$ & \Circle & $\cdot$ & $\cdot$ & $\cdot$ & $\cdot$
    & $\cdot$

    &
    & $\cdot$ & $\cdot$ & $\cdot$ & $\cdot$ & $\cdot$
    & $\cdot$ & \Circle
    & \Circle & $\cdot$ & \RIGHTcircle

    &
    & $\cdot$
    & $\cdot$
    & $\cdot$
    & $\cdot$
    & $\cdot$
    & $\cdot$

    &
    & $\cdot$
    & $\cdot$
    & $\cdot$
    & $\cdot$
    & $\cdot$
    \\

    & Recovered Variables
    &
    & $\cdot$ & $\cdot$ & $\cdot$ & $\cdot$ & $\cdot$ & $\cdot$

    &
    & $\cdot$

    &
    & \Circle & \RIGHTcircle & $\cdot$ & $\cdot$ & $\cdot$ & $\cdot$ & $\cdot$

    &
    & \Circle & $\cdot$ & \RIGHTcircle & $\cdot$ & $\cdot$ & $\cdot$ & $\cdot$ & $\cdot$
    & $\cdot$

    &
    & \Circle & \Circle  & \Circle  & $\cdot$ & $\cdot$
    & $\cdot$ & $\cdot$
    & $\cdot$ & $\cdot$ & $\cdot$

    &
    & $\cdot$
    & $\cdot$
    & $\cdot$
    & $\cdot$
    & $\cdot$
    & \RIGHTcircle

    &
    & $\cdot$
    & $\cdot$
    & $\cdot$
    & $\RIGHTcircle$
    & $\cdot$
    \\

    & Embedded Vector
    &
    & $\cdot$ & $\cdot$ & $\cdot$ & $\cdot$ & $\cdot$ & $\cdot$

    &
    & $\cdot$

    &
    & $\cdot$ & \Circle & $\cdot$ & $\cdot$ & $\cdot$ & $\cdot$ & $\cdot$

    &
    & $\cdot$ & $\cdot$ & \Circle & $\cdot$ & $\cdot$ & $\cdot$ & $\cdot$ & $\cdot$
    & $\cdot$

    &
    & $\cdot$ & $\cdot$ & $\cdot$ & $\cdot$ & $\cdot$
    & \Circle & $\cdot$
    & $\cdot$ & \Circle & \Circle

    &
    & \Circle
    & \Circle
    & \Circle
    & \Circle
    & \Circle
    & \Circle

    &
    & \Circle
    & \Circle
    & $\cdot$
    & \Circle
    & $\cdot$
    \\

    \bottomrule
  \end{tabular}
  
    \begin{tablenotes}
    \item [\RIGHTcircle] This mark denotes a feature that is not directly
        used for similarity comparison but is required for extracting other features used in post-processing.
    \end{tablenotes}
  \end{threeparttable}
  \vspace{-0.2in}
\end{table*}


\PP{Numeric presemantic features}
Counting the occurrences of a particular property of a program is common in BCSA
as such numbers can be directly used as a numeric vector in the similarity
comparison step (S4).
We categorize numeric presemantic features into
three groups based on the granularity of the information required for extracting
them.

First, many researchers extract numeric features from each basic block of a
target code snippet~\cite{shirani2017binshape, shirani2018binarm,eschweiler2016discovre, sun2020hybrid,feng2016scalable, xu2017neural, gao2018vulseeker, sun2020hybrid}.
One may measure the frequency of raw opcodes
(mnemonics)~\cite{shirani2017binshape, shirani2018binarm} or grouped
instructions based on their functionalities (e.g., arithmetic, logical, or control
transfer)~\cite{eschweiler2016discovre, sun2020hybrid}.
This numeric form can also be post-processed through machine
learning~\cite{feng2016scalable, xu2017neural, gao2018vulseeker, sun2020hybrid},
as we further discuss in~\autoref{ss:semantic}.

%
%
%

Similarly, numeric features can be extracted from a CFG as well. CFG-level
numeric features can also reflect structural information that underlies a CFG.
%
For example, a function can be encoded into a numeric vector, which consists of
the number of nodes (i.e., basic blocks) and edges (i.e., control flow), as well
as grouped instructions in its CFG~\cite{eschweiler2016discovre,
nouh2017binsign, sun2020hybrid}.
One may extend such numeric vectors by adding extra features such as the number
of successive nodes or the betweenness centrality of a CFG~\cite{feng2016scalable,
xu2017neural, sun2020hybrid}.
The concept of 3D-CFG~\cite{chen2014achieving}, which places each node in a CFG
onto a 3D space, can be utilized as well.
Here, the distances among the centroids of two 3D-CFGs can represent their
similarity score~\cite{xue2018accurate}.
Other numeric features can be the graph energy, skewness, or cyclomatic
complexity of a CFG~\cite{shirani2017binshape, shirani2018binarm, sun2020hybrid}.
Even loops in a CFG can be converted into numeric features by counting the
number of loop headers and tails, as well as the number of forward and backward
edges~\cite{karamitas2018efficient}.

Finally, previous approaches utilize numeric features obtained from CGs. We
refer to them as CG-level numeric features.
Most of these approaches measure the number of callers and callees in a
CG~\cite{hu:ccs:2009, eschweiler2016discovre, feng2016scalable,
shirani2017binshape, shirani2018binarm, karamitas2018efficient,
liu2018alphadiff, sun2020hybrid}.
When extracting these features, one can selectively apply an inter-procedural
analysis using the ratio of the in-/out- degrees of the internal callees in the
same binary and the external callees of imported
libraries~\cite{chandramohan2016bingo, xue2018accurate, ding2019asm2vec,
sun2020hybrid}.
This is similar to the coupling concept~\cite{henry1981software}, which analyzes
the inter-dependence between software modules.
The extracted features can also be post-processed using machine
learning~\cite{liu2018alphadiff}.
%
%


\PP{Non-numeric presemantic features}
Program properties can also be directly used as a feature.
%
The most straightforward approach involves directly comparing the raw bytes of
binaries~\cite{jang2011bitshred, jang2012redebug, guo2020lightweight}. However,
people tend to not consider this approach because byte-level matching is not
as robust compared to simple code modifications.
For example, anti-malware applications typically make use of manually written
signatures using regular expressions to capture similar, but syntactically
different malware instances~\cite{cha:nsdi:2010}.
Recent approaches have attempted to extract semantic meanings from raw binary
code by utilizing a deep neural network (DNN) to build a feature vector
representation~\cite{liu2018alphadiff, marastoni2018deep}.


Another straightforward approach involves considering the opcodes and operands
of assembly instructions or their intermediate
representations~\cite{khoo2013rendezvous, xue2018accurate}.
Researchers often normalize operands~\cite{huang2017binsequence,
  david2014tracelet, ding2016kam1n0} because their actual values can
significantly vary across different compiler options.
Recent approaches~\cite{david2017similarity, jiang2020similarity} have also
applied re-optimization techniques~\cite{schkufza2013stochastic} for the same
reason.
To compute a similarity score, one can measure the number of matched elements or
the Jaccard distance~\cite{chandramohan2016bingo} between matched groups, within
a comparison unit such as a sliding window~\cite{farhadi:2014}, basic
block~\cite{huang2017binsequence}, or tracelet~\cite{david2014tracelet}.
Here, a tracelet denotes a series of basic blocks.
Although these approaches take different comparison units, one may adjust their
results to compare two procedures, or to find the longest common
subsequence~\cite{huang2017binsequence, ding2016kam1n0} within procedures.
If one converts assembly instructions to a static single assignment (SSA) form,
s/he can compute the tree edit distance between the SSA expression trees as a
similarity score~\cite{pewny2014leveraging}.
Recent approaches have proposed applying popular techniques in natural language
processing (NLP) to represent an assembly instruction or a basic block as an
embedded vector, reflecting their underlying semantics~\cite{zuo2019neural,
ding2019asm2vec, massarelli2019safe, redmond2019cross, duan2020deepbindiff,
massarelli2019investigating}.

Finally, some features can be directly extracted from functions.
These features may include the names of imported functions, and the intersection
of two inputs can show their similarity~\cite{nouh2017binsign,
liu2018alphadiff}. Note that these features can collaborate with other features
as well.
\subsubsection{Semantic Features}
\label{ss:semantic}

We call the features that we can obtain from the semantic analysis phase (S3)
\emph{semantic features}. To obtain semantic features, a complex analysis, such
as symbolic execution~\cite{\paperSymbolic}, dynamic evaluation of code
snippets~\cite{\paperDynamic}, or machine learning-based
embedding~\cite{\paperML} is necessary. There are mainly seven distinct semantic
features used in the previous literature, as listed
in~\autoref{tab:prevfeatures}. It is common to use multiple semantic features
together or combine them with presemantic features.

First, one straightforward method to represent the semantics of a given code
snippet is to use symbolic constraints.
%
The symbolic constraints could express the output variables or states of a basic
block~\cite{luo2014semantics}, a program slice~\cite{david2016statistical,
feng2017extracting, ming2017binsim}, or a path~\cite{zhang2014program,
ming2016deviation, luo2017semantics}.
Therefore, after extracting the symbolic constraints from a target comparison
unit, one can compare them using an SMT solver.

%



Second, one may represent code semantics using I/O
samples~\cite{zhang2014program, chandramohan2016bingo, xue2018accurate,
pewny2015cross}.
The key intuition here is that two identical code snippets produce consistent
I/O samples, and directly comparing them would be time-efficient.
One can generate I/O samples by providing random inputs~\cite{zhang2014program,
pewny2015cross} to a code snippet, or by applying an SMT solver to the symbolic
constraints of the code snippet~\cite{chandramohan2016bingo, xue2018accurate}.
One can also use inter-procedural analysis to precisely model I/O samples if
the target code includes a function call~\cite{chandramohan2016bingo,
xue2018accurate}.


Third, the runtime behavior of a code snippet can directly express its
semantics, as presented by traditional malware analysis~\cite{forrest:1996}.
By executing two target functions with the same execution environment, one can
directly compare the executed instruction sequences~\cite{kargen2017towards} or
visited CFG edges of the target functions~\cite{yuan2018new}.
For comparison, one may focus on specific behaviors observed during the
execution~\cite{egele2014blanket, wang2017memory, hu2017binary, xue2018accurate,
hu2018binmatch, tian2020plagiarism, sun2020hybrid}: the read/write values of
stack and heap memory, return values from function calls, and invoked
system/library function calls during the executions.
To extract such features, one may adopt fuzzing~\cite{wang2017memory,
  manes:tse:2019}, or an emulation-based approach~\cite{hu2018binmatch}.
%
Moreover, one can further check the call names, parameters, or call sequences
for system calls~\cite{hu2016cross, hu2017binary, hu2018binmatch,
ming2017binsim, xue2018accurate}.

The next category is to manually annotate the high-level semantics of a program
or function.
One may categorize library functions by their high-level functionality, such as
whether the function manipulates strings or whether it handles heap
memory~\cite{chandramohan2016bingo, xue2018accurate, nouh2017binsign}.
Annotating cryptographic functions in a target code
snippet~\cite{grobert2011automated} is also helpful because its complex
operations hinder analyzing the symbolic constraints or behavior of the
code~\cite{ming2017binsim}.


The fifth category is extracting features from a program
slice~\cite{horwitz:1988}, because they can represent its data-flow semantics in
an abstract form.
Specifically, one can slice a program into a set of
strands~\cite{david2017similarity, david2018firmup}. Here, a strand is a series
of instructions within the same data flow, which can be obtained from backward
slicing.
Next, these strands can be canonicalized, normalized, or re-optimized for
precise comparison~\cite{david2017similarity, david2018firmup}. Additionally,
one may hash strands for quick comparison~\cite{shalev2018binary} or extract
symbolic constraints from the strands~\cite{david2016statistical}.
One may also extract features from a program dependence graph
(PDG)~\cite{ferrante:1987}, which is essentially a combination of a data-flow
graph and CFG, to represent the convoluted semantics of the target code, including
its structural information~\cite{gao2018vulseeker}.


\begin{table*}[ht!]
  \caption{Summary of the datasets used in previous studies.}
  \label{tab:prevdataset}
  \vspace{-0.1in}
  \centering
  \scriptsize
  \setlength\tabcolsep{0.06cm}
  \def\arraystretch{0.50}
  \begin{threeparttable}
    \begin{tabular}{
  c 
  @{\hspace{0.1cm}}
  r 
  *{2}{c} 
  c
  *{8}{c} 
  c
  *{5}{c} 
  c
  *{6}{c} 
  *{5}{c} 
  c 
  c 
  c
  *{4}{c} 
  c
  *{3}{c} 
}
  \toprule
  \multicolumn{1}{c}{}&
  \multicolumn{1}{c}{}&
  \multicolumn{2}{c}{\textbf{\#Binaries}\tnote{$\ast$}}&
  \multicolumn{1}{c}{}&
  \multicolumn{8}{c}{\rotatebox[origin=l]{0}{\textbf{Architecture}}}&
  \multicolumn{1}{c}{}&
  \multicolumn{5}{c}{\rotatebox[origin=l]{0}{\textbf{Optimization}}}&
  \multicolumn{1}{c}{}&
  \multicolumn{13}{c}{\textbf{Compiler}\tnote{$\dagger$}}&  \multicolumn{1}{c}{}&
  \multicolumn{4}{c}{\textbf{Extra}} &
  \multicolumn{1}{c}{}&
  \multicolumn{3}{c}{\textbf{Info.}}
  \\
  \cmidrule{3-4}    
  \cmidrule{6-13}   
  \cmidrule{15-19}  
  \cmidrule{21-33}  
  \cmidrule(l){34-38}  
  \cmidrule(l){39-42}  

  \multicolumn{1}{c}{\rotatebox[origin=l]{0}{\textbf{Year}}}&
  \multicolumn{1}{r}{\rotatebox[origin=l]{0}{\textbf{Tool [Paper]}}}&
  \multicolumn{1}{c}{\rotatebox[origin=l]{90}{ \textbf{Packages}}}&
  \multicolumn{1}{c}{\rotatebox[origin=l]{90}{ \textbf{Firmware}}}&
  \multicolumn{1}{c}{\rotatebox[origin=l]{0}{}}&
  \multicolumn{1}{c}{\rotatebox[origin=l]{90}{ \textbf{x86}}}&
  \multicolumn{1}{c}{\rotatebox[origin=l]{90}{ \textbf{x64}}}&
  \multicolumn{1}{c}{\rotatebox[origin=l]{90}{ \textbf{arm}}}&
  \multicolumn{1}{c}{\rotatebox[origin=l]{90}{ \textbf{aarch64}}}&
  \multicolumn{1}{c}{\rotatebox[origin=l]{90}{ \textbf{mips}}}&
  \multicolumn{1}{c}{\rotatebox[origin=l]{90}{ \textbf{mips64}}}&
  \multicolumn{1}{c}{\rotatebox[origin=l]{90}{ \textbf{mipseb}}}&
  \multicolumn{1}{c}{\rotatebox[origin=l]{90}{ \textbf{mips64eb}}}&
  \multicolumn{1}{c}{\rotatebox[origin=l]{0}{}}&
  \multicolumn{1}{c}{\rotatebox[origin=l]{90}{ \textbf{O0}}}&
  \multicolumn{1}{c}{\rotatebox[origin=l]{90}{ \textbf{O1}}}&
  \multicolumn{1}{c}{\rotatebox[origin=l]{90}{ \textbf{O2}}}&
  \multicolumn{1}{c}{\rotatebox[origin=l]{90}{ \textbf{O3}}}&
  \multicolumn{1}{c}{\rotatebox[origin=l]{90}{ \textbf{Os}}}&
  \multicolumn{1}{c}{\rotatebox[origin=l]{0}{}}&
  \multicolumn{1}{c}{\rotatebox[origin=l]{90}{ \textbf{GCC 3}}}&
  \multicolumn{1}{c}{\rotatebox[origin=l]{90}{ \textbf{GCC 4}}}&
  \multicolumn{1}{c}{\rotatebox[origin=l]{90}{ \textbf{GCC 5}}}&
  \multicolumn{1}{c}{\rotatebox[origin=l]{90}{ \textbf{GCC 6}}}&
  \multicolumn{1}{c}{\rotatebox[origin=l]{90}{ \textbf{GCC 7}}}&
  \multicolumn{1}{c}{\rotatebox[origin=l]{90}{ \textbf{GCC 8}}}&
  \multicolumn{1}{c}{\rotatebox[origin=l]{90}{ \textbf{Clang 3}}}&
  \multicolumn{1}{c}{\rotatebox[origin=l]{90}{ \textbf{Clang 4}}}&
  \multicolumn{1}{c}{\rotatebox[origin=l]{90}{ \textbf{Clang 5}}}&
  \multicolumn{1}{c}{\rotatebox[origin=l]{90}{ \textbf{Clang 6}}}&
  \multicolumn{1}{c}{\rotatebox[origin=l]{90}{ \textbf{Clang 7}}}&
  \multicolumn{1}{c}{\rotatebox[origin=l]{90}{ \textbf{misc.}}}&
  \multicolumn{1}{c}{\rotatebox[origin=l]{90}{ \textbf{Total \#}}}&
  \multicolumn{1}{c}{\rotatebox[origin=l]{0}{}}&
  \multicolumn{1}{c}{\rotatebox[origin=l]{90}{ \textbf{Noinline}}}&
  \multicolumn{1}{c}{\rotatebox[origin=l]{90}{ \textbf{PIE}}} &
  \multicolumn{1}{c}{\rotatebox[origin=l]{90}{ \textbf{LTO}}} &
  \multicolumn{1}{c}{\rotatebox[origin=l]{90}{ \textbf{Obfus.}}} &
  \multicolumn{1}{c}{\rotatebox[origin=l]{0}{}} &
  \multicolumn{1}{c}{\rotatebox[origin=l]{90}{ \textbf{Code}}} &
  \multicolumn{1}{c}{\rotatebox[origin=l]{90}{ \textbf{Dataset}}} &
  \multicolumn{1}{c}{\rotatebox[origin=l]{90}{ \textbf{IDA}}} \\
  \midrule

  \multirow{6}{*}{2014} &
  TEDEM~\cite{pewny2014leveraging} &
  14 & 
  $\cdot$ & 
  &
  \Circle & 
  $\cdot$ & 
  $\cdot$ & 
  $\cdot$ & 
  $\cdot$ & 
  $\cdot$ & 
  $\cdot$ & 
  $\cdot$ & 
  &
  $\cdot$ & 
  $\cdot$ & 
  $\cdot$ & 
  $\cdot$ & 
  $\cdot$ & 
  &
  $\cdot$ & 
  $\cdot$ & 
  $\cdot$ & 
  $\cdot$ & 
  $\cdot$ & 
  $\cdot$ & 
  $\cdot$ & 
  $\cdot$ & 
  $\cdot$ & 
  $\cdot$ & 
  $\cdot$ & 
  $\cdot$ & 
  $\cdot$ & 
  &
  $\cdot$ & 
  $\cdot$ & 
  $\cdot$ & 
  $\cdot$ & 
  &
  $\cdot$ & 
  $\cdot$ & 
  \Circle 
  \\ 

  &
  Tracy~\cite{david2014tracelet} &
  (115) & 
  $\cdot$ & 
  &
  $\triangle$ & 
  $\triangle$ & 
  $\cdot$ & 
  $\cdot$ & 
  $\cdot$ & 
  $\cdot$ & 
  $\cdot$ & 
  $\cdot$ & 
  &
  $\cdot$ & 
  $\cdot$ & 
  $\cdot$ & 
  $\cdot$ & 
  $\cdot$ & 
  &
  $\cdot$ & 
  $\cdot$ & 
  $\cdot$ & 
  $\cdot$ & 
  $\cdot$ & 
  $\cdot$ & 
  $\cdot$ & 
  $\cdot$ & 
  $\cdot$ & 
  $\cdot$ & 
  $\cdot$ & 
  $\cdot$ & 
  $\cdot$ & 
  &
  $\cdot$ & 
  $\cdot$ & 
  $\cdot$ & 
  $\cdot$ & 
  &
  \Circle & 
  $\cdot$ & 
  \Circle 
  \\ 

  &
  CoP~\cite{luo2014semantics} &
  (214) & 
  $\cdot$ & 
  &
  $\triangle$ & 
  $\cdot$ & 
  $\cdot$ & 
  $\cdot$ & 
  $\cdot$ & 
  $\cdot$ & 
  $\cdot$ & 
  $\cdot$ & 
  &
  \Circle & 
  \Circle & 
  \Circle & 
  \Circle & 
  \Circle & 
  &
  $\cdot$ & 
  1 & 
  $\cdot$ & 
  $\cdot$ & 
  $\cdot$ & 
  $\cdot$ & 
  $\cdot$ & 
  $\cdot$ & 
  $\cdot$ & 
  $\cdot$ & 
  $\cdot$ & 
  1 & 
  2 & 
  &
  $\cdot$ & 
  $\cdot$ & 
  $\cdot$ & 
  \Circle & 
  &
  $\cdot$ & 
  $\cdot$ & 
  \Circle 
  \\ 

  &
  LoPD~\cite{zhang2014program} &
  48 & 
  $\cdot$ & 
  &
  \Circle & 
  $\cdot$ & 
  $\cdot$ & 
  $\cdot$ & 
  $\cdot$ & 
  $\cdot$ & 
  $\cdot$ & 
  $\cdot$ & 
  &
  \Circle & 
  \Circle & 
  \Circle & 
  \Circle & 
  \Circle & 
  &
  $\cdot$ & 
  1 & 
  $\cdot$ & 
  $\cdot$ & 
  $\cdot$ & 
  $\cdot$ & 
  $\cdot$ & 
  $\cdot$ & 
  $\cdot$ & 
  $\cdot$ & 
  $\cdot$ & 
  1 & 
  2 & 
  &
  $\cdot$ & 
  $\cdot$ & 
  \Circle & 
  \Circle & 
  &
  $\cdot$ & 
  $\cdot$ & 
  $\cdot$ 
  \\ 

  &
  BLEX~\cite{egele2014blanket} &
  \num{1140} & 
  $\cdot$ & 
  &
  $\cdot$ & 
  \Circle & 
  $\cdot$ & 
  $\cdot$ & 
  $\cdot$ & 
  $\cdot$ & 
  $\cdot$ & 
  $\cdot$ & 
  &
  \Circle & 
  \Circle & 
  \Circle & 
  \Circle & 
  $\cdot$ & 
  &
  $\cdot$ & 
  1 & 
  $\cdot$ & 
  $\cdot$ & 
  $\cdot$ & 
  $\cdot$ & 
  1 & 
  $\cdot$ & 
  $\cdot$ & 
  $\cdot$ & 
  $\cdot$ & 
  1 & 
  3 & 
  &
  $\cdot$ & 
  $\cdot$ & 
  $\cdot$ & 
  $\cdot$ & 
  &
  $\cdot$ & 
  $\cdot$ & 
  \Circle 
  \\ 

  &
  BinClone~\cite{farhadi:2014} &
  90 & 
  $\cdot$ & 
  &
  $\triangle$ & 
  $\cdot$ & 
  $\cdot$ & 
  $\cdot$ & 
  $\cdot$ & 
  $\cdot$ & 
  $\cdot$ & 
  $\cdot$ & 
  &
  $\cdot$ & 
  $\cdot$ & 
  $\cdot$ & 
  $\cdot$ & 
  $\cdot$ & 
  &
  $\cdot$ & 
  $\cdot$ & 
  $\cdot$ & 
  $\cdot$ & 
  $\cdot$ & 
  $\cdot$ & 
  $\cdot$ & 
  $\cdot$ & 
  $\cdot$ & 
  $\cdot$ & 
  $\cdot$ & 
  $\cdot$ & 
  $\cdot$ & 
  &
  $\cdot$ & 
  $\cdot$ & 
  $\cdot$& 
  $\cdot$ & 
  &
  \Circle & 
  $\cdot$ & 
  \Circle 
  \\ \midrule

  2015 &
  Multi-k-MH~\cite{pewny2015cross} &
  60 & 
  6 & 
  &
  \Circle & 
  $\cdot$ & 
  \Circle & 
  $\cdot$ & 
  \Circle & 
  $\cdot$ & 
  $\cdot$ & 
  $\cdot$ & 
  &
  \Circle & 
  \Circle & 
  \Circle & 
  \Circle & 
  $\cdot$ & 
  &
  $\cdot$ & 
  2 & 
  $\cdot$ & 
  $\cdot$ & 
  $\cdot$ & 
  $\cdot$ & 
  1 & 
  $\cdot$ & 
  $\cdot$ & 
  $\cdot$ & 
  $\cdot$ & 
  $\cdot$ & 
  3 & 
  &
  $\cdot$ & 
  $\cdot$ & 
  $\cdot$& 
  $\cdot$ & 
  &
  $\cdot$ & 
  $\cdot$ & 
  \Circle 
  \\ \midrule

  \multirow{7}{*}{\rotatebox[origin=c]{0}{2016}} &
  discovRE~\cite{eschweiler2016discovre} &
  593 & 
  3 & 
  &
  \Circle & 
  $\cdot$ & 
  \Circle & 
  $\cdot$ & 
  \Circle & 
  $\cdot$ & 
  $\cdot$ & 
  $\cdot$ & 
  &
  \Circle & 
  \Circle & 
  \Circle & 
  \Circle & 
  \Circle & 
  &
  $\cdot$ & 
  1 & 
  $\cdot$ & 
  $\cdot$ & 
  $\cdot$ & 
  $\cdot$ & 
  1 & 
  $\cdot$ & 
  $\cdot$ & 
  $\cdot$ & 
  $\cdot$ & 
  2 & 
  4 & 
  &
  \Circle & 
  $\cdot$ & 
  $\cdot$ & 
  $\cdot$ & 
  &
  $\cdot$ & 
  \RIGHTcircle & 
  \Circle 
  \\ 

  &
  Genius~\cite{feng2016scalable} &
  (\num{7848}) & 
  \num{8128} & 
  &
  \Circle & 
  $\cdot$ & 
  \Circle & 
  $\cdot$ & 
  \Circle & 
  $\cdot$ & 
  $\cdot$ & 
  $\cdot$ & 
  &
  \Circle & 
  \Circle & 
  \Circle & 
  \Circle & 
  $\cdot$ & 
  &
  $\cdot$ & 
  2 & 
  $\cdot$ & 
  $\cdot$ & 
  $\cdot$ & 
  $\cdot$ & 
  1 & 
  $\cdot$ & 
  $\cdot$ & 
  $\cdot$ & 
  $\cdot$ & 
  $\cdot$ & 
  3 & 
  &
  $\cdot$ & 
  $\cdot$ & 
  $\cdot$ & 
  $\cdot$ & 
  &
  $\cdot$ & 
  $\cdot$ & 
  \Circle 
  \\ 

  &
  Esh~\cite{david2016statistical} &
  (833) & 
  $\cdot$ & 
  &
  $\cdot$ & 
  \Circle & 
  $\cdot$ & 
  $\cdot$ & 
  $\cdot$ & 
  $\cdot$ & 
  $\cdot$ & 
  $\cdot$ & 
  &
  $\cdot$ & 
  $\cdot$ & 
  $\cdot$ & 
  $\cdot$ & 
  $\cdot$ & 
  &
  $\cdot$ & 
  3 & 
  $\cdot$ & 
  $\cdot$ & 
  $\cdot$ & 
  $\cdot$ & 
  2 & 
  $\cdot$ & 
  $\cdot$ & 
  $\cdot$ & 
  $\cdot$ & 
  2 & 
  7 & 
  &
  $\cdot$ & 
  $\cdot$ & 
  $\cdot$ & 
  $\cdot$ & 
  &
  \Circle & 
  \Circle & 
  \Circle 
  \\ 

  &
  BinGo~\cite{chandramohan2016bingo} &
  (\num{5143}) & 
  $\cdot$ & 
  &
  \Circle & 
  \Circle & 
  \Circle & 
  $\cdot$ & 
  $\cdot$ & 
  $\cdot$ & 
  $\cdot$ & 
  $\cdot$ & 
  &
  \Circle & 
  \Circle & 
  \Circle & 
  \Circle & 
  $\cdot$ & 
  &
  $\cdot$ & 
  3 & 
  $\cdot$ & 
  $\cdot$ & 
  $\cdot$ & 
  $\cdot$ & 
  1 & 
  $\cdot$ & 
  $\cdot$ & 
  $\cdot$ & 
  $\cdot$ & 
  1 & 
  5 & 
  &
  $\cdot$ & 
  $\cdot$ & 
  $\cdot$ & 
  $\cdot$ & 
  &
  $\cdot$ & 
  $\cdot$ & 
  \Circle 
  \\ 

  &
  MockingBird~\cite{hu2016cross} &
  80 & 
  $\cdot$ & 
  &
  \Circle & 
  $\cdot$ & 
  \Circle & 
  $\cdot$ & 
  \Circle & 
  $\cdot$ & 
  $\cdot$ & 
  $\cdot$ & 
  &
  \Circle & 
  $\cdot$ & 
  \Circle & 
  \Circle & 
  $\cdot$ & 
  &
  $\cdot$ & 
  1 & 
  $\cdot$ & 
  $\cdot$ & 
  $\cdot$ & 
  $\cdot$ & 
  1 & 
  $\cdot$ & 
  $\cdot$ & 
  $\cdot$ & 
  $\cdot$ & 
  $\cdot$ & 
  2 & 
  &
  $\cdot$ & 
  $\cdot$ & 
  $\cdot$ & 
  $\cdot$ & 
  &
  $\cdot$ & 
  $\cdot$ & 
  \Circle 
  \\ 

  &
  Kam1n0~\cite{ding2016kam1n0} &
  96 & 
  $\cdot$ & 
  &
  \Circle & 
  \Circle & 
  $\cdot$ & 
  $\cdot$ & 
  $\cdot$ & 
  $\cdot$ & 
  $\cdot$ & 
  $\cdot$ & 
  &
  $\cdot$ & 
  $\cdot$ & 
  $\cdot$ & 
  $\cdot$ & 
  $\cdot$ & 
  &
  $\cdot$ & 
  $\cdot$ & 
  $\cdot$ & 
  $\cdot$ & 
  $\cdot$ & 
  $\cdot$ & 
  $\cdot$ & 
  $\cdot$ & 
  $\cdot$ & 
  $\cdot$ & 
  $\cdot$ & 
  $\cdot$ & 
  $\cdot$ & 
  &
  $\cdot$ & 
  $\cdot$ & 
  $\cdot$& 
  $\cdot$ & 
  &
  \Circle & 
  $\cdot$ & 
  \Circle 
  \\ 

  &
  BinDNN~\cite{lageman2016bindnn} &
  (\num{2072}) & 
  $\cdot$ & 
  &
  \Circle & 
  \Circle & 
  \Circle & 
  $\cdot$ & 
  $\cdot$ & 
  $\cdot$ & 
  $\cdot$ & 
  $\cdot$ & 
  &
  \Circle & 
  \Circle & 
  \Circle & 
  \Circle & 
  $\cdot$ & 
  &
  $\cdot$ & 
  1 & 
  $\cdot$ & 
  $\cdot$ & 
  $\cdot$ & 
  $\cdot$ & 
  $\cdot$ & 
  $\cdot$ & 
  $\cdot$ & 
  $\cdot$ & 
  $\cdot$ & 
  1 & 
  2 & 
  &
  $\cdot$ & 
  $\cdot$ & 
  $\cdot$ & 
  $\cdot$ & 
  &
  $\cdot$ & 
  $\cdot$ & 
  \Circle 
  \\ \midrule

  \multirow{9}{*}{\rotatebox[origin=c]{0}{2017}} &
  BinSign~\cite{nouh2017binsign} &
  (31) & 
  $\cdot$ & 
  &
  $\triangle$ & 
  $\cdot$ & 
  $\cdot$ & 
  $\cdot$ & 
  $\cdot$ & 
  $\cdot$ & 
  $\cdot$ & 
  $\cdot$ & 
  &
  $\cdot$ & 
  $\cdot$ & 
  $\cdot$ & 
  $\cdot$ & 
  $\cdot$ & 
  &
  $\cdot$ & 
  $\cdot$ & 
  $\cdot$ & 
  $\cdot$ & 
  $\cdot$ & 
  $\cdot$ & 
  $\cdot$ & 
  $\cdot$ & 
  $\cdot$ & 
  $\cdot$ & 
  $\cdot$ & 
  2 & 
  2 & 
  &
  $\cdot$ & 
  $\cdot$ & 
  $\cdot$ & 
  \Circle & 
  &
  $\cdot$ & 
  $\cdot$ & 
  \Circle 
  \\ 

  &
  Xmatch~\cite{feng2017extracting} &
  72 & 
  1 & 
  &
  \Circle & 
  $\cdot$ & 
  $\cdot$ & 
  $\cdot$ & 
  \Circle & 
  $\cdot$ & 
  $\cdot$ & 
  $\cdot$ & 
  &
  $\cdot$ & 
  $\cdot$ & 
  $\cdot$ & 
  $\cdot$ & 
  $\cdot$ & 
  &
  $\cdot$ & 
  2 & 
  $\cdot$ & 
  $\cdot$ & 
  $\cdot$ & 
  $\cdot$ & 
  1 & 
  $\cdot$ & 
  $\cdot$ & 
  $\cdot$ & 
  $\cdot$ & 
  $\cdot$ & 
  3 & 
  &
  \Circle & 
  $\cdot$ & 
  $\cdot$ & 
  $\cdot$ & 
  &
  $\cdot$ & 
  $\cdot$ & 
  \Circle 
  \\ 

  &
  Gemini~\cite{xu2017neural} &
  \num{18269} & 
  \num{8128} & 
  &
  \Circle & 
  $\cdot$ & 
  \Circle & 
  $\cdot$ & 
  \Circle & 
  $\cdot$ & 
  $\cdot$ & 
  $\cdot$ & 
  &
  \Circle & 
  \Circle & 
  \Circle & 
  \Circle & 
  $\cdot$ & 
  &
  $\cdot$ & 
  $\cdot$ & 
  1 & 
  $\cdot$ & 
  $\cdot$ & 
  $\cdot$ & 
  $\cdot$ & 
  $\cdot$ & 
  $\cdot$ & 
  $\cdot$ & 
  $\cdot$ & 
  $\cdot$ & 
  1 & 
  &
  $\cdot$ & 
  $\cdot$ & 
  $\cdot$ & 
  $\cdot$ & 
  &
  \Circle & 
  $\cdot$ & 
  \Circle 
  \\ 

  &
  GitZ~\cite{david2017similarity} &
  44 & 
  $\cdot$ & 
  &
  $\cdot$ & 
  \Circle & 
  $\cdot$ & 
  \Circle & 
  $\cdot$ & 
  $\cdot$ & 
  $\cdot$ & 
  $\cdot$ & 
  &
  \Circle & 
  \Circle & 
  \Circle & 
  \Circle & 
  \Circle & 
  &
  $\cdot$ & 
  3 & 
  $\cdot$ & 
  $\cdot$ & 
  $\cdot$ & 
  $\cdot$ & 
  2 & 
  1 & 
  $\cdot$ & 
  $\cdot$ & 
  $\cdot$ & 
  2 & 
  8 & 
  &
  $\cdot$ & 
  $\cdot$ & 
  $\cdot$ & 
  $\cdot$ & 
  &
  $\cdot$ & 
  $\cdot$ & 
  $\cdot$ 
  \\ 

  &
  BinSim~\cite{ming2017binsim} &
  \num{1062} & 
  $\cdot$ & 
  &
  \Circle & 
  $\cdot$ & 
  $\cdot$ & 
  $\cdot$ & 
  $\cdot$ & 
  $\cdot$ & 
  $\cdot$ & 
  $\cdot$ & 
  &
  $\cdot$ & 
  $\cdot$ & 
  $\cdot$ & 
  $\cdot$ & 
  $\cdot$ & 
  &
  $\cdot$ & 
  $\cdot$ & 
  $\cdot$ & 
  $\cdot$ & 
  $\cdot$ & 
  $\cdot$ & 
  $\cdot$ & 
  $\cdot$ & 
  $\cdot$ & 
  $\cdot$ & 
  $\cdot$ & 
  $\cdot$ & 
  $\cdot$ & 
  &
  $\cdot$ & 
  $\cdot$ & 
  $\cdot$ & 
  \Circle & 
  &
  $\cdot$ & 
  $\cdot$ & 
  \Circle 
  \\ 

  &
  BinSequence~\cite{huang2017binsequence} &
  (\num{1718}) & 
  $\cdot$ & 
  &
  $\triangle$ & 
  $\cdot$ & 
  $\cdot$ & 
  $\cdot$ & 
  $\cdot$ & 
  $\cdot$ & 
  $\cdot$ & 
  $\cdot$ & 
  &
  $\cdot$ & 
  $\cdot$ & 
  $\cdot$ & 
  $\cdot$ & 
  $\cdot$ & 
  &
  $\cdot$ & 
  $\cdot$ & 
  $\cdot$ & 
  $\cdot$ & 
  $\cdot$ & 
  $\cdot$ & 
  $\cdot$ & 
  $\cdot$ & 
  $\cdot$ & 
  $\cdot$ & 
  $\cdot$ & 
  $\cdot$ & 
  $\cdot$ & 
  &
  $\cdot$ & 
  $\cdot$ & 
  $\cdot$ & 
  $\cdot$ & 
  &
  $\cdot$ & 
  $\cdot$ & 
  \Circle 
  \\ 

  &
  IMF-sim~\cite{wang2017memory} &
  \num{1140} & 
  $\cdot$ & 
  &
  $\cdot$ & 
  \Circle & 
  $\cdot$ & 
  $\cdot$ & 
  $\cdot$ & 
  $\cdot$ & 
  $\cdot$ & 
  $\cdot$ & 
  &
  \Circle & 
  $\cdot$ & 
  \Circle & 
  \Circle & 
  $\cdot$ & 
  &
  $\cdot$ & 
  1 & 
  $\cdot$ & 
  $\cdot$ & 
  $\cdot$ & 
  $\cdot$ & 
  1 & 
  $\cdot$ & 
  $\cdot$ & 
  $\cdot$ & 
  $\cdot$ & 
  1 & 
  3 & 
  &
  $\cdot$ & 
  $\cdot$ & 
  $\cdot$ & 
  \Circle & 
  &
  $\cdot$ & 
  $\cdot$ & 
  $\cdot$ 
  \\ 

  &
  CACompare~\cite{hu2017binary} &
  72 & 
  $\cdot$ & 
  &
  \Circle & 
  $\cdot$ & 
  \Circle & 
  $\cdot$ & 
  \Circle & 
  $\cdot$ & 
  $\cdot$ & 
  $\cdot$ & 
  &
  \Circle & 
  $\cdot$ & 
  \Circle & 
  \Circle & 
  $\cdot$ & 
  &
  $\cdot$ & 
  1 & 
  $\cdot$ & 
  $\cdot$ & 
  $\cdot$ & 
  $\cdot$ & 
  1 & 
  $\cdot$ & 
  $\cdot$ & 
  $\cdot$ & 
  $\cdot$ & 
  $\cdot$ & 
  2 & 
  &
  $\cdot$ & 
  $\cdot$ & 
  $\cdot$& 
  $\cdot$ & 
  &
  $\cdot$ & 
  $\cdot$ & 
  \Circle 
  \\

  &
  ASE17~\cite{kargen2017towards} &
  55 & 
  $\cdot$ & 
  &
  \Circle & 
  \Circle & 
  $\cdot$ & 
  $\cdot$ & 
  $\cdot$ & 
  $\cdot$ & 
  $\cdot$ & 
  $\cdot$ & 
  &
  \Circle & 
  $\cdot$ & 
  \Circle & 
  \Circle & 
  $\cdot$ & 
  &
  $\cdot$ & 
  1 & 
  $\cdot$ & 
  $\cdot$ & 
  $\cdot$ & 
  $\cdot$ & 
  1 & 
  $\cdot$ & 
  $\cdot$ & 
  $\cdot$ & 
  $\cdot$ & 
  $\cdot$ & 
  2 & 
  &
  $\cdot$ & 
  $\cdot$ & 
  $\cdot$& 
  \Circle & 
  &
  $\cdot$ & 
  $\cdot$ & 
  $\cdot$ 
  \\ \midrule

  \multirow{10}{*}{\rotatebox[origin=c]{0}{2018}} &
  BinArm~\cite{shirani2018binarm} &
  $\cdot$ & 
  \num{2628} & 
  &
  $\cdot$ & 
  $\cdot$ & 
  $\triangle$ & 
  $\cdot$ & 
  $\cdot$ & 
  $\cdot$ & 
  $\cdot$ & 
  $\cdot$ & 
  &
  $\cdot$ & 
  $\cdot$ & 
  $\cdot$ & 
  $\cdot$ & 
  $\cdot$ & 
  &
  $\cdot$ & 
  $\cdot$ & 
  $\cdot$ & 
  $\cdot$ & 
  $\cdot$ & 
  $\cdot$ & 
  $\cdot$ & 
  $\cdot$ & 
  $\cdot$ & 
  $\cdot$ & 
  $\cdot$ & 
  $\cdot$ & 
  $\cdot$ & 
  &
  $\cdot$ & 
  $\cdot$ & 
  $\cdot$ & 
  $\cdot$ & 
  &
  $\cdot$ & 
  $\cdot$ & 
  \Circle 
  \\ 

  &
  SANER18~\cite{karamitas2018efficient} &
  7 & 
  $\cdot$ & 
  &
  \Circle & 
  \Circle & 
  $\cdot$ & 
  $\cdot$ & 
  $\cdot$ & 
  $\cdot$ & 
  $\cdot$ & 
  $\cdot$ & 
  &
  $\cdot$ & 
  $\cdot$ & 
  $\cdot$ & 
  $\cdot$ & 
  $\cdot$ & 
  &
  $\cdot$ & 
  1 & 
  1 & 
  1 & 
  $\cdot$ & 
  $\cdot$ & 
  1 & 
  $\cdot$ & 
  $\cdot$ & 
  $\cdot$ & 
  $\cdot$ & 
  1 & 
  5 & 
  &
  $\cdot$ & 
  $\cdot$ & 
  $\cdot$ & 
  $\cdot$ & 
  &
  $\cdot$ & 
  $\cdot$ & 
  \Circle 
  \\ 

  &
  BinGo-E~\cite{xue2018accurate} &
  (\num{5145}) & 
  $\cdot$ & 
  &
  \Circle & 
  \Circle & 
  \Circle & 
  $\cdot$ & 
  $\cdot$ & 
  $\cdot$ & 
  $\cdot$ & 
  $\cdot$ & 
  &
  \Circle & 
  \Circle & 
  \Circle & 
  \Circle & 
  $\cdot$ & 
  &
  $\cdot$ & 
  3 & 
  $\cdot$ & 
  $\cdot$ & 
  $\cdot$ & 
  $\cdot$ & 
  1 & 
  $\cdot$ & 
  $\cdot$ & 
  $\cdot$ & 
  $\cdot$ & 
  1 & 
  5 & 
  &
  $\cdot$ & 
  $\cdot$ & 
  $\cdot$ & 
  $\cdot$ & 
  &
  $\cdot$ & 
  $\cdot$ & 
  \Circle 
  \\ 

  &
  WSB~\cite{yuan2018new} &
  (173) & 
  $\cdot$ & 
  &
  $\triangle$ & 
  $\cdot$ & 
  $\cdot$ & 
  $\cdot$ & 
  $\cdot$ & 
  $\cdot$ & 
  $\cdot$ & 
  $\cdot$ & 
  &
  $\cdot$ & 
  \Circle & 
  \Circle & 
  \Circle & 
  $\cdot$ & 
  &
  $\cdot$ & 
  1 & 
  $\cdot$ & 
  $\cdot$ & 
  $\cdot$ & 
  $\cdot$ & 
  1 & 
  $\cdot$ & 
  $\cdot$ & 
  $\cdot$ & 
  $\cdot$ & 
  $\cdot$ & 
  2 & 
  &
  $\cdot$ & 
  $\cdot$ & 
  $\cdot$ & 
  \Circle & 
  &
  $\cdot$ & 
  $\cdot$ & 
  $\cdot$ 
  \\ 

  &
  BinMatch~\cite{hu2018binmatch} &
  (82) & 
  $\cdot$ & 
  &
  \Circle & 
  $\cdot$ & 
  $\cdot$ & 
  $\cdot$ & 
  $\cdot$ & 
  $\cdot$ & 
  $\cdot$ & 
  $\cdot$ & 
  &
  \Circle & 
  $\cdot$ & 
  \Circle & 
  \Circle & 
  $\cdot$ & 
  &
  $\cdot$ & 
  1 & 
  $\cdot$ & 
  $\cdot$ & 
  $\cdot$ & 
  $\cdot$ & 
  1 & 
  $\cdot$ & 
  $\cdot$ & 
  $\cdot$ & 
  $\cdot$ & 
  $\cdot$ & 
  2 & 
  &
  $\cdot$ & 
  $\cdot$ & 
  $\cdot$ & 
  \Circle & 
  &
  $\cdot$ & 
  $\cdot$ & 
  \Circle 
  \\ 

  &
  MASES18~\cite{marastoni2018deep} &
  47 & 
  $\cdot$ & 
  &
  $\triangle$ & 
  $\cdot$ & 
  $\cdot$ & 
  $\cdot$ & 
  $\cdot$ & 
  $\cdot$ & 
  $\cdot$ & 
  $\cdot$ & 
  &
  $\cdot$ & 
  $\cdot$ & 
  $\cdot$ & 
  $\cdot$ & 
  $\cdot$ & 
  &
  $\cdot$ & 
  $\cdot$ & 
  $\cdot$ & 
  $\cdot$ & 
  $\cdot$ & 
  $\cdot$ & 
  $\cdot$ & 
  $\cdot$ & 
  $\cdot$ & 
  $\cdot$ & 
  $\cdot$ & 
  $\cdot$ & 
  $\cdot$ & 
  &
  $\cdot$ & 
  $\cdot$ & 
  $\cdot$ & 
  \Circle & 
  &
  $\cdot$ & 
  $\cdot$ & 
  $\cdot$ 
  \\ 

  &
  Zeek~\cite{shalev2018binary} &
  (\num{20680}) & 
  $\cdot$ & 
  &
  $\cdot$ & 
  \Circle & 
  $\cdot$ & 
  \Circle & 
  $\cdot$ & 
  $\cdot$ & 
  $\cdot$ & 
  $\cdot$ & 
  &
  \Circle & 
  \Circle & 
  \Circle & 
  \Circle & 
  \Circle & 
  &
  $\cdot$ & 
  3 & 
  $\cdot$ & 
  $\cdot$ & 
  $\cdot$ & 
  $\cdot$ & 
  4 & 
  1 & 
  $\cdot$ & 
  $\cdot$ & 
  $\cdot$ & 
  2 & 
  10 & 
  &
  $\cdot$ & 
  $\cdot$ & 
  $\cdot$ & 
  $\cdot$ & 
  &
  $\cdot$ & 
  $\cdot$ & 
  $\cdot$ 
  \\ 

  &
  FirmUp~\cite{david2018firmup} &
  $\cdot$ & 
  \num{2000} & 
  &
  $\triangle$ & 
  $\cdot$ & 
  $\triangle$ & 
  $\cdot$ & 
  $\triangle$ & 
  $\cdot$ & 
  $\cdot$ & 
  $\cdot$ & 
  &
  $\cdot$ & 
  $\cdot$ & 
  $\cdot$ & 
  $\cdot$ & 
  $\cdot$ & 
  &
  $\cdot$ & 
  $\cdot$ & 
  $\cdot$ & 
  $\cdot$ & 
  $\cdot$ & 
  $\cdot$ & 
  $\cdot$ & 
  $\cdot$ & 
  $\cdot$ & 
  $\cdot$ & 
  $\cdot$ & 
  $\cdot$ & 
  $\cdot$ & 
  &
  $\cdot$ & 
  $\cdot$ & 
  $\cdot$ & 
  $\cdot$ & 
  &
  $\cdot$ & 
  $\cdot$ & 
  \Circle 
  \\ 

  &
  $\alpha$Diff~\cite{liu2018alphadiff} &
  (\num{69989}) & 
  2 & 
  &
  \Circle & 
  \Circle & 
  \Circle & 
  $\cdot$ & 
  $\cdot$ & 
  $\cdot$ & 
  $\cdot$ & 
  $\cdot$ & 
  &
  \Circle & 
  \Circle & 
  \Circle & 
  \Circle & 
  $\cdot$ & 
  &
  $\cdot$ & 
  2 & 
  1 & 
  $\cdot$ & 
  $\cdot$ & 
  $\cdot$ & 
  2 & 
  $\cdot$ & 
  $\cdot$ & 
  $\cdot$ & 
  $\cdot$ & 
  $\cdot$ & 
  5 & 
  &
  $\cdot$ & 
  $\cdot$ & 
  $\cdot$ & 
  $\cdot$ & 
  &
  \RIGHTcircle & 
  \RIGHTcircle & 
  \Circle 
  \\ 

  &
  VulSeeker~\cite{gao2018vulseeker} &
  (\num{10512}) & 
  \num{4643} & 
  &
  \Circle & 
  \Circle & 
  \Circle & 
  \Circle & 
  \Circle & 
  \Circle & 
  $\cdot$ & 
  $\cdot$ & 
  &
  \Circle & 
  \Circle & 
  \Circle & 
  \Circle & 
  $\cdot$ & 
  &
  $\cdot$ & 
  1 & 
  1 & 
  $\cdot$ & 
  $\cdot$ & 
  $\cdot$ & 
  $\cdot$ & 
  $\cdot$ & 
  $\cdot$ & 
  $\cdot$ & 
  $\cdot$ & 
  $\cdot$ & 
  2 & 
  &
  $\cdot$ & 
  $\cdot$ & 
  $\cdot$ & 
  $\cdot$ & 
  &
  \Circle & 
  $\cdot$ & 
  \Circle 
  \\ \midrule

  \multirow{6}{*}{\rotatebox[origin=c]{0}{2019}} &
  InnerEye~\cite{zuo2019neural} &
  (844) & 
  $\cdot$ & 
  &
  $\cdot$ & 
  \Circle & 
  \Circle & 
  $\cdot$ & 
  $\cdot$ & 
  $\cdot$ & 
  $\cdot$ & 
  $\cdot$ & 
  &
  $\cdot$ & 
  \Circle & 
  \Circle & 
  \Circle & 
  $\cdot$ & 
  &
  $\cdot$ & 
  $\cdot$ & 
  $\cdot$ & 
  $\cdot$ & 
  $\cdot$ & 
  $\cdot$ & 
  $\cdot$ & 
  $\cdot$ & 
  $\cdot$ & 
  1 & 
  $\cdot$ & 
  $\cdot$ & 
  1 & 
  &
  $\cdot$ & 
  $\cdot$ & 
  $\cdot$ & 
  $\cdot$ & 
  &
  \RIGHTcircle & 
  \RIGHTcircle & 
  $\cdot$ 
  \\ 

  &
  Asm2Vec~\cite{ding2019asm2vec} &
  68 & 
  $\cdot$ & 
  &
  $\cdot$ & 
  \Circle & 
  $\cdot$ & 
  $\cdot$ & 
  $\cdot$ & 
  $\cdot$ & 
  $\cdot$ & 
  $\cdot$ & 
  &
  \Circle & 
  \Circle & 
  \Circle & 
  \Circle & 
  $\cdot$ & 
  &
  $\cdot$ & 
  1 & 
  1 & 
  $\cdot$ & 
  $\cdot$ & 
  $\cdot$ & 
  2 & 
  $\cdot$ & 
  $\cdot$ & 
  $\cdot$ & 
  $\cdot$ & 
  2 & 
  6 & 
  &
  $\cdot$ & 
  $\cdot$ & 
  $\cdot$ & 
  \Circle & 
  &
  \Circle & 
  $\cdot$ & 
  \Circle 
  \\

  &
  SAFE~\cite{massarelli2019safe} &
  (\num{5001}) & 
  $\cdot$ & 
  &
  $\cdot$ & 
  \Circle & 
  \Circle & 
  $\cdot$ & 
  $\cdot$ & 
  $\cdot$ & 
  $\cdot$ & 
  $\cdot$ & 
  &
  \Circle & 
  \Circle & 
  \Circle & 
  \Circle & 
  $\cdot$ & 
  &
  1 & 
  3 & 
  1 & 
  1 & 
  1 & 
  $\cdot$ & 
  2 & 
  1 & 
  1 & 
  1 & 
  $\cdot$ & 
  $\cdot$ & 
  12 & 
  &
  $\cdot$ & 
  $\cdot$ & 
  $\cdot$ & 
  $\cdot$ & 
  &
  \Circle & 
  \RIGHTcircle & 
  \Circle 
  \\

  &
  BAR19i~\cite{redmond2019cross} &
  (804) & 
  $\cdot$ & 
  &
  $\cdot$ & 
  \Circle & 
  \Circle & 
  $\cdot$ & 
  $\cdot$ & 
  $\cdot$ & 
  $\cdot$ & 
  $\cdot$ & 
  &
  $\cdot$ & 
  \Circle & 
  \Circle & 
  \Circle & 
  $\cdot$ & 
  &
  $\cdot$ & 
  $\cdot$ & 
  $\cdot$ & 
  $\cdot$ & 
  $\cdot$ & 
  $\cdot$ & 
  $\cdot$ & 
  $\cdot$ & 
  $\cdot$ & 
  1 & 
  $\cdot$ & 
  $\cdot$ & 
  1 & 
  &
  $\cdot$ & 
  $\cdot$ & 
  $\cdot$ & 
  $\cdot$ & 
  &
  \Circle & 
  $\cdot$ & 
  $\cdot$ 
  \\

  &
  BAR19ii~\cite{massarelli2019investigating} &
  (11,244) & 
  $\cdot$ & 
  &
  \Circle & 
  \Circle & 
  \Circle & 
  $\cdot$ & 
  $\cdot$ & 
  $\cdot$ & 
  $\cdot$ & 
  $\cdot$ & 
  &
  \Circle & 
  \Circle & 
  \Circle & 
  \Circle & 
  $\cdot$ & 
  &
  1 & 
  3 & 
  1 & 
  $\cdot$ & 
  $\cdot$ & 
  $\cdot$ & 
  2 & 
  1 & 
  1 & 
  $\cdot$ & 
  $\cdot$ & 
  2 & 
  11 & 
  &
  $\cdot$ & 
  $\cdot$ & 
  $\cdot$ & 
  $\cdot$ & 
  &
  $\cdot$ & 
  $\cdot$ & 
  \Circle 
  \\

  &
  FuncNet~\cite{luo2019funcnet} & 
  (180) & 
  $\cdot$ & 
  &
  \Circle & 
  $\cdot$ & 
  \Circle & 
  $\cdot$ & 
  \Circle & 
  $\cdot$ & 
  $\cdot$ & 
  $\cdot$ & 
  &
  \Circle & 
  \Circle & 
  \Circle & 
  \Circle & 
  \Circle & 
  &
  $\cdot$ & 
  $\cdot$ & 
  $\cdot$ & 
  1 & 
  $\cdot$ & 
  $\cdot$ & 
  $\cdot$ & 
  $\cdot$ & 
  $\cdot$ & 
  $\cdot$ & 
  $\cdot$ & 
  $\cdot$ & 
  1 & 
  &
  $\cdot$ & 
  $\cdot$ & 
  $\cdot$ & 
  $\cdot$ & 
  &
  $\cdot$ & 
  $\cdot$ & 
  \Circle 
  \\

  \midrule

  \multirow{5}{*}{\rotatebox[origin=c]{0}{2020}} &
  DeepBinDiff~\cite{duan2020deepbindiff} &
  (\num{2206}) & 
  $\cdot$ & 
  &
  $\cdot$ & 
  $\triangle$ & 
  $\cdot$ & 
  $\cdot$ & 
  $\cdot$ & 
  $\cdot$ & 
  $\cdot$ & 
  $\cdot$ & 
  &
  \Circle & 
  \Circle & 
  \Circle & 
  \Circle & 
  $\cdot$ & 
  &
  $\cdot$ & 
  1 & 
  $\cdot$ & 
  $\cdot$ & 
  $\cdot$ & 
  $\cdot$ & 
  $\cdot$ & 
  $\cdot$ & 
  $\cdot$ & 
  $\cdot$ & 
  $\cdot$ & 
  $\cdot$ & 
  1 & 
  &
  $\cdot$ & 
  $\cdot$ & 
  $\cdot$ & 
  $\cdot$ & 
  &
  \Circle & 
  \Circle & 
  \Circle 
  \\

  &
  ImOpt~\cite{jiang2020similarity} &
  18 & 
  $\cdot$ & 
  &
  $\cdot$ & 
  \Circle & 
  $\cdot$ & 
  $\cdot$ & 
  $\cdot$ & 
  $\cdot$ & 
  $\cdot$ & 
  $\cdot$ & 
  &
  \Circle & 
  $\cdot$ & 
  \Circle & 
  \Circle & 
  $\cdot$ & 
  &
  $\cdot$ & 
  $\cdot$ & 
  1 & 
  $\cdot$ & 
  $\cdot$ & 
  $\cdot$ & 
  $\cdot$ & 
  $\cdot$ & 
  $\cdot$ & 
  $\cdot$ & 
  $\cdot$ & 
  $\cdot$ & 
  1 & 
  &
  $\cdot$ & 
  $\cdot$ & 
  $\cdot$ & 
  \Circle & 
  &
  $\cdot$ & 
  $\cdot$ & 
  $\cdot$ 
  \\

  &
  ACCESS20~\cite{guo2020lightweight} &
  12,000 & 
  $\cdot$ & 
  &
  $\triangle$ & 
  $\triangle$& 
  $\cdot$ & 
  $\cdot$ & 
  $\cdot$ & 
  $\cdot$ & 
  $\cdot$ & 
  $\cdot$ & 
  &
  $\cdot$ & 
  $\cdot$ & 
  $\cdot$ & 
  $\cdot$ & 
  $\cdot$ & 
  &
  $\cdot$ & 
  $\cdot$ & 
  $\cdot$ & 
  $\cdot$ & 
  $\cdot$ & 
  $\cdot$ & 
  $\cdot$ & 
  $\cdot$ & 
  $\cdot$ & 
  $\cdot$ & 
  $\cdot$ & 
  $\cdot$ & 
  $\cdot$ & 
  &
  $\cdot$ & 
  $\cdot$ & 
  $\cdot$ & 
  $\cdot$ & 
  &
  $\cdot$ & 
  $\cdot$ & 
  $\cdot$ 
  \\

  &
  Patchecko~\cite{sun2020hybrid} &
  2,108 & 
  2 & 
  &
  \Circle & 
  \Circle & 
  \Circle & 
  \Circle & 
  $\cdot$ & 
  $\cdot$ & 
  $\cdot$ & 
  $\cdot$ & 
  &
  \Circle & 
  \Circle & 
  \Circle & 
  \Circle & 
  \Circle & 
  &
  $\cdot$ & 
  $\cdot$ & 
  $\cdot$ & 
  $\cdot$ & 
  $\cdot$ & 
  $\cdot$ & 
  $\cdot$ & 
  $\cdot$ & 
  $\cdot$ & 
  $\cdot$ & 
  $\cdot$ & 
  $\cdot$ & 
  $\cdot$ & 
  &
  $\cdot$ & 
  $\cdot$ & 
  $\cdot$ & 
  $\cdot$ & 
  &
  $\cdot$ & 
  $\cdot$ & 
  \Circle 
  \\

  &
  \sys $\bigstar$ &
  \numBin & 
  $\cdot$ & 
  &
  \Circle & 
  \Circle & 
  \Circle & 
  \Circle & 
  \Circle & 
  \Circle & 
  \Circle & 
  \Circle & 
  &
  \Circle & 
  \Circle & 
  \Circle & 
  \Circle & 
  \Circle & 
  &
  $\cdot$ & 
  1 & 
  1 & 
  1 & 
  1 & 
  1 & 
  $\cdot$ & 
  1 & 
  1 & 
  1 & 
  1 & 
  $\cdot$ & 
  9 & 
  &
  \Circle & 
  \Circle & 
  \Circle & 
  \Circle & 
  &
  \Circle & 
  \Circle & 
  \Circle 
  \\


  \bottomrule

\end{tabular}

\begin{tablenotes}

\item [$\ast$]
  We only mark items that are stated explicitly in the paper.
  Due to the lack of details about firmware images,
  we were not able to mark optimization options or compilers used to create them.
  For papers that do not explicitly state the number of binaries in their
  dataset, we estimated the number and marked it with parentheses.

\item [$\dagger$] This table focuses on two major compilers:
  \gcc and \clang, as other compilers only support a limited number of
  architectures.

\item [$\triangle$] We infer the target architectures of the dataset as they are
  not stated explicitly in the paper.


  %
  %
  %
  %
\item [\RIGHTcircle] This indicates that only a portion of the code and dataset is available.
  %
  For example, discovRE~\cite{eschweiler2016discovre} makes available only their
  firmware images, and $\alpha$Diff~\cite{liu2018alphadiff} opens transformed
  function images but not the actual dataset.

\end{tablenotes}
\end{threeparttable}
\vspace{-0.2in}
\end{table*}


Recovered program variables can also be semantic features.
For example, one can compare the similarity of string literals referenced in
code snippets~\cite{eschweiler2016discovre, feng2016scalable, xu2017neural,
nouh2017binsign, shirani2017binshape, shirani2018binarm, karamitas2018efficient,
sun2020hybrid}.
One can also utilize the size of local variables, function parameters, or the
return type of functions~\cite{eschweiler2016discovre, nouh2017binsign,
sun2020hybrid, luo2019funcnet}.
One can further check registers or local variables that store the return values
of functions~\cite{xue2018accurate}.


Recently, several approaches have been utilizing embedding vectors, adopting
various machine learning techniques.
%
After building an attributed control-flow graph (ACFG)~\cite{feng2016scalable},
which is a CFG containing numeric presemantic features in its basic blocks, one
can apply spectral clustering~\cite{ng2002spectral} to group multiple ACFGs or
popular encoding methods~\cite{yang2007evaluating, arandjelovic2013all,
dai2016discriminative} to embed them into a vector~\cite{xu2017neural}.
The same technique can also be applied to PDGs~\cite{gao2018vulseeker}.
Meanwhile, recent NLP techniques, such as Word2Vec~\cite{mikolov2013efficient}
or convolutional neural network models~\cite{kim2014convolutional}, can be
utilized for embedding raw bytes or assembly instructions into numeric
vectors~\cite{zuo2019neural, liu2018alphadiff, marastoni2018deep,
ding2019asm2vec, massarelli2019safe, redmond2019cross, duan2020deepbindiff,
massarelli2019investigating}.
For this embedding, one can also consider a higher-level
granularity~\cite{zuo2019neural, ding2019asm2vec} by applying other NLP
techniques, such as sentence embedding~\cite{hochreiter1997long} or paragraph
embedding~\cite{le2014distributed}.
Note that one may apply machine learning to compare embedding vectors rather
than generating them~\cite{lageman2016bindnn, shalev2018binary}, and
\autoref{tab:prevfeatures} does \emph{not} mark them to use embedded vectors.

\subsubsection{Key Assumptions from Past Research}
During our literature study, we found that most of the approaches highly rely on
semantic features extracted in (S3), assuming that they should not change across
compilers nor target architectures.
However, none of them clearly justifies the necessity of such complex
semantics-based analyses. They focus only on the end results without considering
the precise reasoning behind their approaches.


This is indeed the key motivation for our research. Although most existing
approaches focus on complex analyses, there may exist elementary features that
we have overlooked.
For example, there may exist effective presemantic features, which can beat
semantic features regardless of target architectures and compilers.
It can be the case that those known features have not been thoroughly evaluated
on the right benchmark as there has been no comprehensive study on them.


Furthermore, existing research assumes the correctness of the underlying binary
analysis framework, such as IDA Pro~\cite{ida}, which is indeed the most popular
tool used, as shown in the rightmost column of~\autoref{tab:prevdataset}.
However, CFGs derived from those tools may be inherently wrong. They may miss
some important basic blocks, for instance, which can directly affect the
precision of BCSA features.

Indeed, both (S1) and (S2) are challenging research problems by themselves:
there are abundant research efforts to improve the precision of both analyses.
For example,
disassembling binary code itself is an undecidable
problem~\cite{andriesse2016depth}, and writing an efficient and accurate binary
lifter is significantly challenging in practice~\cite{kim2017testing,
  jung:bar:2019}. Identifying functions from binaries~\cite{\paperFuncID} and
recovering control-flow edges~\cite{kinder:cav:2008} for indirect branches are
still active research fields.
All these observations lead us to research questions in~\autoref{ss:rq}.

%
%
%
%

\subsection{Benchmarks Used in Prior Works}
\label{ss:prevbench}

It is imperative to use the right benchmark to evaluate a BCSA technique.
Therefore, we studied the benchmarks used in the past literature, as shown
in~\autoref{tab:prevdataset}.
However, during the study, we found that it is radically difficult to properly
evaluate a new BCSA technique using the previous benchmarks.
%
%
%

First, we were not able to find a single pair of papers that use the same
benchmark.  Some of them share packages such as GNU
\tc{coreutils}~\cite{chandramohan2016bingo, egele2014blanket, wang2017memory},
but the exact binaries, versions, and compiler options are not the
same. Although there is no known standard for evaluating BCSA, it is surprising
to observe that none of the papers use the same dataset.
We believe this is partly because of the difficulty in preparing the same
benchmark. For example, even if we can download the same version of the source
code used in a paper, it is extraordinarily difficult to cross-compile the
program for various target architectures with varying compiler options; it
requires significant effort to set up the environment.
However, \textit{only two out of \numAnalyzedPapers papers we studied
fully open their dataset}. Even in that case, it is hard to rebuild or extend
the benchmark
because of the absence of a public compilation script for the benchmark.

Second, the number of binaries used in each paper is limited and may not be
enough for analytics.
The \emph{\#Binaries} column of~\autoref{tab:prevdataset} summarizes the number
of program binaries obtained from two different sources: application packages
and firmware images.
Since a single package can contain multiple binaries, we manually extracted the
packages used in each paper and counted the number of binaries in each
package. We counted only the binaries after a successful compilation, such that
the object files that were generated during the compilation process were not
counted.
If a paper does not explicitly mention package versions, we used
the most recent package versions at the time of writing and marked them with
parentheses.
Note that only 6 out of \numAnalyzedPapers papers have more than 10,000
binaries, and none reaches 100,000 binaries.
Firmware may include numerous binaries, but it cannot be directly used for BCSA
because one cannot generate the ground truth without having the source code.


Finally, previous benchmarks only cover a few compilers, compiler options, and
target architectures. Some papers do not even describe their tested compiler
options or package versions.
The \emph{Compiler} column of the table presents the number of minor versions
used for each major version of the compilers.
Notably, all the benchmarks except one consider less than five different major
compiler versions.
The \emph{Extra} column of the table shows the use of extra compiler options for
each benchmark. Only a few consider function inlining and
Link-Time Optimization (LTO). None of them deal with the Position Independent
Executable (PIE) option, although, currently, it is widely
used~\cite{sec2016pie}.

All these observations lead us to the research questions outlined in the next
subsection (\autoref{ss:rq}) and eventually motivate us to create our own
benchmark that we call \sys, which is shown in the last row
of~\autoref{tab:prevdataset}.

\subsection{Research Problems and Questions}
\label{ss:rq}

We now summarize several key problems observed from the previous literature and introduce research questions derived from these problems.
First, none of the papers uses the same benchmark for their evaluation, and the
way they evaluate their techniques significantly differs.
Second, only a few of the studies release their source code and data,
which makes it radically difficult
to reproduce or improve upon existing works.
Furthermore, most papers use manually chosen ground truth data
for their evaluation, which are easily error-prone.
Finally, current state-of-the-art approaches in BCSA focus on extracting
semantic features with complex analysis techniques (from
\autoref{ss:presemantic} and \autoref{ss:semantic}).
These observations naturally lead us to the below research questions.
Note that some of the questions are indeed open-ended, and we only address them
in part.

\smallskip
\PP{RQ1.} How should we establish a large-scale benchmark and ground truth data?

One may build benchmarks by manually compiling application source code. However,
there are so many different compiler versions, optimization levels, and options
to consider when building binaries. Therefore, it is desirable to automate this
process to build a large-scale benchmark for BCSA.
%
%
It should be noted that many of the existing studies have also attempted to
build ground truth from source code.
However, the number of binaries and compiler options used in those studies is
limited and is not enough for data-driven research. Furthermore, those studies
release neither their source code nor dataset (\autoref{ss:prevbench}).
On the contrary, we present a script that can automatically build large-scale
ground truth data from a given set of source packages with clear descriptions
(\autoref{sec:dataset}).

\smallskip
\PP{RQ2.} Is the effectiveness of presemantic features limited to the target
architectures and compiler options used?

We note that most previous studies assume that presemantic features are
significantly less effective than semantic features, as they can largely vary
depending on the underlying architectures and compiler optimizations used. For
example, compilers may perform target-specific optimization techniques for a
specific architecture.
Indeed, 36 out of the \numAnalyzedPapers papers ($\approx 84\%$) we studied
focus on new semantic features in their analysis, as shown
in~\autoref{tab:prevfeatures}.
To determine whether this assumption is valid,
we investigate it through a series of rigorous experimental studies. Although
byte-level information significantly varies depending on the target and the
optimization techniques, we found that some presemantic features, such as
structural information obtained from CFGs, are broadly similar across different
binaries of the same program.
Additionally, we demonstrated that utilizing such presemantic features without a
complex semantic analysis can achieve an accuracy that is comparable to that of
a recent deep learning-based approach with a semantic analysis
(\autoref{sec:rq2}).
%

\smallskip
\PP{RQ3.} Can debugging information help BCSA achieve a high accuracy rate?

We are not aware of any quantitative study on how much debugging information
affects the accuracy of BCSA. Most prior works simply assume that debugging
information is not available, but how much does it help? How would decompilation
techniques affect the accuracy of BCSA?
To answer this question, we extracted a list of function types from our
benchmark and used them to perform BCSA on our dataset. Surprisingly, we were
able to achieve a higher accuracy rate than any other existing works on BCSA
without using any sophisticated method~(\autoref{sec:rq3}).
%

\smallskip
\PP{RQ4.} Can we benefit from analyzing failure cases of BCSA?

Most existing works do not analyze their failure cases as they rely on
uninterpretable machine learning techniques. However, our goal is to use a
simple and interpretable model to learn from failure and gain insights for
future research.
Therefore, we manually examined failure cases using our interpretable method
and observed three common causes for failure, which have been mostly overlooked
by the previous literature. First, COTS binary analysis tools indeed return
false results. Second, different compiler back-ends for the same architecture
can be substantially different from each other. Third, there are
architecture-specific code snippets for the same function. We believe that all
these observations help in setting directions for future
studies~(\autoref{sec:rq4}).
%

\smallskip
\PP{Analysis Scope.}
In this paper, we focus on function-level similarity analyses because functions
are a fundamental unit of binary analysis, and function-level BCSA is widely
used in previous literature~\cite{\paperFunc}.
We believe one can easily extend our work to support whole-binary-level
similarity analyses as in the previous papers~\cite{\paperBinary}.

\section{Establishing Large-Scale Benchmark and Ground Truth for BCSA (RQ1)}
\label{sec:dataset}

Building a large-scale benchmark for BCSA and establishing its ground truth is
challenging.
One potential approach for generating the ground truth data is to manually
identify similar functions from existing binaries or firmware
images~\cite{pewny2014leveraging, david2014tracelet, david2016statistical}.
However, this requires domain expertise and is often error-prone and
time-consuming.

Another approach for obtaining the ground truth is to compile binaries from
existing source code with varying compiler options and target
architectures~\cite{feng2016scalable, chandramohan2016bingo, gao2018vulseeker,
  feng2017extracting}.
If we compile multiple binaries (with different compiler options) from the same
source code, one can determine which function corresponds to which source lines.
Unfortunately, most existing approaches do not open their benchmarks nor the
compilation scripts used to produce them (\autoref{tab:prevdataset}).
%
%

Therefore, we present \sys,
which is a comprehensive benchmark for BCSA,
along with automated compilation scripts
that help reproduce and extend it
for various research purposes.
The rest of this section details \sys and
discusses how we establish the ground truth (RQ1).


\subsection{\sys: Large-Scale BCSA Benchmark}
\label{ss:sys}

%

\sys is a comprehensive BCSA benchmark
that comprises \numBin binaries
compiled from \numPack packages of source code
with \numOpt distinct combinations of
compilers, compilation options, and target architectures.
Therefore, \sys covers most of the benchmarks
used in existing approaches,
as shown in~\autoref{tab:prevdataset}.
\sys includes binaries compiled for
\numArch different architectures.
For example,
we use both little- and big-endian binaries for MIPS
to investigate the effect of endianness.
It uses \numComp different versions of compilers:
\gcc v\{4.9.4, 5.5.0, 6.4.0, 7.3.0, 8.2.0\} and
\clang v\{4.0, 5.0, 6.0, 7.0\}.
We also consider \numOpti optimization levels
from \tc{O0} to \tc{O3} as well as \tc{Os},
which is the code size optimization.
Finally,
we take PIE, LTO, and obfuscation options into account,
which are less explored in BCSA.

We select GNU software packages~\cite{gnupacks}
as our compilation target
because of their popularity and accessibility:
they are real applications
that are widely used on Linux systems,
and their source code is publicly available.
We successfully compiled \numPack GNU packages
for all our target architectures and compiler options.

To better support targeted comparisons,
we divide \sys into six datasets:
\normaldataset, \sizeoptdataset, \noinlinedataset,
\piedataset, \ltodataset, and \obfuscationdataset.
The summary of each dataset is
shown in~\autoref{tab:dataset}.
Each dataset contains binaries obtained by
compiling the GNU packages with different combinations
of compiler options and targets.
There is \emph{no} intersection among the datasets.

\normaldataset includes binaries compiled for
\numArch different architectures
with different compilers and optimization levels.
We did not use other extra options
such as PIE, LTO, and no-inline for this dataset.

\sizeoptdataset is the same as \normaldataset
except that it uses only the \tc{Os} optimization option
instead of \tc{O0}--\tc{O3}.

Similarly, \piedataset, \noinlinedataset, \ltodataset, and \obfuscationdataset are
no different from \normaldataset except that
they are generated by using an additional flag
to enable PIE,
to disable inline optimization,
to enable LTO, and
to enable compile-time obfuscation, respectively.

PIE makes memory references in binary relative
to support ASLR.
On some architectures, e.g., x86,
compilers inject additional code snippets
to achieve relative addressing.
As a result, the compiled output can differ severely.
Although PIE became the default
on most Linux systems~\cite{sec2016pie},
it has not been well studied for BCSA.
Note we were not able to compile all \numPack packages
with the PIE option enabled.
Therefore, we have fewer binaries in \piedataset
than \normaldataset.

Function inlining embeds callee functions
into the body of the caller.
This can make presemantic features largely vary.
Therefore,
we investigate the effect of function inlining on BCSA
by explicitly turning off the inline optimization
with the \tc{fno-inline} option.

LTO is an optimization technique that operates at link time.
It removes unnecessary code blocks,
thereby reducing the number of presemantic features.
However, it has also been less studied in BCSA.
We were only able to successfully compile 29 packages
when the LTO option was enabled.

Finally, the \obfuscationdataset dataset uses
Obfuscator-LLVM~\cite{junod2015obfuscator}
to obfuscate the target binaries.
We chose Obfuscator-LLVM
among various other tools previously used~\cite{\toolObfus}
because it is the most commonly used~\cite{\paperObfusLLVM},
and we can directly compare the effect of obfuscation
using the vanilla LLVM compiler.
We use Obfuscator-LLVM's latest version
with four obfuscation options:
instruction substitution (SUB),
bogus control flow (BCF),
control flow flattening (FLA), and
a combination of all the options.
We regard each option as a distinct compiler,
as shown in the \emph{Comp} column of~\autoref{tab:dataset}.
One can obfuscate a single binary multiple times.
However, we only applied it once.
This is because obfuscating a binary multiple times
could emit a significantly large binary,
which becomes time-consuming for IDA Pro to preprocess.
For example,
when we obfuscate \tc{a2ps} twice with all three options,
the compiled binary reaches over 30 MB,
which is 30 times larger than the normal one.

\begin{table}[t]
  \scriptsize
  \centering
  \setlength\tabcolsep{0.06cm}
  \def\arraystretch{0.50}

  \caption{Summary of \sys.}
  \label{tab:dataset}
\vspace{-0.1in}
\begin{threeparttable}
\begin{tabular}{@{}lccccrrr@{}}
\toprule

  \multicolumn{1}{c}{\textbf{Dataset}} &
  \textbf{\# of} &
  \textbf{\# of} &
  \textbf{\# of} &
  \textbf{\# of} &
  \multicolumn{1}{c}{\textbf{\# of}} &
  \multicolumn{1}{c}{\textbf{\# of Orig.}} &
  \multicolumn{1}{c}{\textbf{\# of Final}}
  \\

  \multicolumn{1}{c}{\textbf{Name}} &
  \textbf{Pkgs} &
  \textbf{Archs} &
  \textbf{Optis} &
  \textbf{Comps} &
  \multicolumn{1}{c}{\textbf{Binaries}} &
  \multicolumn{1}{c}{\textbf{Functions}} &
  \multicolumn{1}{c}{\textbf{Functions}\tnote{$\ast$}}
  \\ \midrule
  
  \normaldataset        & \num{51}
  & \num{8}  & \num{4}  & \num{9}
  & \num{67680}
  & \num{34355824}
  & \num{8708459}    \\
  
  \sizeoptdataset       & \num{51}
  & \num{8}  & \num{1}\tnote{$\dagger$}  & \num{9}
  & \num{16920}
  & \num{8350442}
  & \num{2060625}    \\
  
  \piedataset           & \num{46}\tnote{$\dagger$}
  & \num{8}  & \num{4}  & \num{9}
  & \num{36000}
  & \num{23090676}
  & \num{7766235}    \\
  
  \noinlinedataset      & \num{51}
  & \num{8}  & \num{4}  & \num{9}
  & \num{67680}
  & \num{38617186}
  & \num{10291001}    \\
  
  \ltodataset           & \num{29}\tnote{$\dagger$}
  & \num{8}  & \num{4}  & \num{9}
  & \num{24768}
  & \num{12279982}
  & \num{3375308}     \\
  
  \obfuscationdataset   & \num{51}
  & \num{8}  & \num{4}  & \num{4}\tnote{$\ddagger$}
  & \num{30080}
  & \num{15809489}
  & \num{4054694}     \\
  \midrule
  
  \textbf{Total}        & \num{51}
  & \multicolumn{3}{c}{\num{1352} options}
  & \num{243128}
  & \num{132503599}
  & \num{36256322}    \\
  \midrule
\end{tabular}

\begin{tablenotes}
\item [$\ast$] The target functions are selected in the manner described
    in~\autoref{ss:groundtruth}.

\item [$\dagger$] The number of packages and compiler options varies
    because some packages can be compiled only with a specific set of compile
    options.

\item [$\ddagger$] We count each of the four obfuscation options as a distinct
    compiler (\autoref{ss:sys}).

\end{tablenotes}
\end{threeparttable}

\vspace{-0.2in}
\end{table}

The number of packages and that of compiler options
used in compiling each dataset differ
because some packages can be compiled only with
a specific set of compile options and targets.
Some packages fail to compile
because they have architecture-specific code,
such as inline assemblies, or
because they use compiler-specific grammars.
For example,
\clang does not support
both the LTO option and the \tc{Os} option to be turned on.
There are also cases where packages have conflicting dependencies.
We also excluded the ones that did not compile within 30 min
because some packages require
a considerable amount of time to compile.
For instance,
\tc{smalltalk} took more than 10 h
to compile with the obfuscation option enabled.

To summarize, \sys contains
\numBin binaries and \numFunc functions in total,
which is indeed many orders of magnitude
larger than the other benchmarks
that appear in the previous literature.
The \emph{Source} column of \autoref{tab:prevdataset}
shows the difference clearly.
\sys does not include firmware images
because our goal is to automatically build a benchmark
with clear ground truth.
One may extend our benchmark with firmware images.
However, it would take significant manual effort
to identify their ground truth.
For additional details regarding each package, please refer to~\autoref{tab:fulldataset} in the
Appendix.


Our benchmark and compilation scripts are available on GitHub.
Our compilation environment is
based on Crosstool-NG~\cite{crosstool-ng},
GNU Autoconf~\cite{mackenzie1996autoconf}, and
Linux Parallels~\cite{Tange2011a}.
Through this environment,
we compiled the entire datasets of \sys
in approximately 30 h on our server machine
with 144 Intel Xeon E7-8867v4 cores.

\subsection{Building Ground Truth}
\label{ss:groundtruth}

Next, we establish the ground truth for our dataset.
We first define the criteria for determining the equivalence of two functions.
In particular, we check whether two functions with the same name originated from
the same source files and have the same line numbers.
Additionally, we verify that both functions come from the same package and
have the same name in their binaries to ensure their equivalence.

Based on these criteria, we constructed the ground truth
by performing the following steps.
First, we compiled all the binaries with debugging information
using the \tc{-g} option.
We then leveraged IDA Pro~\cite{ida} to identify functions
in the compiled binaries.
%
%
Next, we labeled each identified function with its name, package name, binary
name, as well as the name of the corresponding source file and line numbers.
To achieve this, we wrote a script that parses the debugging information from
each binary.

Using this information, we then sanitize our dataset to avoid having incorrect
or biased results.
Among the identified functions, we selected only the ones in the \tc{code}
(\tc{.text}) segments, as functions in other segments may not include valid
binary code.
For example, we disregarded functions in the Procedure Linkage Table (\tc{.plt})
sections because these functions are wrappers to call external functions and do
not include actual function bodies.
In our dataset, we filtered out 40\% of the identified functions in this step.

We also disregarded approximately 4\% of the functions that are
generated by the compiler, but not by the application developers. We can easily
identify such compiler intrinsic functions by checking the corresponding source
files and line numbers.
%
%
For example, \gcc utilizes intrinsic functions such as \tc{__udivdi3} in
\tc{libgcc2.c} or \tc{__aeabi_uldivmod} in \tc{bpabi.S} to produce highly
optimized code.



Additionally, we removed duplicate functions within the same project/package.
Two different binaries often share the same source code, especially when they are
in the same project/package. For example, the GNU \tc{coreutils} package
contains 105 different executables that share 80\% of the functions in common.
We removed duplicate functions within each package by checking the source file
names and their line numbers.
Moreover, compilers can also generate multiple copies of the same function
within a single binary due to optimization.
These functions share the same source code but have a difference in their binary
forms. For example, some parts of the binary code are removed or reordered for
optimization purposes. As these functions share a large portion of the code,
considering all of them would produce a biased result. To avoid this, we
selected only one copy for each of the functions in our experiments.
%
This step filtered out approximately 54\% of the remaining functions.
%
%
The last column of~\autoref{tab:dataset} reports the final counting results,
which is the number of unique functions.

%
%
%


By performing all the above steps, we can automatically build large-scale ground
truth data.
The total time spent building the ground truth of all our datasets was
13,300 seconds.
By leveraging this ground truth data, we further investigate the remaining
research questions (\ie, RQ2--RQ4) in the following sections.
To encourage further research, we have released all our datasets and source
code.

\section{Building an Interpretable Model}
\label{sec:method}

Previous BCSA techniques focused on achieving a higher accuracy by leveraging
recent advances in deep learning techniques~\cite{liu2018alphadiff,
marastoni2018deep, xu2017neural, gao2018vulseeker}.
This often requires building a complicated model, which is not straightforward
to understand and hinders researchers from reasoning about the BCSA results and
further answering the fundamental questions regarding BCSA.
%
Therefore, we design an \emph{interpretable} model for BCSA to answer the
research questions and implement \techname, which is a BCSA tool that employs
the model.
This section illustrates how we obtain such a model and how we set up our
experimental environment.

%
%

\subsection{\techname Overview}
\label{ss:overview}


At a high level,
\techname leverages a set of presemantic features
widely used in the previous literature
to reassess the effectiveness of presemantic features (RQ2).
%
%
It evaluates each feature in two input functions,
based on our similarity scoring metric (\autoref{ss:scoring}),
which directly measures the difference between each feature value.
In other words, it captures how much each feature differs
across different compile options.

Note \techname is intentionally designed to be simple so that we can answer the
research questions presented in~\autoref{ss:rq}.
Despite the simplicity of our approach, \techname still produces a high accuracy
rate that is comparable to state-of-the-art
tools~(\autoref{ss:comparison-state-of-the-art}).
We are \emph{not} arguing here that \techname is the best BCSA algorithm.

\begin{table}[t]
  \centering
  \caption{Summary of numeric presemantic features used in \techname.}
  \label{tbl:featureset}
  \vspace{-6pt}
  \scriptsize
  \centering
  \setlength\tabcolsep{0.04cm}
  \def\arraystretch{0.50}
\begin{tabular}{l l @{\extracolsep{\fill}} r}
  \toprule
  \textbf{Category} &
  \multicolumn{1}{c}{\textbf{Features}}
  &
  \textbf{Count} \\
  \midrule

  \multirow{11}[10]{*}{CFG}      &
  \# of basic blocks, edges, loops, SCCs, and back edges  &
  \multirow{11}{*}{41} \\
    &
    \# of all, arith, data transfer, cmp, and logic instrs. & \\
    &
    \# of shift, bit-manipulating, float, misc instrs. & \\
    &
    \# of arith + shift, and data transfer + misc instrs. & \\
    &
    \# of all/unconditional/conditional control transfer instrs. & \\
    &
    Avg. \# of edges per a basic block & \\
    &
    Avg./Sum of basic block, loop, and SCC sizes & \\
    &
    Avg. \# of all, arith, data transfer, cmp, and logic instrs. & \\
    &
    Avg. \# of shift, bit-manipulating, float, misc instrs. & \\
    &
    Avg. \# of arith + shift, and data transfer + misc instrs. & \\
    &
    Avg. \# of all/unconditional/conditional control transfer instrs. & \\
  \midrule

  \multirow{2}{*}{CG}       &
    \# of callers, callees, imported callees &
    \multirow{2}{*}{6} \\
    &
    \# of incoming/outgoing/imported calls & \\
  \midrule


  \multicolumn{2}{c}{\textbf{Total}} & 47 \\
  \bottomrule
\end{tabular}
  \vspace{-12pt}
\end{table}

\subsection{Features Used in \techname}
\label{ss:features}


Recall from RQ2, one of our goals is to reconsider the capability of presemantic
features. Therefore, we focus on choosing various presemantic features used in
the previous BCSA literature instead of inventing novel ones.

However, creating a comprehensive feature set is not straightforward because of
the following two reasons. First, there are numerous existing features that
are similar to one another, as discussed in~\autoref{sec:back}. Second, some
features require domain-specific knowledge, which is \emph{not} publicly
available. For example, several existing papers~\cite{\paperInst} categorize
instructions into semantic groups. However, grouping instructions is largely a
subjective task, and there is no known standard for it. Furthermore, most
existing works do not make their grouping algorithms public.

We address these challenges by (1) manually extracting representative
presemantic features and (2) open-sourcing our feature extraction
implementation.
Specifically, we focus on numeric presemantic features. Because these features
are represented as numbers, the relationship among their values across
different compile options can be easily observed.

Table~\ref{tbl:featureset} summarizes the selected features.
Our feature set consists of CFG- and CG-level numeric features as they can
effectively reveal structural changes in the target code. In particular, we
utilize features related to basic blocks, CFG edges, natural loops, and strongly
connected components (SCCs) from CFGs,
by leveraging NetworkX~\cite{hagberg2008exploring}.
We also categorize instructions into several semantic groups based on our
careful judgment by referring to the reference manuals~\cite{intelmanual,
seal2001arm, mips2001mips} and leveraging Capstone~\cite{capstone}'s internal
grouping. Next, we count the number of instructions in each semantic group per
each function (i.e., CFG).
Additionally, we take six features from CGs. The number of callers and
callees represents a unique number of outgoing and incoming edges from CGs,
respectively.

To extract these features, we conducted the following steps.
First, we pre-processed the binaries in \sys with IDA Pro~\cite{ida}.
We then generated the ground truth of these binaries as we described
in~\autoref{ss:groundtruth}.
For those functions of which we have the ground truth,
we extracted the aforementioned features.
\autoref{tab:time} shows the time spent for each of these steps.
%
The IDA pre-processing took most of the time as IDA performs various internal
analyses.
Meanwhile, the feature extraction took much less time as it merely operates on
the precomputed results from the pre-processing step.
%





\subsection{Scoring Metric}
\label{ss:scoring}

Our scoring metric is based on the computation of the relative
difference~\cite{wiki:reldiff} between feature values.
Given two functions $A$ and $B$, let us denote a value of feature $f$ for each
function as $A_f$ and $B_f$, respectively. Recall that any feature in \techname
can be represented as a number.
We can compute the relative difference $\delta$ between the two feature values as
follows:
\begin{equation}
\scriptsize
  \delta(A_f, B_f) = \frac{|A_f - B_f|}{|\text{max}(A_f, B_f)|}
\end{equation}

Let us suppose we have $N$ distinct features $(f_1, f_2, \cdots, f_N)$ in our
feature set. We can then define our similarity score $s$ between two functions
$A$ and $B$ by taking the average of relative differences for all the features
as follows:

\begin{equation}
\scriptsize
  s(A, B) =
  1 -%
  \frac{%
    \left(%
    \delta(A_{f_1}, B_{f_1}) + \cdots +%
    \delta(A_{f_N}, B_{f_N})
    \right)
  }{N}
\end{equation}

Although each numeric feature can have a different range of values, \techname
can effectively handle them using relative differences by representing the
difference of each feature with a value between 0 and 1. Therefore, the score
$s$ is always within the range of 0 to 1.

Furthermore, we can intuitively understand and interpret the BCSA results using
our scoring metric. For example, suppose there are two functions $A$ and $B$
derived from the same source code with and without compiler option $X$,
respectively. If the relative difference of the feature value $f$ between the
two functions is small, it implies that $f$ is a robust feature against compiler
option $X$.

In this paper, we focus only on simple relative differences, rather than
exploring complex relationships among the features for interpretability.
However, we believe that our approach could be a stepping-stone toward
fabricating more improved interpretable models to understand such complex
relationships.

%

\begin{table}[t]
\centering
  \caption{Breakdown of the feature extracting time for \binkit.}
  \label{tab:time}
  \vspace{-6pt}
  \setlength\tabcolsep{0.1cm}
  \def\arraystretch{0.50}
  \scriptsize

\begin{threeparttable}
\begin{tabular}{@{}lrrrc@{}}
\toprule
 \multicolumn{1}{c}{\begin{tabular}[c]{@{}c@{}}Dataset\\ Name\end{tabular}} & \multicolumn{1}{c}{\begin{tabular}[c]{@{}c@{}}IDA\\ Pre-processing\\ (s)\end{tabular}} & \multicolumn{1}{c}{\begin{tabular}[c]{@{}c@{}}Ground Truth\\ Building\\ (s)\end{tabular}} & \multicolumn{1}{c}{\begin{tabular}[c]{@{}c@{}}Feature\\ Extraction\\ (s)\end{tabular}} & \multicolumn{1}{c}{\begin{tabular}[c]{@{}c@{}}Avg. Feature\tnote{$\dagger$}\\ Extraction\\
 (ms)\end{tabular}} \\ \midrule
\normaldataset   & 14,968.42  & 3,380.01  & 661.81 & 0.08 \\
\sizeoptdataset  & 2,171.70   & 353.13    & 649.57 & 0.32 \\
\piedataset      & 13,893.92  & 2,601.74  & 133.60 & 0.02 \\
\noinlinedataset & 14,780.06  & 3,883.88  & 579.82 & 0.06 \\
\ltodataset      & 5,263.97   & 1,314.94  & 392.48 & 0.12 \\
\obfuscationdataset & 97,723.47  & 1,766.60  & 4,189.44       & 1.03 \\ \bottomrule
\end{tabular}

\begin{tablenotes}
\item [$\dagger$] The average time spent for extracting features from a function,
  which is computed by dividing the total time (the fourth
  column of this table) by the number of functions (the last column
  of~\autoref{tab:dataset}).

\end{tablenotes}
\end{threeparttable}

\vspace{-15pt}
\end{table}

\subsection{Feature Selection}
\label{ss:featureselection}

Based on our scoring metric, we perform lightweight preprocessing to select
useful features for BCSA as some features may not help in making a distinction
between functions.
To measure the quality of a given feature set, we compute the area under the
receiver operating characteristic (ROC) curve (i.e., the ROC AUC) of generated
models.

Suppose we are given a dataset in \sys, which is generated from source code
containing $N$ unique functions. In total, we have a maximum of $N \cdot M$ functions
in our dataset, where $M$ is the number of combinations of compiler options used
to generate the dataset. The actual number of functions can be less than $N
\cdot M$ due to function inlining.
%
%
For each unique function $\lambda$, we randomly select two other functions with
the following conditions. (1) A true positive (TP) function, \functp, is
generated from the same source code as in $\lambda$, with different compiler
options, and (2) a true negative (TN) function, \functn, is generated from
source code that is different from the one used to generate $\lambda$, with the
same compiler options as for \functp. We generate such pairs for each unique
function, thereby acquiring around $2 \cdot N$ function pairs.
We then compute the similarity scores for the functions in each pair and their
AUC.

We note that the same methodology has been used in prior
works~\cite{xu2017neural, gao2018vulseeker}.
We chose the method as it allows us to efficiently analyze the tendency over a
large-scale dataset.
%
One may also consider
top-k~\cite{xu2017neural,david2018firmup,ding2019asm2vec,gao2018vulseeker} or
precision@k~\cite{xu2017neural,ding2019asm2vec} as an evaluation metric, but
this approach has too much computational overhead: $O((N \cdot M)^2)$
operations.

Unfortunately, there is no efficient algorithm for selecting an optimal feature
subset to use; it is indeed a well-known NP-hard
problem~\cite{guyon2003introduction}. Therefore, we leverage a greedy feature
selection algorithm~\cite{caruana94greedyattribute}.
%
%
Starting from an empty set $\mathbb{F}$, we determine whether we can add a
feature to $\mathbb{F}$ to increase its AUC. For every possible feature, we make
a union with $\mathbb{F}$ and compute the corresponding AUC. We then select one
that maximizes the AUC and update $\mathbb{F}$ to include the selected feature.
We repeat this process until the AUC does not increase further by adding a new
feature.
Although our approach does not guarantee finding an optimal solution, it still
provides empirically meaningful results, as we describe in the following
sections.


\subsection{Experimental Setup}



For all experiments in this study, we perform 10-fold cross-validation on each
test.
When we split a test dataset, we ensure functions that share the same source
code (\ie, source file name and line number) are either in a training or testing
set, but not in both.
For each fold, during the learning phase, \ie, the feature selection phase, we
select up to 200K functions from a training set and conduct feature selection,
as training millions of functions would take a significant amount of time.
Limiting the number of functions for training may degrade the final results.
However, when we tested the number of functions from 100K to 1000K, the results
remained almost consistent.
%
%
In the validation phase, we test all the functions in the testing set
without any sampling. Thus, after 10-fold validation, all the functions in the
target dataset are tested at least once.


We ran all our experiments on a server equipped with four Intel Xeon E7-8867v4
2.40 GHz CPUs (total 144 cores), 896 GB DDR4 RAM, and 8 TB SSD. We set up Ubuntu
18.04.5 LTS with IDA Pro v6.95~\cite{ida}
on the server. For feature selection and similarity comparison, we utilized
Python scikit-learn~\cite{pedregosa2011scikit}, SciPy~\cite{scipy}, and
NumPy~\cite{walt2011numpy}.

\section{Presemantic Feature Analysis (RQ2)}
\label{sec:rq2}

We now present our experimental results using \techname on the presemantic
features (\autoref{ss:features}) to answer RQ2 (\autoref{ss:rq}). With our
comprehensive analysis of these features, we obtained several useful insights
for future research. In this section, we discuss our findings and lessons
learned.

\begin{table*}[t]
  \centering
  \caption{In-depth analysis results of presemantic features obtained by running \tiknib on \sys.}

  \label{tab:result}
  \vspace{-6pt}
  \setlength\tabcolsep{0.7pt}
  \def\arraystretch{0.50}
  \scriptsize

  \begin{threeparttable}
\begin{tabular}{@{}clllccccccccccccccccccccccccccccccccccc@{}}
\toprule
\multirow{5}{*}{\rotatebox[origin=l]{90}{\textbf{Index}}}
    &
     \multicolumn{3}{c}{
     \multirow{5}{*}{
     \textbf{Description}}}
     &  &
     &  & \multicolumn{3}{c}{\textbf{Opt Level}}
     &  & \multicolumn{4}{c}{\textbf{Compiler}} 
     &  & \multicolumn{6}{c}{\textbf{Arch}}
     &  & \multicolumn{3}{c}{\textbf{vs. SizeOpt}\tnote{$\dagger$}}
     &  & \multicolumn{3}{c}{\textbf{vs. Extra}\tnote{$\dagger$}}
     &  & \multicolumn{4}{c}{\textbf{vs. Obfus.}\tnote{$\dagger$}}
     &  & \multicolumn{2}{c}{\textbf{Bad}\tnote{$\ddagger$}}
     \\
     
     \cmidrule(lr){7-10}
     \cmidrule(lr){11-15}
     \cmidrule(lr){16-22}
     \cmidrule(lr){23-26}
     \cmidrule(lr){27-30}
     \cmidrule(lr){31-35}
     \cmidrule(l){36-38}
&  &  & 
  &  &
  {\tiny \textbf{Rand.}} &
  &
  {\tiny \textbf{\begin{tabular}[c]{@{}c@{}}O0\\ vs.\\ O3\end{tabular}}} &
  {\tiny \textbf{\begin{tabular}[c]{@{}c@{}}O2\\ vs.\\ O3\end{tabular}}} &
  {\tiny \textbf{Rand.}} &
  &
  {\tiny \textbf{\begin{tabular}[c]{@{}c@{}}\gcc v4\\ vs.\\ \gcc v8\end{tabular}}} &
  {\tiny \textbf{\begin{tabular}[c]{@{}c@{}}\clang v4\\ vs.\\ \clang v7\end{tabular}}} &
  {\tiny \textbf{\begin{tabular}[c]{@{}c@{}}\gcc\\ vs.\\ \clang\end{tabular}}} &
  {\tiny \textbf{Rand.}} &
  &
  {\tiny \textbf{\begin{tabular}[c]{@{}c@{}}x86\\ vs.\\ ARM\end{tabular}}} &
  {\tiny \textbf{\begin{tabular}[c]{@{}c@{}}x86\\ vs.\\ MIPS\end{tabular}}} &
  {\tiny \textbf{\begin{tabular}[c]{@{}c@{}}ARM\\ vs.\\ MIPS\end{tabular}}} &
  {\tiny \textbf{\begin{tabular}[c]{@{}c@{}}32\\ vs.\\ 64\end{tabular}}} &
  {\tiny \textbf{\begin{tabular}[c]{@{}c@{}}LE\\ vs.\\ BE\end{tabular}}} &
  {\tiny \textbf{Rand.}} &
  &
  {\tiny \textbf{\begin{tabular}[c]{@{}c@{}}O0\\ vs.\\ Os\end{tabular}}} &
  {\tiny \textbf{\begin{tabular}[c]{@{}c@{}}O1\\ vs.\\ Os\end{tabular}}} &
  {\tiny \textbf{\begin{tabular}[c]{@{}c@{}}O3\\ vs.\\ Os\end{tabular}}} &
  &
  {\tiny \textbf{PIE}} &
  {\tiny \textbf{NoInline}} &
  {\tiny \textbf{LTO}} &
  &
  {\tiny \textbf{BCF}} &
  {\tiny \textbf{FLA}} &
  {\tiny \textbf{SUB}} &
  {\tiny \textbf{All}} &
  &
  {\tiny \textbf{Norm.}} &
  {\tiny \textbf{\begin{tabular}[c]{@{}c@{}}Norm.\\ vs.\\ Obfus.\end{tabular}}\tnote{$\dagger$}} \\
  \midrule

\multirow{2}{*}{\CC{1}} & \multicolumn{3}{l}{\# of Train Pairs ($10^6$)}     &      & 0.40       &      & 0.13       & 0.19       & 0.36       &      & 0.19     & 0.19 & 0.19  & 0.40       &      & 0.17 & 0.16  & 0.17  & 0.18       & 0.20       & 0.39       &      & 0.14       & 0.17       & 0.18       &      & 6.04\tnote{$\uparrow$}  & 0.17     & 0.19   &      & 0.19   & 0.19   & 0.20   & 0.18   &      & 3.63\tnote{$\uparrow$}      & 4.56\tnote{$\uparrow$}   \\
 & \multicolumn{3}{l}{\# of Test Pairs ($10^6$)}      &      & 1.58       &      & 0.26       & 0.33       & 1.43       &      & 0.16     & 0.17 & 0.75  & 1.57       &      & 0.17 & 0.16  & 0.17  & 0.71       & 0.40       & 1.55       &      & 0.29       & 0.34       & 0.32       &      & 0.24   & 1.09     & 1.37   &      & 0.17   & 0.17   & 0.18   & 0.17   &      & 0.40\tnote{$\uparrow$}      & 0.51\tnote{$\uparrow$}   \\ \midrule
\multirow{2}{*}{\CC{2}} & \multicolumn{3}{l}{Train Time (sec)\tnote{$\ast$}}     &      & 60.0       &      & 24.6       & 25.3       & 77.3       &      & 28.9     & 24.6 & 35.7  & 76.9       &      & 29.0 & 28.8  & 28.0  & 19.0       & 22.4       & 48.2       &      & 20.8       & 24.4       & 15.6       &      & 10.5   & 12.6     & 42.0   &      & 25.0   & 28.8   & 44.1   & 17.6   &      & 5.9& 5.4     \\
& \multicolumn{3}{l}{Test Time (sec)\tnote{$\ast$}}    &      & 59.1       &      & 23.3       & 23.8       & 49.7       &      & 12.0     & 11.8 & 40.9  & 54.4       &      & 12.3 & 12.5  & 12.0  & 39.9       & 22.8       & 52.2       &      & 24.8       & 23.2       & 22.6       &      & 32.0   & 62.5     & 60.7   &      & 11.9   & 11.3   & 10.6   & 11.2   &      & 1.6& 1.6     \\ \midrule

\multirow{30}[25]{*}{\CC{3}} & \multirow{30}[25]{*}{\rotatebox[origin=c]{90}{TP-TN Gap of Features}} & CFG  & \# of edges per BB       &      & 0.42       &      & 0.37       & 0.45       & 0.42       &      & 0.45     & 0.45 & 0.41  & \cellcolor[HTML]{C0C0C0}0.44      &      & 0.43 & 0.41  & 0.41  & 0.44       & 0.46       & 0.43       &      & 0.40       & 0.43       & 0.44       &      & 0.38   & 0.48     & 0.47   &      & 0.31   & \cellcolor[HTML]{C0C0C0}0.37  & 0.46   & \cellcolor[HTML]{C0C0C0}0.29  &      & 0.38       & 0.22    \\
& & CFG  & \# of edges      &      & 0.57       &      & 0.50       & 0.68       & 0.58       &      & 0.68     & 0.69 & 0.57  & 0.64       &      & 0.67 & \cellcolor[HTML]{C0C0C0}0.62 & \cellcolor[HTML]{C0C0C0}0.63 & 0.67       & 0.72       & \cellcolor[HTML]{C0C0C0}0.64      &      & 0.54       & 0.63       & 0.60       &      & 0.63   & 0.66     & 0.72   &      & 0.40   & 0.37   & 0.72   & 0.31   &      & \cellcolor[HTML]{C0C0C0}0.52      & 0.25    \\
& & CFG  & \# of loops      &      & 0.45       &      & 0.46       & 0.49       & 0.45       &      & 0.50     & 0.47 & 0.46  & 0.48       &      & \cellcolor[HTML]{C0C0C0}0.50& \cellcolor[HTML]{C0C0C0}0.49 & \cellcolor[HTML]{C0C0C0}0.50 & 0.50       & \cellcolor[HTML]{C0C0C0}0.51      & 0.49       &      & 0.48       & 0.49       & 0.45       &      & 0.51   & 0.45     & 0.51   &      & 0.46   & 0.36   & 0.50   & 0.29   &      & 0.44       & 0.22    \\
& & CFG  & \# of inter loops &      & \cellcolor[HTML]{C0C0C0}0.46      &      & \cellcolor[HTML]{C0C0C0}0.47      & 0.49       & \cellcolor[HTML]{C0C0C0}0.46      &      & \cellcolor[HTML]{C0C0C0}0.50    & 0.48 & \cellcolor[HTML]{C0C0C0}0.47 & \cellcolor[HTML]{C0C0C0}0.48      &      & \cellcolor[HTML]{C0C0C0}0.50& \cellcolor[HTML]{C0C0C0}0.49 & \cellcolor[HTML]{C0C0C0}0.50 & \cellcolor[HTML]{C0C0C0}0.50      & 0.50       & \cellcolor[HTML]{C0C0C0}0.49      &      & \cellcolor[HTML]{C0C0C0}0.48      & 0.49       & 0.45       &      & 0.51   & 0.45     & \cellcolor[HTML]{C0C0C0}0.50  &      & 0.46   & 0.38   & \cellcolor[HTML]{C0C0C0}0.50  & 0.32   &      & \cellcolor[HTML]{C0C0C0}0.45      & 0.27    \\
& & CG  & \# of callees    &      & 0.57       &      & \cellcolor[HTML]{C0C0C0}0.52      & \cellcolor[HTML]{C0C0C0}0.63      & \cellcolor[HTML]{C0C0C0}0.58      &      & \cellcolor[HTML]{C0C0C0}0.64    & \cellcolor[HTML]{C0C0C0}0.63& \cellcolor[HTML]{C0C0C0}0.57 & \cellcolor[HTML]{C0C0C0}0.61      &      & \cellcolor[HTML]{C0C0C0}0.64& \cellcolor[HTML]{C0C0C0}0.57 & 0.58  & 0.60       & \cellcolor[HTML]{C0C0C0}0.64      & 0.59       &      & 0.54       & \cellcolor[HTML]{C0C0C0}0.62      & \cellcolor[HTML]{C0C0C0}0.60      &      & \cellcolor[HTML]{C0C0C0}0.56  & \cellcolor[HTML]{C0C0C0}0.61    & 0.62   &      & \cellcolor[HTML]{C0C0C0}0.61  & 0.61   & \cellcolor[HTML]{C0C0C0}0.64  & \cellcolor[HTML]{C0C0C0}0.60  &      & \cellcolor[HTML]{C0C0C0}0.52      & \cellcolor[HTML]{C0C0C0}0.56   \\
& & CG  & \# of callers    &      & 0.55       &      & \cellcolor[HTML]{C0C0C0}0.55      & \cellcolor[HTML]{C0C0C0}0.59      & \cellcolor[HTML]{C0C0C0}0.56      &      & \cellcolor[HTML]{C0C0C0}0.60    & \cellcolor[HTML]{C0C0C0}0.58& \cellcolor[HTML]{C0C0C0}0.54 & \cellcolor[HTML]{C0C0C0}0.58      &      & 0.58 & 0.51  & 0.53  & 0.56       & \cellcolor[HTML]{C0C0C0}0.60      & 0.57       &      & \cellcolor[HTML]{C0C0C0}0.58      & \cellcolor[HTML]{C0C0C0}0.60      & 0.58       &      & 0.54   & 0.58     & 0.57   &      & \cellcolor[HTML]{C0C0C0}0.56  & 0.56   & \cellcolor[HTML]{C0C0C0}0.58  & \cellcolor[HTML]{C0C0C0}0.55  &      & \cellcolor[HTML]{C0C0C0}0.54      & \cellcolor[HTML]{C0C0C0}0.54   \\
& & CG  & \# of imported callees   &      & 0.55       &      & \cellcolor[HTML]{C0C0C0}0.56      & 0.63       & 0.58       &      & \cellcolor[HTML]{C0C0C0}0.62    & \cellcolor[HTML]{C0C0C0}0.59& \cellcolor[HTML]{C0C0C0}0.55 & \cellcolor[HTML]{C0C0C0}0.59      &      & \cellcolor[HTML]{C0C0C0}0.63& 0.49  & 0.50  & 0.59       & 0.57       & 0.57       &      & 0.58       & 0.61       & 0.58       &      & 0.58   & 0.59     & \cellcolor[HTML]{C0C0C0}0.59  &      & \cellcolor[HTML]{C0C0C0}0.56  & \cellcolor[HTML]{C0C0C0}0.57  & \cellcolor[HTML]{C0C0C0}0.59  & \cellcolor[HTML]{C0C0C0}0.56  &      & 0.52       & 0.55    \\
& & CG  & \# of imported calls     &      & 0.56       &      & 0.57       & 0.65       & 0.59       &      & 0.65     & 0.62 & 0.56  & 0.61       &      & 0.67 & 0.50  & 0.51  & 0.62       & \cellcolor[HTML]{C0C0C0}0.60      & 0.59       &      & \cellcolor[HTML]{C0C0C0}0.59      & \cellcolor[HTML]{C0C0C0}0.62      & 0.59       &      & 0.61   & 0.60     & 0.62   &      & 0.53   & 0.59   & \cellcolor[HTML]{C0C0C0}0.62  & 0.52   &      & 0.52       & \cellcolor[HTML]{C0C0C0}0.51   \\
& & CG  & \# of incoming calls     &      & \cellcolor[HTML]{C0C0C0}0.56      &      & \cellcolor[HTML]{C0C0C0}0.55      & \cellcolor[HTML]{C0C0C0}0.61      & \cellcolor[HTML]{C0C0C0}0.58      &      & \cellcolor[HTML]{C0C0C0}0.63    & \cellcolor[HTML]{C0C0C0}0.60& \cellcolor[HTML]{C0C0C0}0.55 & \cellcolor[HTML]{C0C0C0}0.59      &      & \cellcolor[HTML]{C0C0C0}0.61& \cellcolor[HTML]{C0C0C0}0.52 & \cellcolor[HTML]{C0C0C0}0.54 & \cellcolor[HTML]{C0C0C0}0.59      & \cellcolor[HTML]{C0C0C0}0.63      & \cellcolor[HTML]{C0C0C0}0.59      &      & \cellcolor[HTML]{C0C0C0}0.60      & \cellcolor[HTML]{C0C0C0}0.62      & 0.59       &      & \cellcolor[HTML]{C0C0C0}0.57  & \cellcolor[HTML]{C0C0C0}0.61    & \cellcolor[HTML]{C0C0C0}0.60  &      & 0.54   & \cellcolor[HTML]{C0C0C0}0.58  & \cellcolor[HTML]{C0C0C0}0.61  & 0.53   &      & \cellcolor[HTML]{C0C0C0}0.54      & 0.50    \\
& & CG  & \# of outgoing calls     &      & 0.59       &      & 0.53       & \cellcolor[HTML]{C0C0C0}0.67      & 0.60       &      & 0.68     & 0.67 & 0.58  & 0.64       &      & \cellcolor[HTML]{C0C0C0}0.69& \cellcolor[HTML]{C0C0C0}0.60 & \cellcolor[HTML]{C0C0C0}0.60 & 0.63       & 0.69       & 0.63       &      & 0.56       & 0.65       & 0.62       &      & 0.60   & 0.64     & 0.66   &      & 0.57   & \cellcolor[HTML]{C0C0C0}0.64  & \cellcolor[HTML]{C0C0C0}0.68  & 0.55   &      & \cellcolor[HTML]{C0C0C0}0.54      & 0.53    \\
& & Inst & Avg. \# of arith+shift   &      & 0.35       &      & \cellcolor[HTML]{C0C0C0}0.35      & \cellcolor[HTML]{C0C0C0}0.55      & 0.43       &      & 0.52     & 0.52 & 0.36  & 0.47       &      & 0.39 & 0.25  & \cellcolor[HTML]{C0C0C0}0.26 & 0.40       & 0.53       & 0.36       &      & 0.36       & \cellcolor[HTML]{C0C0C0}0.49      & \cellcolor[HTML]{C0C0C0}0.49      &      & 0.43   & \cellcolor[HTML]{C0C0C0}0.55    & 0.53   &      & 0.35   & 0.35   & 0.50   & 0.30   &      & 0.27       & 0.24    \\
& & Inst & Avg. \# of compare       &      & 0.44       &      & 0.40       & 0.52       & 0.46       &      & 0.51     & 0.52 & 0.44  & 0.49       &      & 0.45 & 0.44  & 0.37  & 0.50       & 0.55       & 0.46       &      & 0.43       & \cellcolor[HTML]{C0C0C0}0.50      & 0.49       &      & 0.46   & 0.54     & 0.55   &      & 0.33   & 0.37   & 0.54   & 0.29   &      & 0.39       & 0.21    \\
& & Inst & Avg. \# of ctransfer &      & 0.34       &      & 0.33       & \cellcolor[HTML]{C0C0C0}0.45      & 0.37       &      & 0.42     & 0.42 & 0.33  & 0.40       &      & 0.39 & 0.36  & 0.38  & 0.39       & 0.46       & 0.37       &      & 0.35       & 0.39       & 0.39       &      & 0.38   & 0.43     & \cellcolor[HTML]{C0C0C0}0.46  &      & 0.30   & 0.28   & 0.43   & 0.24   &      & 0.31       & 0.18    \\
& & Inst & Avg. \# of ctransfer+cond.  &      & 0.27       &      & \cellcolor[HTML]{C0C0C0}0.25      & 0.32       & \cellcolor[HTML]{C0C0C0}0.28      &      & 0.32     & 0.31 & 0.26  & 0.30       &      & 0.29 & 0.28  & 0.29  & 0.29       & 0.33       & 0.28       &      & \cellcolor[HTML]{C0C0C0}0.28      & 0.29       & 0.29       &      & 0.28   & 0.33     & 0.34   &      & 0.23   & 0.20   & 0.32   & 0.17   &      & 0.22       & 0.11    \\
& & Inst & Avg. \# of dtransfer     &      & \cellcolor[HTML]{C0C0C0}0.34      &      & 0.30       & \cellcolor[HTML]{C0C0C0}0.49      & \cellcolor[HTML]{C0C0C0}0.38      &      & \cellcolor[HTML]{C0C0C0}0.46    & \cellcolor[HTML]{C0C0C0}0.49& 0.35  & 0.43       &      & 0.35 & \cellcolor[HTML]{C0C0C0}0.31 & \cellcolor[HTML]{C0C0C0}0.33 & 0.38       & 0.52       & 0.36       &      & 0.31       & \cellcolor[HTML]{C0C0C0}0.43      & 0.43       &      & 0.37   & 0.49     & 0.49   &      & 0.35   & 0.33   & 0.47   & 0.27   &      & 0.31       & 0.19    \\
& & Inst & Avg. \# of dtransfer+misc&      & 0.32       &      & 0.28       & 0.48       & 0.36       &      & 0.44     & 0.47 & 0.34  & 0.41       &      & 0.35 & 0.28  & 0.30  & 0.37       & 0.49       & 0.34       &      & 0.29       & 0.42       & 0.42       &      & 0.35   & \cellcolor[HTML]{C0C0C0}0.48    & \cellcolor[HTML]{C0C0C0}0.48  &      & 0.34   & 0.32   & 0.45   & 0.26   &      & 0.29       & 0.18    \\
& & Inst & Avg. \# of float instrs. &      & 0.25       &      & 0.30       & 0.34       & 0.28       &      & 0.25     & 0.34 & 0.28  & 0.26       &      & 0.28 & 0.31  & 0.29  & 0.31       & 0.40       & 0.26       &      & 0.28       & 0.31       & 0.31       &      & 0.29   & 0.28     & \cellcolor[HTML]{C0C0C0}0.33  &      & 0.25   & 0.31   & 0.35   & 0.25   &      & 0.44       & 0.20    \\
& & Inst & Avg. \# of total instrs. &      & 0.30       &      & \cellcolor[HTML]{C0C0C0}0.27      & 0.42       & 0.34       &      & 0.40     & 0.42 & 0.32  & 0.38       &      & \cellcolor[HTML]{C0C0C0}0.33& 0.28  & 0.28  & 0.34       & 0.45       & 0.32       &      & 0.28       & 0.38       & 0.38       &      & \cellcolor[HTML]{C0C0C0}0.33  & 0.42     & 0.43   &      & \cellcolor[HTML]{C0C0C0}0.31  & \cellcolor[HTML]{C0C0C0}0.29  & 0.40   & \cellcolor[HTML]{C0C0C0}0.24  &      & \cellcolor[HTML]{C0C0C0}0.28      & \cellcolor[HTML]{C0C0C0}0.17   \\
& & Inst & \# of arith      &      & 0.40       &      & 0.43       & 0.64       & 0.51       &      & 0.62     & 0.61 & 0.46  & 0.55       &      & 0.43 & 0.28  & 0.29  & 0.48       & 0.62       & 0.41       &      & \cellcolor[HTML]{C0C0C0}0.44      & 0.58       & 0.56       &      & 0.53   & 0.61     & 0.60   &      & 0.40   & 0.50   & 0.58   & 0.35   &      & 0.33       & 0.27    \\
& & Inst & \# of arith+shift&      & 0.40       &      & 0.43       & 0.64       & \cellcolor[HTML]{C0C0C0}0.51      &      & 0.62     & 0.61 & \cellcolor[HTML]{C0C0C0}0.46 & \cellcolor[HTML]{C0C0C0}0.55      &      & 0.44 & 0.27  & 0.28  & 0.47       & 0.63       & 0.41       &      & 0.43       & 0.58       & 0.56       &      & 0.53   & 0.61     & 0.61   &      & 0.40   & \cellcolor[HTML]{C0C0C0}0.50  & 0.59   & 0.35   &      & 0.32       & 0.27    \\
& & Inst & \# of bit-manipulating   &      & 0.26       &      & 0.29       & 0.35       & 0.28       &      & 0.33     & 0.32 & 0.26  & 0.30       &      & 0.19 & 0.13  & 0.20  & 0.33       & 0.17       & 0.25       &      & 0.27       & 0.32       & 0.30       &      & 0.33   & 0.31     & \cellcolor[HTML]{C0C0C0}0.33  &      & 0.25   & 0.23   & 0.24   & 0.18   &      & 0.41       & 0.03    \\
& & Inst & \# of compare    &      & 0.56       &      & \cellcolor[HTML]{C0C0C0}0.53      & 0.67       & \cellcolor[HTML]{C0C0C0}0.61      &      & \cellcolor[HTML]{C0C0C0}0.69    & 0.69 & 0.60  & 0.65       &      & 0.50 & \cellcolor[HTML]{C0C0C0}0.60 & 0.44  & 0.65       & 0.72       & 0.60       &      & \cellcolor[HTML]{C0C0C0}0.59      & 0.64       & 0.60       &      & \cellcolor[HTML]{C0C0C0}0.64  & 0.65     & \cellcolor[HTML]{C0C0C0}0.71  &      & 0.39   & 0.40   & 0.70   & 0.30   &      & 0.50       & 0.22    \\
& & Inst & \# of ctransfer  &      & \cellcolor[HTML]{C0C0C0}0.50      &      & \cellcolor[HTML]{C0C0C0}0.45      & 0.61       & \cellcolor[HTML]{C0C0C0}0.52      &      & 0.60     & 0.62 & 0.49  & 0.56       &      & \cellcolor[HTML]{C0C0C0}0.60& \cellcolor[HTML]{C0C0C0}0.54 & \cellcolor[HTML]{C0C0C0}0.57 & \cellcolor[HTML]{C0C0C0}0.58      & 0.65       & \cellcolor[HTML]{C0C0C0}0.56      &      & \cellcolor[HTML]{C0C0C0}0.45      & 0.55       & 0.54       &      & \cellcolor[HTML]{C0C0C0}0.57  & 0.58     & \cellcolor[HTML]{C0C0C0}0.63  &      & 0.39   & 0.38   & \cellcolor[HTML]{C0C0C0}0.64  & 0.33   &      & 0.41       & 0.26    \\
& & Inst & \# of cond. ctransfer &      & \cellcolor[HTML]{C0C0C0}0.60      &      & 0.54       & 0.68       & 0.61       &      & 0.69     & \cellcolor[HTML]{C0C0C0}0.70& \cellcolor[HTML]{C0C0C0}0.63 & \cellcolor[HTML]{C0C0C0}0.67      &      & \cellcolor[HTML]{C0C0C0}0.69& \cellcolor[HTML]{C0C0C0}0.66 & \cellcolor[HTML]{C0C0C0}0.67 & \cellcolor[HTML]{C0C0C0}0.69      & 0.72       & \cellcolor[HTML]{C0C0C0}0.67      &      & \cellcolor[HTML]{C0C0C0}0.60      & 0.64       & 0.60       &      & 0.64   & 0.65     & \cellcolor[HTML]{C0C0C0}0.72  &      & 0.39   & 0.37   & 0.71   & 0.31   &      & 0.52       & 0.25    \\
& & Inst & \# of dtransfer  &      & 0.42       &      & \cellcolor[HTML]{C0C0C0}0.36      & 0.63       & \cellcolor[HTML]{C0C0C0}0.46      &      & 0.61     & 0.61 & \cellcolor[HTML]{C0C0C0}0.48 & \cellcolor[HTML]{C0C0C0}0.55      &      & 0.48 & \cellcolor[HTML]{C0C0C0}0.37 & \cellcolor[HTML]{C0C0C0}0.41 & 0.50       & 0.64       & \cellcolor[HTML]{C0C0C0}0.46      &      & \cellcolor[HTML]{C0C0C0}0.34      & \cellcolor[HTML]{C0C0C0}0.57      & 0.56       &      & 0.49   & \cellcolor[HTML]{C0C0C0}0.61    & 0.61   &      & \cellcolor[HTML]{C0C0C0}0.41  & 0.39   & 0.61   & 0.33   &      & \cellcolor[HTML]{C0C0C0}0.38      & 0.26    \\
& & Inst & \# of dtransfer+misc     &      & 0.41       &      & 0.35       & 0.63       & 0.45       &      & 0.60     & 0.61 & 0.48  & 0.55       &      & 0.50 & 0.35  & 0.37  & 0.50       & 0.64       & 0.45       &      & 0.33       & 0.56       & 0.55       &      & 0.48   & 0.61     & \cellcolor[HTML]{C0C0C0}0.62  &      & 0.40   & 0.39   & 0.61   & 0.33   &      & 0.39       & 0.25    \\
& & Inst & \# of float instrs.      &      & 0.25       &      & 0.30       & \cellcolor[HTML]{C0C0C0}0.34      & 0.28       &      & 0.25     & 0.35 & 0.28  & 0.27       &      & 0.28 & 0.31  & 0.29  & 0.31       & 0.40       & 0.26       &      & 0.28       & 0.31       & 0.32       &      & 0.29   & 0.28     & \cellcolor[HTML]{C0C0C0}0.33  &      & 0.30   & 0.33   & \cellcolor[HTML]{C0C0C0}0.35  & 0.26   &      & 0.44       & 0.20    \\
& & Inst & \# of misc       &      & 0.30       &      & 0.26       & 0.52       & 0.34       &      & 0.48     & 0.48 & 0.33  & 0.42       &      & 0.11 & 0.31  & 0.27  & 0.46       & 0.67       & 0.34       &      & 0.23       & 0.39       & 0.38       &      & 0.38   & 0.48     & 0.50   &      & 0.30   & 0.31   & \cellcolor[HTML]{C0C0C0}0.50  & 0.26   &      & 0.32       & 0.15    \\
& & Inst & \# of shift      &      & 0.36       &      & 0.38       & 0.45       & 0.38       &      & 0.45     & 0.40 & 0.36  & 0.41       &      & 0.30 & 0.46  & 0.47  & 0.42       & 0.55       & 0.38       &      & 0.34       & 0.39       & 0.38       &      & 0.42   & 0.39     & \cellcolor[HTML]{C0C0C0}0.47  &      & \cellcolor[HTML]{C0C0C0}0.41  & \cellcolor[HTML]{C0C0C0}0.40  & 0.44   & \cellcolor[HTML]{C0C0C0}0.38  &      & 0.48       & 0.49    \\
& & Inst & \# of total instrs.      &      & 0.43       &      & 0.38       & 0.62       & 0.47       &      & \cellcolor[HTML]{C0C0C0}0.60    & \cellcolor[HTML]{C0C0C0}0.61& \cellcolor[HTML]{C0C0C0}0.50 & \cellcolor[HTML]{C0C0C0}0.56      &      & \cellcolor[HTML]{C0C0C0}0.54& 0.37  & 0.38  & 0.52       & \cellcolor[HTML]{C0C0C0}0.63      & 0.47       &      & 0.35       & \cellcolor[HTML]{C0C0C0}0.56      & 0.54       &      & 0.50   & 0.59     & \cellcolor[HTML]{C0C0C0}0.61  &      & 0.39   & 0.39   & 0.59   & 0.32   &      & 0.42       & 0.25    \\ \midrule
\multirow{2}{*}{\CC{4}} & \multicolumn{3}{l}{Avg. TP-TN Gap}&      & 0.42       &      & 0.41       & 0.53       & 0.45       &      & 0.52     & 0.52 & 0.44  & 0.49       &      & 0.46 & 0.42  & 0.43  & 0.49       & 0.55       & 0.46       &      & 0.42       & 0.49       & 0.48       &      & 0.48   & 0.51     & 0.54   &      & 0.39   & 0.38   & 0.53   & 0.32   &      & 0.42       & 0.27    \\
& \multicolumn{3}{l}{Avg. TP-TN Gap of Grey}&      & 0.49       &      & 0.44       & 0.54       & 0.49       &      & 0.59     & 0.60 & 0.52  & 0.56       &      & 0.57 & 0.52  & 0.50  & 0.59       & 0.60       & 0.57       &      & 0.49       & 0.56       & 0.54       &      & 0.53   & 0.57     & 0.53   &      & 0.48   & 0.48   & 0.57   & 0.44   &      & 0.47       & 0.45    \\ \midrule
\CC{5} & \multicolumn{3}{l}{Avg. \# of Selected Features}  &      & 8.5&      & 13.9       & 9.3& 12.9       &      & 10.1     & 8.7  & 11.7  & 11.0       &      & 11.0 & 12.0  & 11.0  & 7.1& 8.4& 7.3&      & 11.0       & 10.0       & 5.7&      & 14.3   & 5.3      & 16.5   &      & 8.3    & 9.9    & 15.7   & 6.1    &      & 11.6       & 8.1     \\ \midrule
\multirow{2}{*}{\CC{6}} & \multicolumn{3}{l}{ROC AUC}       &      & 0.95       &      & 0.93       & 0.99       & 0.96       &      & 1.00     & 1.00 & 0.97  & 0.98       &      & 1.00 & 0.98  & 0.98  & 0.99       & 1.00       & 0.98       &      & 0.96       & 0.99       & 0.97       &      & 0.97   & 0.98     & 1.00   &      & 0.98   & 0.98   & 1.00   & 0.95   &      & 0.93       & 0.91    \\
& \multicolumn{3}{l}{Std. of ROC AUC}       &      & 0.00       &      & 0.00       & 0.00       & 0.00       &      & 0.00     & 0.00 & 0.00  & 0.00       &      & 0.00 & 0.00  & 0.00  & 0.00       & 0.00       & 0.00       &      & 0.00       & 0.00       & 0.00       &      & 0.00   & 0.00     & 0.00   &      & 0.00   & 0.00   & 0.00   & 0.00   &      & 0.00       & 0.00    \\
\midrule
\multirow{2}{*}{\CC{7}} & \multicolumn{3}{l}{Average Precision (AP)}&      & 0.95       &      & 0.93       & 0.99       & 0.97       &      & 1.00     & 1.00 & 0.97  & 0.99       &      & 0.99 & 0.97  & 0.98  & 0.99       & 1.00       & 0.98       &      & 0.96       & 0.99       & 0.98       &      & 0.98   & 0.98     & 1.00   &      & 0.98   & 0.98   & 1.00   & 0.95   &      & 0.93       & 0.90    \\
& \multicolumn{3}{l}{Std. of AP}    &      & 0.00       &      & 0.00       & 0.00       & 0.00       &      & 0.00     & 0.00 & 0.00  & 0.00       &      & 0.00 & 0.00  & 0.00  & 0.00       & 0.00       & 0.00       &      & 0.00       & 0.00       & 0.00       &      & 0.00   & 0.00     & 0.00   &      & 0.00   & 0.00   & 0.00   & 0.00   &      & 0.00       & 0.01    \\
\bottomrule

\end{tabular}

\begin{tablenotes}
\item 
  All values in the table are 10-fold cross validation averages. We
  color a cell gray if a feature was consistently selected (\ie, 10 times) during
  the 10-fold validation.
  Due to the space constraints, we only display features that have been selected
  at least once during the 10-fold validation.

\item [$\dagger$] We compare a function from the \normaldataset to the corresponding function in each target dataset.

\item [$\ddagger$] We match functions whose compiler options are largely distant
  to test for bad cases. Please refer to~\autoref{sss:combining} for additional information.

\item [$\uparrow$]
  These in the first rows (\CC{1}) are divided by $10^4$ instead
  of $10^6$.

\item [$\ast$]
  The train and test times in the seconds rows (\CC{2}) do not include
  the time for data loading.
\end{tablenotes}
\end{threeparttable}

\vspace{-15pt}
\end{table*}


\subsection{Analysis Result}
\label{ss:result}

To analyze the impact of various compiler options and target architectures on
BCSA, we conducted a total of \numTest tests
using \techname.
We conducted the tests on our benchmark, \sys,
with the ground truth that we built in~\autoref{sec:dataset}.
\autoref{tab:result} describes the experimental results where each column
corresponds to a test we performed.
Note that we present only
\numTestShown out of \numTest tests because of the space limit.
Unless otherwise specified, all the tests were performed on the \normaldataset
dataset.
As described in~\autoref{ss:featureselection}, we prepared 10-fold sets for each
test. We divided the tests into seven groups according to their purposes, as
shown in the top row of the table. For example, the \emph{Arch} group contains a
set of tests to evaluate each feature against varying target architectures.

For each test, we select function pairs for training and testing as described
in~\autoref{ss:featureselection}. That is, for a function $\lambda$, we select
its corresponding functions (\ie, \functp and \functn). Therefore, $N$ functions
produce $2 \cdot N$ functions pairs.
The first row (\CC{1}) of~\autoref{tab:result} shows the number of function
pairs for each test. When selecting these pairs, we deliberately choose the
target options based on the goal of each test.
For instance, we test the influence of
varying the target architecture from x86 to ARM
(\emph{x86 vs. ARM} column of~\autoref{tab:result}).
For each function $\lambda$ in the x86 binaries of our dataset,
we select both \functp and \functn from the ARM binaries
compiled with the same compiler option as in $\lambda$.
In other words, we fix all the other options,
except for the target architecture for
choosing \functp and \functn so we can focus on our testing goal.
The same rule applies to other columns.
For the \emph{Rand.} columns,
we alter all the compiler options in the group randomly
to generate function pairs.

The second row (\CC{2}) of~\autoref{tab:result} presents the time spent for
training and testing in each test, which excludes the time for loading the
function data on the memory. The average time spent for a single function was
less than 1 ms.
%
%
%

Each cell in the third row (\CC{3}) of~\autoref{tab:result}
represents the average of
$\delta(\lambda_f, \functn_f) - \delta(\lambda_f, \functp_f)$
for feature $f$, which we call the \emph{TP-TN gap} of $f$.
%
This TP-TN gap measures the similarity between $\functp$ and $\lambda$, as well
as the difference between $\functn$ and $\lambda$, in terms of the target
feature. Thus, when the gap of a feature is larger, its discriminative
capability for BCSA is higher.
As we conduct 10-fold validation for each test, we highlight the cells with gray
when the corresponding feature is chosen in all ten trials. Such features
show relatively higher TP-TN gaps than the others do in each test.
We also present the average TP-TN gaps in the fourth row (\CC{4}) of the table.

The average number of the selected features in each test is shown in the fifth
row (\CC{5}) of~\autoref{tab:result}.
A few presemantic features could achieve high AUCs and average precisions (APs),
as shown in the sixth row (\CC{6}) and seventh row (\CC{7}) of the same table,
respectively.
We now summarize our observations as follows.

\subsubsection{Optimization is largely influential}
%
Many researchers have focused on designing a model for \emph{cross-architecture}
BCSA~\cite{pewny2015cross, chandramohan2016bingo, eschweiler2016discovre,
hu2016cross, xue2018accurate}.
However, our experimental results show that architecture may not be the most
critical factor for BCSA. Instead, optimization level was the most influential
factor in terms of the relative difference between presemantic features.
In particular, we measured the average TP-TN gap of all the presemantic features
for each test (\emph{Avg. of TP-TN Gap} row of the table) and found that the
average gap of the \tc{O0} vs. \tc{O3} test (0.41) is less than that of the x86
vs. ARM test (0.46) and the x86 vs. MIPS test (0.42).
Furthermore, the optimization level random test (\emph{Rand.} column of the
\emph{Opt Level} group) shows the lowest AUC (0.96) compared to that of the
architecture and compiler group (0.98).
These results confirm that compilers can produce largely distinct binaries
depending on the optimization techniques used;
hence, the variation among the binaries
due to the optimization is
considerably greater than
that due to the target architecture on our dataset.

\subsubsection{Compiler version has a small impact}
%
Approximately one-third of the previous benchmarks shown
in~\autoref{tab:prevdataset} employ multiple versions of the same compiler.
%
%
However, we found that even the major versions of the same compiler produce
similar binaries. In other words, compiler versions do not heavily affect
presemantic features. Although~\autoref{tab:result} does not include all the
tests we performed because of the space constraints, it is apparent from the
\emph{Compiler} column that the two tests between two different versions of the
same compiler, i.e., \gcc v4 vs. \gcc v8 and \clang v4 vs. \clang v7, have much
higher TP-TN gaps (0.52) than other tests, and their AUCs are close to 1.0.

\subsubsection{\gcc and \clang have diverse characteristics}
%
Conversely, the \gcc vs. \clang test resulted in the lowest TP-TN gap (0.44) and AUC (0.97)
among the tests in the \emph{Compiler} group.
This can be because each compiler employs a different back-end, thereby
producing different binaries.
Another potential problem is that the techniques inside each optimization level
can vary depending on the compiler. We detail this
in~\autoref{ss:compiler-diversity}.

\subsubsection{ARM binaries are closer to x86 binaries than MIPS}
%
The tests in the \emph{Arch} group measure the influence of target architectures
with the \normaldataset dataset. Overall, the target architecture did not have
much of an effect on the accuracy rate. The AUCs were over 0.98 in all the
cases. Surprisingly, the x86 vs. ARM test had the highest
TP-TN gap (0.46) and AUC (1.0), indicating
that the presemantic features of the x86 and ARM binaries are similar to each
other, despite being distinct architectures.
The ARM vs. MIPS test
showed a lower TP-TN gap (0.43) and AUC (0.98)  although both of them are RISC architectures.
Additionally, the effect of the word size (\ie, bits) and endianness was relatively small.
Nevertheless, we cannot rule out the possibility that our feature extraction for
MIPS binaries is erroneous. We further discuss this issue
in~\autoref{ss:toolerror}.

%

\subsubsection{Closer optimization levels show similar results}
%
We also measured the effect of size optimization (\tc{Os}) by matching function
$\lambda$ in the \normaldataset dataset with a function (\functp and \functn) in
the \sizeoptdataset dataset.
Subsequently, the binaries compiled with the \tc{Os} option were similar to the
ones compiled with the \tc{O1} and \tc{O2} options.
This is not surprising because \tc{Os} enables most of the \tc{O2} techniques in
both \gcc and \clang~\cite{gccopt, clangopt}.
Furthermore, we observe that the \tc{O1} and \tc{O2} options produce similar
binaries, although this is not shown in~\autoref{tab:result} due to the space
limit.

\subsubsection{Extra options have less impact}
\label{sss:extra}
%
To assess the influence of the PIE, no-inline, and LTO options, we compared
functions in the \normaldataset dataset with those in the \piedataset,
\noinlinedataset, and \ltodataset datasets, respectively. For the no-inline test,
we limit the optimization level from \tc{O1} to \tc{O3} as function inlining is
applied from \tc{O1}.
It was observed that the influence of such extra options is not significant.
Binaries with and without the PIE option were similar to each other because it
only changes the instructions to use relative addresses; hence, it does not
affect our presemantic features. Function inlining also does not affect several
features, such as the number of incoming calls, which results in a high AUC
(0.97). LTO does not exhibit any notable effects either.

However, by analyzing each test case, we found that some options affect the AUC
more than others. For example, in the no-inline test, the AUC largely
decreases as the optimization level increases: \tc{O1} (0.995), \tc{O2} (0.981),
and \tc{O3} (0.967). This is because as more optimization techniques are applied,
more functions are inlined and transformed
in the \normaldataset,
while their corresponding functions in the \noinlinedataset are not inlined.
On the other hand, in the LTO test, the AUC increases as the version of \clang
increases: v4 (0.956), v5 (0.968), v6 (0.986), and v7 (0.986). In contrast, \gcc
shows stable AUCs (0.987--0.988) across all versions, and all the AUCs are higher than those of
\clang.
This result indicates that varying multiple options would significantly affect
the success rate, which we describe below.

\subsubsection{Obfuscator-LLVM does not affect CG features}
%
Many previous studies~\cite{\paperObfusLLVM} chose
Obfuscator-LLVM~\cite{junod2015obfuscator} for their obfuscation tests as it
significantly varies the binary code~\cite{ding2019asm2vec}. However, applying all
of its three obfuscation options shows an AUC of 0.95 on our dataset, which is
relatively higher than that of the optimization level tests.
Obfuscation severely decreases the average TP-TN gaps except for CG
features. This is because Obfuscator-LLVM applies intra-procedural obfuscation.
The \tc{SUB} obfuscation substitutes arithmetic instructions while preserving
the semantics; the \tc{BCF} obfuscation notably affects CFG features by adding
bogus control flows; the \tc{FLA} obfuscation changes the predicates of
control structures~\cite{laszlo2009obfuscating}. However, none of them conducts
inter-procedural obfuscation, which modifies the function call relationship.
Thus, we encourage future studies to use other obfuscators, such as
Themida~\cite{themida} or VMProtect~\cite{vmprotect}, for evaluating their
techniques against inter-procedural obfuscation.




\subsubsection{Comparison target option does matter}
\label{sss:combining}
%
Based on the experimental results thus far, we perform extra tests to understand
the influence of comparing multiple compiler options by intentionally selecting
\functp and \functn from binaries that could provide the lowest TP-TN gap. In
this study, we present two of them because of the space limit.
Specifically, for the first test, we selected functions from 32-bit ARM binaries
compiled using GCC v4 with the \tc{O0} option, and the corresponding \functp and
\functn functions from 64-bit MIPS big-endian binaries compiled using Clang v7
with the \tc{O3} option.
For the second test, we changed the \clang compiler to the Obfuscator-LLVM with
all three obfuscation options turned on.
The \emph{Bad} column of the table summarizes the results.
The AUC in both cases was approximately 0.93 and 0.91, respectively.
Their average TP-TN gaps were also
significantly lower (0.42 and 0.27) than those in the other tests.
%
%
This signifies the importance of choosing the comparison targets for evaluating
BCSA techniques. Existing BCSA research compares functions for all possible
targets in a dataset, as shown in the \emph{Rand.} tests in this study. However,
our results suggest that researchers should carefully choose evaluation targets
to avoid overlooking the influence of bad cases.



\subsection{Comparison Against State-of-the-Art Techniques}
\label{ss:comparison-state-of-the-art}

\begin{table}[t]
  \scriptsize
  \centering
  \setlength\tabcolsep{0.05cm}
  \def\arraystretch{0.50}

  \caption{Summary of datasets for comparing \tiknib to VulSeeker (\ie, \syscomp datasets).}
  \label{tab:syscomp}
  \vspace{-6pt}

  \begin{threeparttable}
 \begin{tabular}{@{}ccccrr@{}}
\toprule
       \textbf{Name} & \textbf{Package}  & \textbf{Architecture}     & \begin{tabular}[c]{@{}c@{}}\textbf{Compiler}\\ \textbf{(GCC)}\end{tabular}     & \begin{tabular}[c]{@{}c@{}}\textbf{\# of Orig.}\\ \textbf{Funcs}\end{tabular} & \begin{tabular}[c]{@{}c@{}}\textbf{\# of Final}\\ \textbf{Funcs}\end{tabular} \\ \midrule
ASE1  & \multicolumn{1}{c}{OpenSSL v1.0.1\{f,u\}}& \multicolumn{1}{c}{\{x86,arm,mips\}\_32} & \multicolumn{1}{c}{v5.5.0}   & 152K      & 126K      \\ \midrule
\aseii  & \multicolumn{1}{c}{\begin{tabular}[c]{@{}l@{}}OpenSSL v1.0.1\{f,u\}\\ BusyBox v1.21\\ Coreutils v6.\{5,7\}\end{tabular}} & ''       & ''   & 704K      & 183K      \\ \midrule
\aseiii & ''       & \multicolumn{1}{c}{\begin{tabular}[c]{@{}l@{}}\{x86,arm,mips\}\_32,\\ \{x86,arm,mips\}\_64\end{tabular}} & \multicolumn{1}{c}{\begin{tabular}[c]{@{}l@{}}v4.9.4,\\ v5.5.0\end{tabular}} & 2,777K    & 735K      \\ \midrule
\aseiv  & ''       & \multicolumn{2}{c}{Same as \normaldataset options}       & 16,799K   & 4,467K    \\ \bottomrule
\end{tabular}

\begin{tablenotes}
  \item 
  As the index of the dataset grows, the number of packages, architectures, and compiler options increases
  \asei and \aseiii are the datasets used in VulSeeker~\cite{gao2018vulseeker}.
  For all datasets, the optimization levels are \tc{O0}--\tc{O3}.
\end{tablenotes}
\end{threeparttable}

\vspace{-12pt}
\end{table}

\begin{figure}[t]
  \begin{multicols}{2}
    \begin{subfigure}[h]{\linewidth}
      \includegraphics[width=\linewidth]{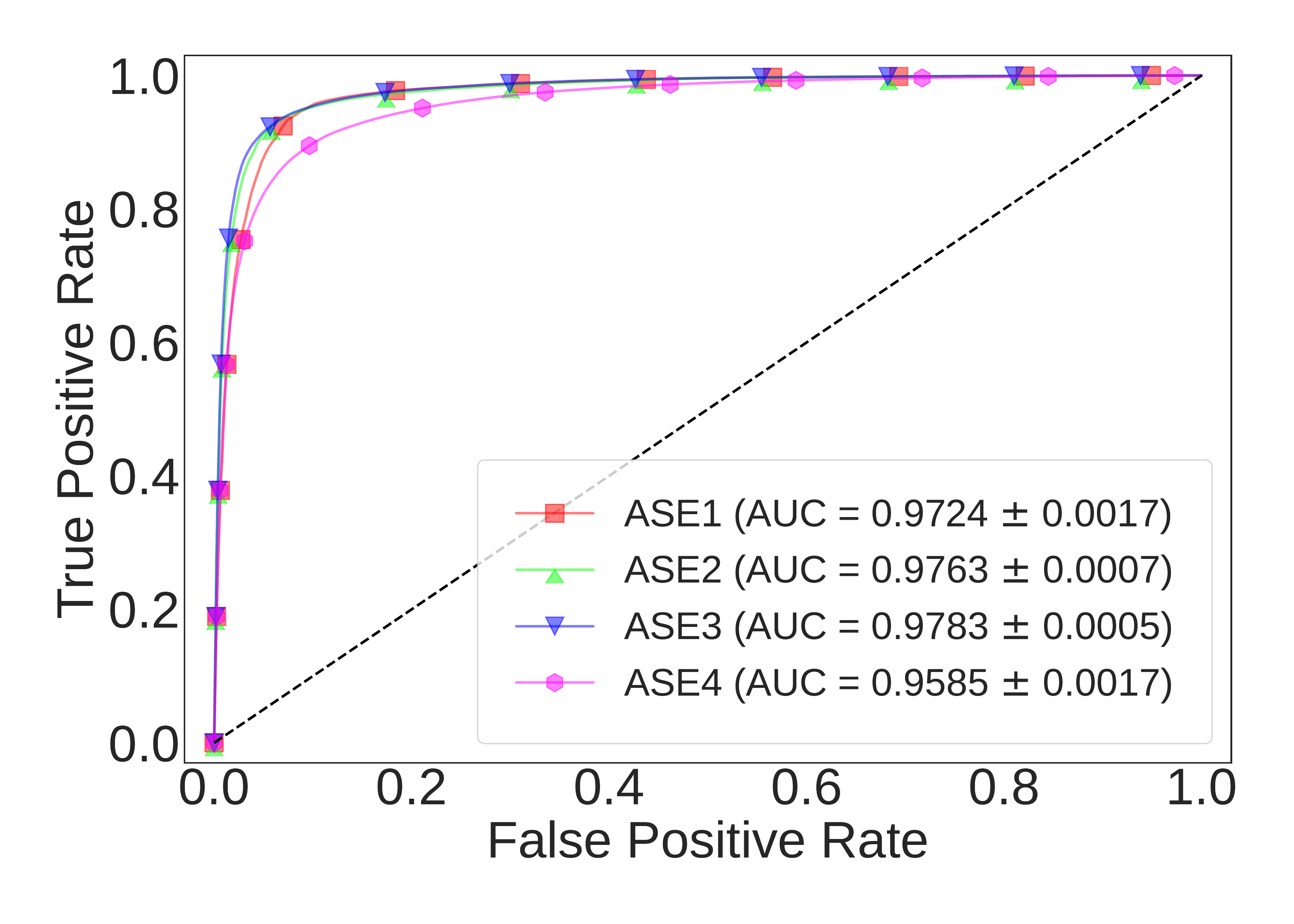}
      \vspace{-15pt}
      \caption{ROC AUCs for all \syscomp datasets.}
      \label{fig:syscompall}
    \end{subfigure}

    \begin{subfigure}[h]{\linewidth}
      \includegraphics[width=\linewidth]{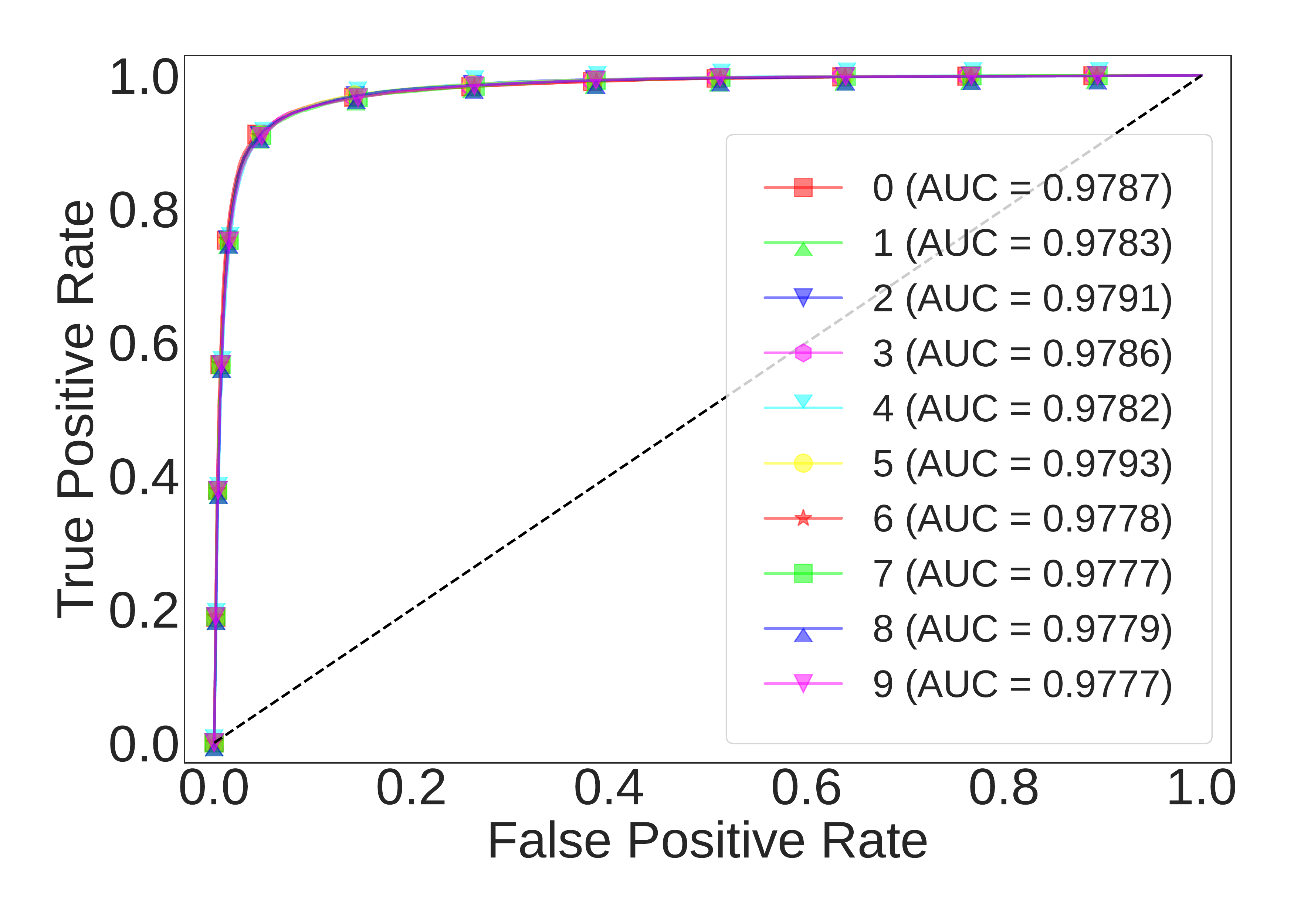}
      \vspace{-15pt}
      \caption{ROC AUC of each fold for \aseiii.}
      \label{fig:syscompfold}
    \end{subfigure}
  \end{multicols}
  \vspace{-14pt}



  \caption{Results obtained by running \techname on \syscomp datasets.}
  \label{fig:syscomp}
  \vspace{-15pt}
\end{figure}

From our experiments in~\autoref{ss:result}, we show that using only presemantic
features with a simple linear model (i.e., \techname) is enough to obtain high
AUC values. Next, we compare \techname with state-of-the-art techniques.

To accomplish this, we chose one of the latest approaches,
VulSeeker~\cite{gao2018vulseeker}, as our target because it utilizes both
presemantic and semantic features in a numeric form by leveraging neural
network-based post-processing. Thus, we can directly evaluate our simple model
using numeric presemantic features.
Note that \textit{our goal is not to claim that our approach is better, but to
demonstrate that the proper engineering of presemantic features can achieve
results that are comparable to those of state-of-the-art techniques}.

For this experiment, we prepared the datasets of VulSeeker, along with the
additional ones as listed in~\autoref{tab:syscomp}. We refer to these datasets
as \asei through \aseiv. \asei and \aseiii are the ones used in VulSeeker, and
\aseii and \aseiv are extra ones with more packages, target architectures, and
compilers. Note that the number of packages, architectures, and compiler options
increases as the index of the dataset increases. The optimization levels for all
datasets are \tc{O0}--\tc{O3}.
We intentionally omitted firmware images used in the original paper, as they do
not provide solid ground truth.
For each dataset, we established the ground truth in the same way described
in~\autoref{ss:groundtruth}.
The time spent for IDA pre-processing, ground truth building, and feature
extracting was 2197 s, 889 s, and 239 s, respectively.
We then conducted experiments with the methodology explained
in~\autoref{sec:method}; note that the same methodology was used in the original
paper.

\autoref{fig:syscomp} depicts the results.
\autoref{fig:syscompall} shows that
the AUCs of \techname on \asei and \aseiii are 0.9724 and 0.9783, respectively.
However, those of VulSeeker were 0.99 and 0.8849 as reported by the
authors~\cite{gao2018vulseeker}.
\autoref{fig:syscompfold} illustrates that
the AUC of each fold in \aseiii
ranged from 0.9777 to 0.9793,
which is higher than that of VulSeeker (0.8849).
Therefore, \techname was more robust than VulSeeker in terms of the size and
compile options in the dataset.
\techname also exhibits stable results, even for \aseii and \aseiv.


From these results, we conclude that presemantic features combined with proper feature
engineering can achieve results that are comparable to those of state-of-the-art
BCSA techniques.
Although our current focus is on comparing feature values, it is possible to
extend our work to analyze the complex relationships among the features by
utilizing advanced machine learning techniques~\cite{\paperML}.

\begin{table*}[t]
  \centering
  \caption{Real-world vulnerability (Heartbleed, CVE-2014-0160) analysis result using \tiknib (top-k and precision@1).}
\vspace{-0.1in}
  \label{tab:resulttopk}
  \setlength\tabcolsep{0.11cm}
  \def\arraystretch{0.50}
  \scriptsize

\begin{threeparttable}
  \centering

\begin{tabular}{@{}lccccccccccccccccc@{}}
\toprule
\multicolumn{1}{c}{\textbf{Source option}} & \multicolumn{1}{c}{\textbf{All}} & \multicolumn{1}{c}{\textbf{ARM}} & \multicolumn{1}{c}{\textbf{ARM}} & \multicolumn{1}{c}{\textbf{ARM}} & \multicolumn{1}{c}{\textbf{MIPS}} & \multicolumn{1}{c}{\textbf{MIPS}} & \multicolumn{1}{c}{\textbf{MIPS}} & \multicolumn{1}{c}{\textbf{x86}} & \multicolumn{1}{c}{\textbf{x86}} & \multicolumn{1}{c}{\textbf{x86}} & \multicolumn{1}{c}{\textbf{O2}} & \multicolumn{1}{c}{\textbf{O3}} & \multicolumn{1}{c}{\textbf{\gcc}} & \multicolumn{1}{c}{\textbf{\gcc v4}} & \multicolumn{1}{c}{\textbf{\gcc v8}} & \multicolumn{1}{c}{\textbf{\clang v4}} & \multicolumn{1}{c}{\textbf{\clang v7}} \\
 \multicolumn{1}{c}{\textbf{\tiny{to}}} & \multicolumn{1}{c}{\textbf{\tiny{to}}} & \multicolumn{1}{c}{\textbf{\tiny{to}}} & \multicolumn{1}{c}{\textbf{\tiny{to}}} & \multicolumn{1}{c}{\textbf{\tiny{to}}} & \multicolumn{1}{c}{\textbf{\tiny{to}}} & \multicolumn{1}{c}{\textbf{\tiny{to}}} & \multicolumn{1}{c}{\textbf{\tiny{to}}} & \multicolumn{1}{c}{\textbf{\tiny{to}}} & \multicolumn{1}{c}{\textbf{\tiny{to}}} & \multicolumn{1}{c}{\textbf{\tiny{to}}} & \multicolumn{1}{c}{\textbf{\tiny{to}}} & \multicolumn{1}{c}{\textbf{\tiny{to}}} & \multicolumn{1}{c}{\textbf{\tiny{to}}} & \multicolumn{1}{c}{\textbf{\tiny{to}}} & \multicolumn{1}{c}{\textbf{\tiny{to}}} & \multicolumn{1}{c}{\textbf{\tiny{to}}} & \multicolumn{1}{c}{\textbf{\tiny{to}}} \\
 \multicolumn{1}{c}{\textbf{Target option}} & \multicolumn{1}{c}{\textbf{All}} & \multicolumn{1}{c}{\textbf{ARM}} & \multicolumn{1}{c}{\textbf{MIPS}} & \multicolumn{1}{c}{\textbf{x86}} & \multicolumn{1}{c}{\textbf{MIPS}} & \multicolumn{1}{c}{\textbf{ARM}} & \multicolumn{1}{c}{\textbf{x86}} & \multicolumn{1}{c}{\textbf{x86}} & \multicolumn{1}{c}{\textbf{ARM}} & \multicolumn{1}{c}{\textbf{MIPS}} & \multicolumn{1}{c}{\textbf{O3}} & \multicolumn{1}{c}{\textbf{O2}} & \multicolumn{1}{c}{\textbf{\clang}} & \multicolumn{1}{c}{\textbf{\gcc v8}} & \multicolumn{1}{c}{\textbf{\gcc v4}} & \multicolumn{1}{c}{\textbf{\clang v7}} & \multicolumn{1}{c}{\textbf{\clang v4}} \\ \midrule
  \# of Option Pairs & 552 & 56 & 64 & 64 & 56 & 64 & 64 & 56 & 64 & 64 & 144 & 144 & 144 & 36 & 36 & 36 & 36 \\
  \midrule
  Rank (tls, vuln)\tnote{$\ast$} & 1.19 & 1.14 & 1.66 & 1 & 1 & 1.62 & 1 & 1 & 1.25 & 1 & 1.18 & 1.19 & 1 & 1.44 & 1.06 & 1 & 1 \\
  Precision@1 (tls, vuln)\tnote{$\ast$} & 0.89 & 0.86 & 0.66 & 1 & 1 & 0.75 & 1 & 1 & 0.75 & 1 & 0.9 & 0.89 & 1 & 0.78 & 0.94 & 1 & 1 \\
  \midrule
  Rank (dtls, vuln)\tnote{$\dagger$} & 4.54 & 9.82 & 11.81 & 3.06 & 2 & 4.72 & 2 & 2.07 & 1.75 & 3.62 & 4.5 & 4.38 & 2.72 & 3.11 & 5.06 & 3.61 & 3.33 \\
  Rank (tls, patched)\tnote{$\ddagger$} & 29.16 & 12.12 & 57.69 & 3.56 & 3.82 & 51.62 & 43.94 & 4.29 & 6.38 & 70.59 & 27.5 & 28.96 & 27.68 & 32.89 & 40.89 & 20.22 & 22.67 \\
  Rank (dtls, patched)\tnote{$\ddagger$} & 76.47 & 46.95 & 145.75 & 7.25 & 8.21 & 128 & 128.94 & 9.57 & 11.94 & 181.03 & 73.04 & 75.41 & 87.31 & 66.28 & 87.33 & 68.44 & 78 \\
 \bottomrule
  \end{tabular}
\begin{tablenotes}
%
%
\item We tested three 64-bit architectures (aarch64, x86-64, and mips64el) that
  are widely used in software packages.

\item All rank and precision@1 values are averaged; because each test
  involves multiple combinations of options (the first row), their results are
  averaged.

\item [$\ast$] These are the results of the vulnerable
    \tc{tls1_process_heartbeat} function in \tc{OpenSSL} v1.0.1f.

\item [$\dagger$] This is the result of the vulnerable \tc{dtls1_process_heartbeat}
    function in \tc{OpenSSL} v1.0.1f, which is similar to but distinct from
    the \tc{tls1_process_heartbeat} function.

\item [$\ddagger$] These are the results of their patched versions in
    \tc{OpenSSL} v1.0.1u.


\end{tablenotes}
\end{threeparttable}
\vspace{-15pt}

\end{table*}



\subsection{Analysis Case Study: Heartbleed (CVE-2014-0160)}
\label{ss:vuln}

To further assess the effectiveness of presemantic features,
we apply \techname to vulnerability discovery,
which is a common practical application of BCSA~\cite{\paperVuln}.
We investigate whether \techname can effectively identify
a vulnerable function across various compiler options and architectures.

We chose the \tc{tls1_process_heartbeat} function in the \tc{OpenSSL} package as
our target function because it contains the infamous Heartbleed vulnerability
(\ie, CVE-2014-0160), which has been widely used in prior studies for
evaluation~\cite{eschweiler2016discovre, xu2017neural,
  feng2016scalable,ding2019asm2vec}.
We utilized two versions of \tc{OpenSSL} in the \aseiv dataset shown in
\autoref{tab:syscomp}: v1.0.1f contains the vulnerable function, while v1.0.1u
contains the patched version. As the dataset was compiled with 288 distinct
combinations of compiler options and architectures, each function has 576
samples: 288 (the number of possible combinations) $\times$ 2 (the number of
available \tc{OpenSSL} versions) $\approx$ 576.

Notably, testing all possible combinations of options entails a significant
computational overhead; it requires 288 (the number of options for our target
function) $\times$ 287 (the number of options for a function in \tc{OpenSSL})
$\times$ 2 (the number of \tc{OpenSSL} versions) $\times$ 5K (the number of
functions in \tc{OpenSSL}) $\approx$ 826M operations.  Therefore, we focused on
architectures and compiler options that are widely used in software packages.
Specifically, we chose three 64-bit architectures (aarch64, x86-64, and
mips64el) and two levels of optimization (\tc{O2}--\tc{O3}).  This setup
reflects real-world scenarios, as many software packages use \tc{O2}--\tc{O3} by
default: \tc{coreutils} uses \tc{O2}, while \tc{OpenSSL} uses \tc{O3}.  Previous
studies~\cite{ding2019asm2vec, david2016statistical} also used the same setup
(\tc{O2}--\tc{O3}) except that they only tested x86 binaries. Additionally, we
selected four compilers (Clang v4.0, Clang, v7.0, GCC v4.9.4, and GCC v8.2.0) to
consider extreme cases. Consequently, there were 24 possible combinations of
these architectures and compiler options.

We conducted a total of 552 tests on these 24 option combinations:
24 (the number of options for our target function) $\times$
23 (the number of options for a function in \tc{OpenSSL}).
For each test, we simply computed the similarity scores for all function pairs
using \tiknib and checked the rank of the vulnerable function.
To reflect real-world scenarios,
we assumed in all tests that we were not aware of the precise optimization level,
compiler type, or compiler version of the testing binary.
On the other hand,
we assumed that we could recognize the architecture of the testing binary
as it is straightforward.
Therefore, when we train \tiknib,
we chose a feature set that achieved the best performance across
all possible combinations of optimization levels, compiler types,
and compiler versions, while setting the source and target architectures fixed.
For training, we used the \normaldataset dataset (\autoref{tab:dataset})
as it does not include \tc{OpenSSL};
thus, the training and testing datasets are completely distinct.

\autoref{tab:resulttopk} summarizes the results, with each column corresponding
to the tests for the specified options. We organized the results by the option group
specified in each column after running all 522 tests.  The first row of the
table (\emph{\# of Option Pairs}) indicates the total number of option pairs,
which is the same as that of true positive pairs.  The remaining rows of the
table show the averaged values obtained by the option pair tests.  For example,
the \emph{All to All} column represents the averaged results of all possible
combinations ($24 \times 23$).  The \emph{ARM to MIPS} column, on the other
hand, represents the averaged results of all combinations with the source and
target architectures set to ARM and MIPS, respectively.  That is, we queried the
vulnerable functions compiled with ARM and searched for their true positives
compiled with MIPS while varying the other options.

In the majority of the tests,
\techname successfully identified the vulnerable function
with a rank close to 1.0 and a precision@1 close to 1.0,
demonstrating its effectiveness in vulnerability discovery.
Meanwhile, it performed marginally worse in the tests for MIPS.
This result corroborates our observation in~\autoref{ss:result}
that feature extraction for MIPS binaries can be erroneous.
We further discuss this issue in~\autoref{ss:toolerror}.
Additionally, the last three rows of~\autoref{tab:resulttopk} display the ranks
of additional functions worth noting.  The \tc{dtls} represents the DTLS
implementation of our target function (\ie, \tc{dtls1_process_heartbeat}), which
also contains the same vulnerability. Due to its similarity to our target
function, it was ranked highly in all tests.
The last two rows of the table present the ranks of the patched versions of these
two functions in \tc{OpenSSL} v1.0.1u. Notably, the patch of the vulnerability
affects the presemantic features of these functions, particularly the number
of control transfer and arithmetic instructions.
Consequently, the patched functions had a low rank.

%
%
%
%
\subsection{Analyzing Real-World Vulnerabilities on Firmware Images
of IoT Devices}
\label{ss:firm}

We further evaluate the efficacy of presemantic features by identifying
vulnerable functions in real-world firmware images of IoT devices using \tiknib.
For the firmware images, we utilized the firmware dataset of
FirmAE~\cite{kim:2020:firmae}, which is one of the industry-leading large-scale
firmware emulation frameworks. The dataset consists of 1,124 firmware images of
wireless routers and IP cameras from the top eight vendors.

Particularly, we search for another infamous vulnerability (CVE-2015-1791) from
\tc{OpenSSL}, which has a race condition error in the
\tc{ssl3_get_new_session_ticket()} function. This vulnerability has also been
extensively used in previous
studies~\cite{xu2017neural,gao2018vulseeker,feng2016scalable}.
Since there exist numerous functions ($\approx$52M) in the firmware images,
identifying the vulnerable function among them is sufficient to evaluate the
impact of presemantic features.


We evaluate our system, \tiknib, against state-of-the-art techniques that
support both ARM and MIPS architectures~\cite{xu2017neural,gao2018vulseeker}.
Notably, these two architectures are prevalent in IoT
devices~\cite{kim:2020:firmae}. However, while analyzing the repositories of
these tools, we found that they did not include their complete source code nor
datasets. As a result, we were unable to directly compare our system to theirs.
Instead, we compared the results to the ones stated in the
paper~\cite{gao2018vulseeker}.
Specifically, we compiled the vulnerable version of \tc{OpenSSL} (\ie, v1.0.1f)
using a variety of compiler options and architectures, including six
architectures (x86, ARM, and MIPS, each with 32 and 64 bits), two compilers (GCC
v4.9.4 and v5.5.0), and two optimization levels (\tc{O2}--\tc{O3}).  Here, we
used two optimization levels (\tc{O2}--\tc{O3}) because many real-world software
packages use them by default, as described in~\autoref{ss:vuln}.
Consequently, we obtained 24 samples of the vulnerable function.  Notably, this
dataset is essentially a subset of the \aseiii dataset, which is introduced
in~\autoref{tab:syscomp}.
Then, we queried each sample vulnerable function against all 52M functions in
the 1,124 firmware images.  This resulted in 24 similarity scores for each of
the 52M functions.  We then calculated the top-k result by averaging the
similarity scores for each function.  Finally, we manually counted the number of
functions that were actually vulnerable in the top-100 results.


\begin{table}[t]
  \scriptsize
  \centering
  \setlength\tabcolsep{0.5cm}
  \def\arraystretch{0.50}

  \caption{Top-k results of identifying CVE-2015-1791 for 52M functions in 1,124
  IoT firmware images using \tiknib.}
  \label{tab:firmkit:compare}
  \vspace{-6pt}

\begin{threeparttable}

\begin{tabular}{@{}rrrrr@{}}
\toprule
\multicolumn{1}{c}{Top-k}
& \multicolumn{1}{c}{Gemini\tnote{$\dagger$}}
& \multicolumn{1}{c}{VulSeeker\tnote{$\dagger$}}
& \multicolumn{1}{c}{\tiknib (Ours)}
 \\ \midrule
1  & 1 (100\%)    & 1 (100\%)  & 1 (100\%) \\
5  & 2 (\ \ 40\%)     & 3 (\ \ 60\%) & 5 (100\%) \\
10      & 4 (\ \ 40\%)     & 6 (\ \ 60\%) & 10 (100\%) \\
50      & 36 (\ \ 72\%)    & 41 (\ \ 82\%)  & 46 (\ \ 92\%) \\
100     & 75 (\ \ 75\%)    & 83 (\ \ 83\%)  & 82 (\ \ 82\%) \\ 
\bottomrule
\end{tabular}

\begin{tablenotes}
\item [$\dagger$]
Among 43 BCSA papers that we studied in~\autoref{sec:back},
10 released their source code, and
two of these 10 support both ARM and MIPS architectures
(Gemini~\cite{xu2017neural} and VulSeeker~\cite{gao2018vulseeker}).
However, we were not able to compare the results directly
because these tools released neither their firmware datasets nor
complete source code.
Here, we present the results stated in the latest one~\cite{gao2018vulseeker};
note that their firmware dataset is different from the one that we used.
%


\end{tablenotes}
\end{threeparttable}

\vspace{-12pt}
\end{table}


\autoref{tab:firmkit:compare} summarizes the top-k results
for the average similarity score for all 52M firmware functions.
While our dataset is distinct from those used in the previous
studies~\cite{xu2017neural,gao2018vulseeker},
\tiknib equipped with presemantic features
achieved a level of performance comparable to that of the state-of-the-art tools.
It should be noted that \textit{our objective is not to assert that
our approach is superior to the state-of-the-art tools, but rather
to demonstrate the efficacy of appropriately utilizing presemantic features.}
Additionally, our experimental results indicate that the real-world IoT firmware
images (at least those that we tested) are highly likely to be compiled with
\tc{O2} or \tc{O3}.

\begin{table*}[t]
  \centering
  \caption{In-depth analysis results of
  presemantic and type features
  obtained by running \tiknib on \sys.}
\vspace{-0.1in}
  \label{tab:resultapp}
  \setlength\tabcolsep{0.9pt}
  \def\arraystretch{0.50}
  \scriptsize

  \begin{threeparttable}
 

\begin{tabular}{@{}lllccccccccccccccccccccccccccccccccccc@{}}
\toprule
 &  & &  &
 &  & \multicolumn{3}{c}{\textbf{Opt Level}}
 &  & \multicolumn{4}{c}{\textbf{Compiler}} 
 &  & \multicolumn{6}{c}{\textbf{Arch}}
 &  & \multicolumn{3}{c}{\textbf{vs. SizeOpt}\tnote{$\dagger$}}
 &  & \multicolumn{3}{c}{\textbf{vs. Extra}\tnote{$\dagger$}}
 &  & \multicolumn{4}{c}{\textbf{vs. Obfus.}\tnote{$\dagger$}}
 &  & \multicolumn{2}{c}{\textbf{Bad}\tnote{$\ddagger$}}
 \\
 \cmidrule(lr){7-9}
 \cmidrule(lr){11-14}
 \cmidrule(lr){16-21}
 \cmidrule(lr){23-25}
 \cmidrule(lr){27-29}
 \cmidrule(lr){31-34}
 \cmidrule(l){36-37}
  &  &  &  &
  {\tiny \textbf{Rand.}} &
  &
  {\tiny \textbf{\begin{tabular}[c]{@{}c@{}}O0\\ vs.\\ O3\end{tabular}}} &
  {\tiny \textbf{\begin{tabular}[c]{@{}c@{}}O2\\ vs.\\ O3\end{tabular}}} &
  {\tiny \textbf{Rand.}} &
  &
  {\tiny \textbf{\begin{tabular}[c]{@{}c@{}}\gcc4\\ vs.\\ \gcc8\end{tabular}}} &
  {\tiny \textbf{\begin{tabular}[c]{@{}c@{}}\clang4\\ vs.\\ \clang7\end{tabular}}} &
  {\tiny \textbf{\begin{tabular}[c]{@{}c@{}}\gcc\\ vs.\\ \clang\end{tabular}}} &
  {\tiny \textbf{Rand.}} &
  &
  {\tiny \textbf{\begin{tabular}[c]{@{}c@{}}x86\\ vs.\\ ARM\end{tabular}}} &
  {\tiny \textbf{\begin{tabular}[c]{@{}c@{}}x86\\ vs.\\ MIPS\end{tabular}}} &
  {\tiny \textbf{\begin{tabular}[c]{@{}c@{}}ARM\\ vs.\\ MIPS\end{tabular}}} &
  {\tiny \textbf{\begin{tabular}[c]{@{}c@{}}32\\ vs.\\ 64\end{tabular}}} &
  {\tiny \textbf{\begin{tabular}[c]{@{}c@{}}LE\\ vs.\\ BE\end{tabular}}} &
  {\tiny \textbf{Rand.}} &
  &
  {\tiny \textbf{\begin{tabular}[c]{@{}c@{}}O0\\ vs.\\ Os\end{tabular}}} &
  {\tiny \textbf{\begin{tabular}[c]{@{}c@{}}O1\\ vs.\\ Os\end{tabular}}} &
  {\tiny \textbf{\begin{tabular}[c]{@{}c@{}}O3\\ vs.\\ Os\end{tabular}}} &
  &
  {\tiny \textbf{PIE}} &
  {\tiny \textbf{NoInline}} &
  {\tiny \textbf{LTO}} &
  &
  {\tiny \textbf{BCF}} &
  {\tiny \textbf{FLA}} &
  {\tiny \textbf{SUB}} &
  {\tiny \textbf{All}} &
  &
  {\tiny \textbf{Norm.}} &
  {\tiny \textbf{\begin{tabular}[c]{@{}c@{}}Norm.\\ vs.\\ Obfus.\end{tabular}}\tnote{$\dagger$}} \\
  \midrule


\multicolumn{3}{l}{Avg. \# of Selected Features} &  & 4.0&  & 4.0& 6.8& 5.3&  & 7.1  & 10.5 & 6.8   & 3.0&  & 12.0 & 6.4   & 7.1   & 9.4& 7.7& 9.0&  & 7.0& 7.0& 7.0&  & 8.4& 6.0  & 7.6&  & 7.0& 7.0& 8.3& 6.0&  & 4.0& 4.5 \\
\multicolumn{3}{l}{Avg. TP-TN Gap of Grey} &  & 0.53   &  & 0.52   & 0.56   & 0.53   &  & 0.58 & 0.59 & 0.53  & 0.56   &  & 0.54 & 0.54  & 0.54  & 0.56   & 0.56   & 0.55   &  & 0.54   & 0.57   & 0.57   &  & 0.56   & 0.56 & 0.55   &  & 0.50   & 0.52   & 0.58   & 0.50   &  & 0.55   & 0.55\\
\multicolumn{3}{l}{ROC AUC}&  & 0.99   &  & 0.99   & 1.00   & 0.99   &  & 1.00 & 1.00 & 0.99  & 0.99   &  & 1.00 & 0.99  & 1.00  & 1.00   & 1.00   & 1.00   &  & 0.99   & 1.00   & 1.00   &  & 0.99   & 1.00 & 1.00   &  & 1.00   & 1.00   & 1.00   & 0.99   &  & 0.99   & 0.99\\ 
\multicolumn{3}{l}{Average Precision (AP)}   &  & 0.99   &  & 0.99   & 1.00   & 0.99   &  & 1.00 & 1.00 & 0.99  & 0.99   &  & 1.00 & 0.99  & 1.00  & 1.00   & 1.00   & 1.00   &  & 0.99   & 1.00   & 1.00   &  & 0.99   & 1.00 & 1.00   &  & 1.00   & 1.00   & 1.00   & 0.99   &  & 0.99   & 0.98\\

\bottomrule
\end{tabular}

\begin{tablenotes}
%
%
\item All values in the table are 10-fold cross validation averages.

\item [$\dagger$] We compare a function from the \normaldataset to the corresponding function in each target dataset.

\item [$\ddagger$] We match functions whose compiler options are largely distant
  to test for bad cases. Please refer to~\autoref{sss:combining} for additional information.

\end{tablenotes}
\end{threeparttable}
\vspace{-15pt}

\end{table*}



\section{Benefit of Type Information (RQ3)}
\label{sec:rq3}

To assess the implication of debugging information on BCSA, we use type
information as a case study on the presumption that they do not vary unless the
source code is changed.
Specifically, we extract three types of features per function: the number of
arguments, the types of arguments, and the return type of a function.
Note that inferring the correct type information is challenging and is actively
researched~\cite{chua2017neural, van2016tough}.
In this context, we only consider basic types: \tc{char}, \tc{short}, \tc{int},
\tc{float}, \tc{enum}, \tc{struct}, \tc{void}, and \tc{void *}.
To extract type information, we create a type map to handle custom types defined
in each package by recursively following definitions using
Ctags~\cite{hiebert1999exuberant}. We then assign a unique prime number as an
identifier to each type. To represent the argument types as a single number, we
multiply their type identifiers.


To investigate the benefit of these type features, we conducted the same
experiments described in~\autoref{sec:rq2}, and \autoref{tab:resultapp} presents
the results. Here, we explain the results by comparing them
with~\autoref{tab:result}, which we obtained without using the type features.
The first row of~\autoref{tab:resultapp} shows that the average number of
selected features, including type features, is smaller than that of selected
features (\CC{5}) in~\autoref{tab:result}. Note that all three type features
were always selected in all tests.
The second row in~\autoref{tab:resultapp} shows that utilizing the type features
could achieve a large TP-TN gap on average (over 0.50); the corresponding values
in \CC{4} of~\autoref{tab:result} are much smaller.
Consequently, the AUC and AP with type features reached over 0.99 in all tests,
as shown in the last two rows of~\autoref{tab:resultapp}.
%
%
%
Additionally, it shows a similar result (\ie, an AUC close to 1.0) on the ASE
datasets that we utilized for the state-of-the-art comparison
(\autoref{ss:comparison-state-of-the-art}).
%

This result confirms that type information indeed benefits BCSA in terms of the
success rate, although recovering such information is a difficult task.
Therefore, we encourage further research on BCSA to take account of recovering
debugging information, such as type recovery or inference, from binary
code~\cite{\paperType}.

\section{Failure Case Inquiry (RQ4)}
\label{sec:rq4}

We carefully analyzed the failure cases in our experiments and found their
causes.
It was possible because our benchmark (\ie, \sys) has the ground
truth and our tool (\ie, \techname) uses an interpretable model.
We first checked the TP-TN gap of each feature for failure cases and further
analyzed them using IDA Pro.
We found that optimization largely affects the BCSA performance, as described
in~\autoref{ss:result}.
In this section, we discuss other failure causes and summarize the lessons
learned; however, many of these causes are closely related to optimization.
We categorized the causes into three cases:
(1) errors in binary analysis tools (\autoref{ss:toolerror}),
(2) differences in compiler back-ends (\autoref{ss:compiler-diversity}),
and (3) architecture-specific code (\autoref{ss:archspecific}).

\subsection{Errors in Binary Analysis Tools}
\label{ss:toolerror}

Most BCSA research heavily relies on COTS binary analysis tools such as IDA
Pro~\cite{ida}. However, we found that IDA Pro can yield false results.
First, IDA Pro fails to analyze indirect branches, especially when handling MIPS
binaries compiled with \clang using the position-independent code (PIC) option.
The PIC option sets the compiler to generate machine code that can be placed in
any address, and it is mainly used for compiling shared libraries or PIE
binaries.
%
Particularly, compilers use register-indirect branch instructions, such as
\tc{jalr}, to invoke functions in a position-independent manner. For example,
when calling a function, \gcc stores the base address of the Global Offset Table
(GOT) in the \tc{gp} register, and uses it to calculate the function addresses
at runtime. In contrast, \clang uses the \tc{s0} or \tc{v0} register to store
such base addresses. This subtle difference confuses IDA Pro and makes it fail
to obtain the base address of the GOT, so that it cannot compute the target
addresses of indirect branches.


Moreover, IDA Pro sometimes generates incomplete CFGs. When there is a
\emph{switch} statement, compilers often make a table that stores a list of jump
target addresses. However, IDA Pro often failed to correctly identify the number
of elements in the table, especially on the ARM architecture, where switch tables
can be placed in a code segment. Sometimes, switch tables are located between
basic blocks, and it is more difficult to distinguish them.

The problem worsens when handling MIPS binaries compiled for \clang with PIC,
because switch tables are typically stored in a read-only data section, which
can be referenced through a GOT. Therefore, if IDA Pro cannot fully analyze the
base address of the GOT, it also fails to identify the jump targets of switch
statements.

As we manually analyzed the errors, we may have missed some. Systematically
finding such errors is a difficult task because the internals of many
disassembly tools are not fully disclosed, and they differ significantly. One
may extend the previous study~\cite{andriesse2016depth} to further analyze the
errors of disassembly tools and extracted features, and we leave this for future
studies.

During the analysis, we found that IDA Pro also failed to fetch some function
names if they had a prefix pre-defined in IDA Pro, such as \tc{off_} or
\tc{sub_}.
%
For example, it failed to fetch the name of the \tc{off_to_chars} function in
the \tc{tar} package.
We used IDA Pro v6.95 in our experiments, but we found that its latest version
(v7.5) does not have this issue.


\subsection{Diversity of Compiler Back-ends}
\label{ss:compiler-diversity}

From~\autoref{ss:result}, the characteristics of binaries largely vary depending
on the underlying compiler back-end. Our study reveals that \gcc and \clang emit
significantly different binaries from the same source code.

\begin{figure}[t]
    \includegraphics[width=\linewidth]{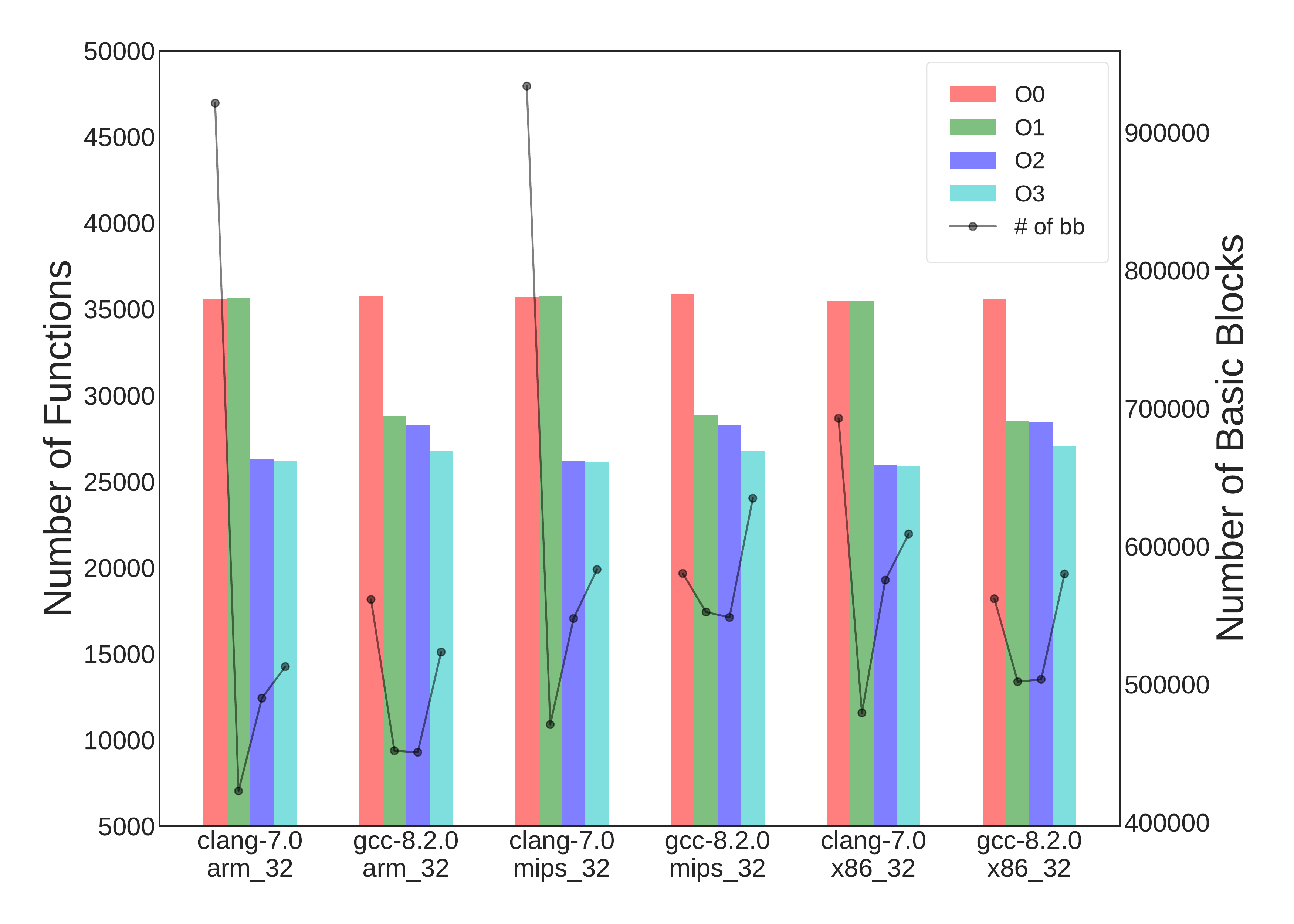}


  \vspace{-13pt}
  \caption{The final number
  of functions and basic blocks in \normaldataset
  (see Appendix for the detailed version).}
  \label{fig:numfuncsbbs}
  \vspace{-15pt}
\end{figure}

First, the number of basic blocks for the two compilers significantly
differs.
To observe how the number changes depending on different compiler options and
target architectures, we counted the number for the \normaldataset dataset.
\autoref{fig:numfuncsbbs} illustrates the number of functions and basic blocks
in the dataset for selected compiler options and architectures (see Appendix for
details).
As shown in the figure, the number of basic blocks in binaries compiled with
\clang is significantly larger than that in binaries compiled with \gcc for
\tc{O0}. We figured out that \clang inserts dummy basic blocks for \tc{O0} on
ARM and MIPS; these dummy blocks have only one branch instruction to the next
block.
%
These dummy blocks are removed when the optimization level increases
(\tc{O1}) as optimization techniques in \clang merge such basic blocks into
their predecessors.

In addition, the two compilers apply different internal techniques for the same
optimization level, while they express the optimization level with the same
terms (\ie, \tc{O0}--\tc{O3} and \tc{Os}).
%
%
In particular, by analyzing the number of caller and callee functions, we
discovered that \gcc applies function inlining from \tc{O1}, whereas \clang
applies it from \tc{O2}.
Consequently, the number of functions for each compiler significantly differs
(see the number of functions in \tc{O1} for \clang and that for \gcc
in~\autoref{fig:numfuncsbbs}).
%
%

%
Moreover, we discovered that two compilers internally leverage different
function-level code for specific operations.
For example, \gcc has functions, such as \tc{__umoddl3} in \tc{libgcc2.c} or
\tc{__aeabi_dadd} in \tc{ieee754-df.S}, to optimize certain arithmetic
operations.
Furthermore, on x86, \gcc generates a special function, such as
\tc{__x86.get_pc_thunk.bx}, to load the current instruction pointer to a
register, whereas \clang inlines this procedure inside the target function.
These functions can largely affect the call-related features, such as the number
of control transfer instructions or that of outgoing calls.
Although we removed these compiler-specific functions so as not to include them in our
experiments (\autoref{ss:groundtruth}), they may have been inlined in their
caller functions in higher optimization levels (\tc{O2}--\tc{O3}).
Considering such functions took approximately 4\% of the identified functions by
IDA Pro, they may have affected the resulting features.

Similarly, the two compilers also utilize different instruction-level codes.
For example, in the case of move instructions for ARM, \gcc uses conditional
instructions, such as \tc{MOVLE}, \tc{MOVGT}, or \tc{MOVNE}, unless the
optimization level is zero (\tc{O0}).
In contrast, \clang utilizes regular move instructions along with branch
instructions.
This significantly affects the number of instructions as well as that of basic
blocks in the resulting binaries.
Consequently, in such special cases, the functions compiled using \gcc have a
relatively smaller number of basic blocks compared with those using \clang.

Finally, compilers sometimes generate multiple copies of the same function for
optimization purposes. For example, they conduct inter-procedural scalar
replacement of aggregates, removal of unused parameters, or optimization of
cache/memory usage.
%
Consequently, a compiled binary can have multiple functions that share the same
source code but have different binary code.
We found that \gcc and \clang operate differently on this. Specifically, we
discovered three techniques in \gcc that produce function copies with special
suffixes, such as \tc{.part}, \tc{.cold}, or \tc{.isra}.
For instance, for the \tc{get_data} function of \tc{readelf} in \tc{binutils}
(in \tc{O3}), \gcc yields three copies with the \tc{.isra} suffix, while \clang
does not produce any such functions.
Similarly, for the \tc{tree_eval} and \tc{expr_eval} functions in \tc{bool} (in
\tc{O3}), \gcc produces two copies with the \tc{.cold} suffix, but \clang does
not.
Although we selected only one such copy in our experiments to avoid biased
results (\autoref{ss:groundtruth}), the other copies can still survive in their
caller functions by inlining.

%
%
%

%

In summary, the diversities of compiler back-ends can largely affect the
performance of BCSA, by making the resulting binaries divergent. Here, we
have introduced the major issues we discovered. We encourage further studies to
investigate the implications of detailed options at each optimization level
across different compilers.




\subsection{Architecture-Specific Code} \label{ss:archspecific}

When manually inspecting failures, we found that some packages have
architecture-specific code snippets guarded with conditional macros such as
\tc{#if} and \tc{#ifdef} directives.
For example, various functions in \tc{OpenSSL}, such as \tc{mul_add} and
\tc{BN_UMULT_HIGH}, are written in architecture-specific inline assembly code to
generate highly optimized binaries.
%
%
This means that a function may correspond to two or more distinct source lines
depending on the target architecture.

Therefore, instruction-level presemantic features can be significantly different
across different architectures when the target programs have
architecture-specific code snippets, and one should consider such code when
designing cross-architecture BCSA techniques.


%


\section{Discussion}
\label{sec:discuss}

Our study identifies several future research directions in BCSA.
First, many BCSA papers have focused on building a general model that can result
in stable outcomes with any compiler option. However, one could train a model
targeting a specific set of compiler options, as shown in our experiment, to
enhance their BCSA techniques. It is evident from our experiment's results that
one can easily increase the success rate of their technique by inferring the
compiler options used to compile the target binaries.
There exists such an inference technique~\cite{rosenblum2011recovering}, and
combining it with existing BCSA methods is a promising research direction.

Second, there are only a few studies on utilizing decompilation techniques for
BCSA. However, our study reveals the importance of such techniques, and thus,
invites further research on leveraging them for BCSA. One could also conduct a
comprehensive analysis on the implication of semantic features along with
decompilation techniques.

Additionally, we investigated fundamental presemantic features in this study.
However, the effectiveness of semantic features is not well-studied yet in this
field.
Therefore, we encourage further research into investigating the effectiveness of
semantic features along with other presemantic features that are not covered in
the study.
In particular, adopting NLP techniques would be another essential study as in many
recent studies.




Our scope is limited to a function-level analysis (\autoref{ss:overview}).
However, one may extend the scope to handle other BCSA scenarios to compare
binaries~\cite{bindiff, ding2019asm2vec, duan2020deepbindiff} or a series of
instructions~\cite{huang2017binsequence, david2014tracelet, ding2016kam1n0}.
Additionally, one can extend our approach for various purposes, such as
vulnerability discovery~\cite{eschweiler2016discovre, feng2016scalable,
xu2017neural, ding2019asm2vec, david2016statistical, tol2020fastspec,
sun2020hybrid},
malware detection~\cite{jang2011bitshred, canali2012quantitative,
comparetti2010identifying, babic2011malware, xiao2016matching, ming2015memoized,
alrabaee2018fossil},
library function identification~\cite{grobert2011automated, calvet2012aligot,
xu2017cryptographic, shirani2017binshape, qiu2015library, jia2020neural},
plagiarism/authorship detection~\cite{zhang2014program, tian2020plagiarism,
alrabaee2020cpa},
or patch identification~\cite{xu2017spain, hu2019automatically,
    zhao2020patchscope}.
%
%
%
However, extending our work to other BCSA tasks may not be directly applicable.
This is because it requires additional domain knowledge to design an appropriate
model that fits the purpose and careful consideration of the trade-offs. We
believe that the reported insights in this study can help in this process.

%
%
%

Recall from \autoref{sec:back}, we did not intend to completely survey the
existing techniques, but instead, we focused on systematizing the fundamental
features used in previous literature. Furthermore, our goal was to investigate
underexplored research questions in the field by conducting a series of rigorous
experiments. For a complete survey, we refer readers to the recent surveys on
BCSA~\cite{haq2019survey, qasem2021automatic}.

Finally, because our focus is on comparing binaries without source code, we
intentionally exclude similarity comparison techniques that require source code.
Nevertheless, it is noteworthy that there has been plentiful literature on
comparing two source code snippets~\cite{kamiya2002ccfinder, schleimer:2003,
li2004cp, jiang2007deckard, dam2017automatic, lahiri2012symdiff,
kim:oakland:2017, wang2016automatically, li2016vulpecker, jang2012redebug} or
comparing source snippets with binary snippets~\cite{miyani2017binpro,
rahimian2012resource, hemel2011finding, ji2021buggraph}.

\section{Conclusion}
\label{s:conclusion}

We studied previous BCSA literature in terms of the features and
benchmarks used.
We discovered that none of the previous BCSA studies used the same benchmark for their
evaluation, and that some of them required manually fabricating the ground truth for
their benchmark.
This observation inspired us to design \sys, the first large-scale public benchmark
for BCSA, along with a set of automated build scripts.
Additionally, we developed a BCSA tool, \techname, that employs an interpretable model.
Using our benchmark and tool, we answered less-explored research questions
regarding the syntactic and structural BCSA features.
%
We discovered that several elementary features can be robust across different
architectures, compiler types, compiler versions, and even intra-procedural
obfuscation.
Further, we proposed potential strategies for enhancing BCSA.
%
We conclude by inviting further research on BCSA using our findings and
benchmark.



\ifCLASSOPTIONcompsoc
 \section*{Acknowledgements}
\else
 \section*{Acknowledgement}
\fi

We appreciate the anonymous reviewers for their thoughtful comments.
This work was supported by Institute of Information \& communications
Technology Planning \& Evaluation (IITP) grant funded by the Korea government
(MSIT) (No.2021-0-01332, Developing Next-Generation Binary Decompiler).


\bibliographystyle{sty/IEEEtran}
\footnotesize
\bibliography{main}

\begin{thebibliography}{100}
\providecommand{\url}[1]{#1}
\csname url@samestyle\endcsname
\providecommand{\newblock}{\relax}
\providecommand{\bibinfo}[2]{#2}
\providecommand{\BIBentrySTDinterwordspacing}{\spaceskip=0pt\relax}
\providecommand{\BIBentryALTinterwordstretchfactor}{4}
\providecommand{\BIBentryALTinterwordspacing}{\spaceskip=\fontdimen2\font plus
\BIBentryALTinterwordstretchfactor\fontdimen3\font minus
  \fontdimen4\font\relax}
\providecommand{\BIBforeignlanguage}[2]{{%
\expandafter\ifx\csname l@#1\endcsname\relax
\typeout{** WARNING: IEEEtran.bst: No hyphenation pattern has been}%
\typeout{** loaded for the language `#1'. Using the pattern for}%
\typeout{** the default language instead.}%
\else
\language=\csname l@#1\endcsname
\fi
#2}}
\providecommand{\BIBdecl}{\relax}
\BIBdecl

\bibitem{reiss:2009}
S.~P. Reiss, ``Semantics-based code search,'' in \emph{Proceedings of the
  International Conference on Software Engineering}, 2009, pp. 243--253.

\bibitem{gplviolation}
\BIBentryALTinterwordspacing
``gpl-violations.org project prevails in court case on gpl violation by
  d-link.'' [Online]. Available:
  \url{https://web.archive.org/web/20141007073104/http://gpl-violations.org/news/20060922-dlink-judgement_frankfurt.html}
\BIBentrySTDinterwordspacing

\bibitem{shin2015recognizing}
E.~C.~R. Shin, D.~Song, and R.~Moazzezi, ``Recognizing functions in binaries
  with neural networks,'' in \emph{Proceedings of the {USENIX} Security
  Symposium}, 2015, pp. 611--626.

\bibitem{bao2014byteweight}
T.~Bao, J.~Burket, M.~Woo, R.~Turner, and D.~Brumley, ``{ByteWeight}: Learning
  to recognize functions in binary code,'' in \emph{Proceedings of the {USENIX}
  Security Symposium}, 2014, pp. 845--860.

\bibitem{comparetti2010identifying}
P.~M. Comparetti, G.~Salvaneschi, E.~Kirda, C.~Kolbitsch, C.~Kruegel, and
  S.~Zanero, ``Identifying dormant functionality in malware programs,'' in
  \emph{Proceedings of the IEEE Symposium on Security and Privacy}, 2010, pp.
  61--76.

\bibitem{jang2011bitshred}
J.~Jang, D.~Brumley, and S.~Venkataraman, ``Bitshred: Feature hashing malware
  for scalable triage and semantic analysis,'' in \emph{Proceedings of the ACM
  Conference on Computer and Communications Security}, 2011, pp. 309--320.

\bibitem{luo2014semantics}
L.~Luo, J.~Ming, D.~Wu, P.~Liu, and S.~Zhu, ``Semantics-based
  obfuscation-resilient binary code similarity comparison with applications to
  software plagiarism detection,'' in \emph{Proceedings of the International
  Symposium on Foundations of Software Engineering}, 2014, pp. 389--400.

\bibitem{zhang2014program}
F.~Zhang, D.~Wu, P.~Liu, and S.~Zhu, ``Program logic based software plagiarism
  detection,'' in \emph{Proceedings of the IEEE International Symposium on
  Software Reliability Engineering}, 2014, pp. 66--77.

\bibitem{meng:esorics:2017}
X.~Meng, B.~P. Miller, and K.-S. Jun, ``Identifying multiple authors in a
  binary program,'' in \emph{Proceedings of the European Symposium on Research
  in Computer Security}, 2017, pp. 286--304.

\bibitem{pewny2014leveraging}
J.~Pewny, F.~Schuster, L.~Bernhard, T.~Holz, and C.~Rossow, ``Leveraging
  semantic signatures for bug search in binary programs,'' in \emph{Proceedings
  of the Annual Computer Security Applications Conference}, 2014, pp. 406--415.

\bibitem{eschweiler2016discovre}
S.~Eschweiler, K.~Yakdan, and E.~Gerhards-Padilla, ``discov{RE}: Efficient
  cross-architecture identification of bugs in binary code,'' in
  \emph{Proceedings of the Network and Distributed System Security Symposium},
  2016.

\bibitem{xu2017neural}
X.~Xu, C.~Liu, Q.~Feng, H.~Yin, L.~Song, and D.~Song, ``Neural network-based
  graph embedding for cross-platform binary code similarity detection,'' in
  \emph{Proceedings of the ACM Conference on Computer and Communications
  Security}, 2017, pp. 363--376.

\bibitem{gao2018vulseeker}
J.~Gao, X.~Yang, Y.~Fu, Y.~Jiang, and J.~Sun, ``{VulSeeker}: A semantic
  learning based vulnerability seeker for cross-platform binary,'' in
  \emph{Proceedings of the ACM/IEEE International Conference on Automated
  Software Engineering}, 2018, pp. 896--899.

\bibitem{david2018firmup}
Y.~David, N.~Partush, and E.~Yahav, ``{FirmUp}: Precise static detection of
  common vulnerabilities in firmware,'' in \emph{Proceedings of the
  International Conference on Architectural Support for Programming Languages
  and Operating Systems}, 2018, pp. 392--404.

\bibitem{chandramohan2016bingo}
M.~Chandramohan, Y.~Xue, Z.~Xu, Y.~Liu, C.~Y. Cho, and H.~B.~K. Tan, ``{BinGo}:
  Cross-architecture cross-os binary search,'' in \emph{Proceedings of the
  International Symposium on Foundations of Software Engineering}, 2016, pp.
  678--689.

\bibitem{feng2017extracting}
Q.~Feng, M.~Wang, M.~Zhang, R.~Zhou, A.~Henderson, and H.~Yin, ``Extracting
  conditional formulas for cross-platform bug search,'' in \emph{Proceedings of
  the {ACM} Symposium on Information, Computer and Communications Security},
  2017, pp. 346--359.

\bibitem{shirani2018binarm}
P.~Shirani, L.~Collard, B.~L. Agba, B.~Lebel, M.~Debbabi, L.~Wang, and
  A.~Hanna, ``Binarm: Scalable and efficient detection of vulnerabilities in
  firmware images of intelligent electronic devices,'' in \emph{International
  Conference on Detection of Intrusions and Malware, and Vulnerability
  Assessment}.\hskip 1em plus 0.5em minus 0.4em\relax Springer, 2018, pp.
  114--138.

\bibitem{xue2018accurate}
Y.~Xue, Z.~Xu, M.~Chandramohan, and Y.~Liu, ``Accurate and scalable
  cross-architecture cross-os binary code search with emulation,'' \emph{IEEE
  Transactions on Software Engineering}, 2018.

\bibitem{liu2018alphadiff}
B.~Liu, W.~Huo, C.~Zhang, W.~Li, F.~Li, A.~Piao, and W.~Zou, ``$\alpha$diff:
  Cross-version binary code similarity detection with {DNN},'' in
  \emph{Proceedings of the ACM/IEEE International Conference on Automated
  Software Engineering}, 2018, pp. 667--678.

\bibitem{ding2019asm2vec}
S.~H. Ding, B.~C. Fung, and P.~Charland, ``{Asm2Vec}: Boosting static
  representation robustness for binary clone search against code obfuscation
  and compiler optimization,'' in \emph{Proceedings of the {IEEE} Symposium on
  Security and Privacy}.\hskip 1em plus 0.5em minus 0.4em\relax IEEE, 2019.

\bibitem{massarelli2019safe}
L.~Massarelli, G.~A. Di~Luna, F.~Petroni, R.~Baldoni, and L.~Querzoni,
  ``{SAFE}: {S}elf-attentive function embeddings for binary similarity,'' in
  \emph{International Conference on Detection of Intrusions and Malware, and
  Vulnerability Assessment}.\hskip 1em plus 0.5em minus 0.4em\relax Springer,
  2019, pp. 309--329.

\bibitem{pewny2015cross}
J.~Pewny, B.~Garmany, R.~Gawlik, C.~Rossow, and T.~Holz, ``Cross-architecture
  bug search in binary executables,'' in \emph{Proceedings of the {IEEE}
  Symposium on Security and Privacy}.\hskip 1em plus 0.5em minus 0.4em\relax
  IEEE, 2015, pp. 709--724.

\bibitem{feng2016scalable}
Q.~Feng, R.~Zhou, C.~Xu, Y.~Cheng, B.~Testa, and H.~Yin, ``Scalable graph-based
  bug search for firmware images,'' in \emph{Proceedings of the ACM Conference
  on Computer and Communications Security}, 2016, pp. 480--491.

\bibitem{zuo2019neural}
F.~Zuo, X.~Li, Z.~Zhang, P.~Young, L.~Luo, and Q.~Zeng, ``Neural machine
  translation inspired binary code similarity comparison beyond function
  pairs,'' in \emph{Proceedings of the Network and Distributed System Security
  Symposium}, 2019.

\bibitem{marastoni2018deep}
N.~Marastoni, R.~Giacobazzi, and M.~Dalla~Preda, ``A deep learning approach to
  program similarity,'' in \emph{Proceedings of the 1st International Workshop
  on Machine Learning and Software Engineering in Symbiosis}.\hskip 1em plus
  0.5em minus 0.4em\relax ACM, 2018, pp. 26--35.

\bibitem{redmond2019cross}
K.~Redmond, L.~Luo, and Q.~Zeng, ``A cross-architecture instruction embedding
  model for natural language processing-inspired binary code analysis,'' in
  \emph{The NDSS Workshop on Binary Analysis Research}, 2019.

\bibitem{duan2020deepbindiff}
Y.~Duan, X.~Li, J.~Wang, and H.~Yin, ``{D}eep{B}in{D}iff: Learning program-wide
  code representations for binary diffing,'' in \emph{Proceedings of the
  Network and Distributed System Security Symposium}, 2020.

\bibitem{sun2020hybrid}
P.~Sun, L.~Garcia, G.~Salles-Loustau, and S.~Zonouz, ``Hybrid firmware analysis
  for known mobile and {IoT} security vulnerabilities,'' in \emph{2020 50th
  Annual IEEE/IFIP International Conference on Dependable Systems and Networks
  (DSN)}.\hskip 1em plus 0.5em minus 0.4em\relax IEEE, 2020, pp. 373--384.

\bibitem{massarelli2019investigating}
L.~Massarelli, G.~A. Di~Luna, F.~Petroni, L.~Querzoni, and R.~Baldoni,
  ``Investigating graph embedding neural networks with unsupervised features
  extraction for binary analysis,'' in \emph{The NDSS Workshop on Binary
  Analysis Research}, 2019.

\bibitem{egele2014blanket}
M.~Egele, M.~Woo, P.~Chapman, and D.~Brumley, ``Blanket execution: Dynamic
  similarity testing for program binaries and components,'' in
  \emph{Proceedings of the USENIX Security Symposium}, 2014, pp. 303--317.

\bibitem{wang2017memory}
S.~Wang and D.~Wu, ``In-memory fuzzing for binary code similarity analysis,''
  in \emph{Proceedings of the IEEE/ACM International Conference on Automated
  Software Engineering}, 2017, pp. 319--330.

\bibitem{ding2016kam1n0}
S.~H. Ding, B.~Fung, and P.~Charland, ``Kam1n0: Mapreduce-based assembly clone
  search for reverse engineering,'' in \emph{Proceedings of the ACM SIGKDD
  International Conference on Knowledge Discovery and Data Mining}, 2016, pp.
  461--470.

\bibitem{hu2016cross}
Y.~Hu, Y.~Zhang, J.~Li, and D.~Gu, ``Cross-architecture binary semantics
  understanding via similar code comparison,'' in \emph{Proceedings of the
  {IEEE} International Conference on Software Analysis, Evolution, and
  Reengineering}, 2016, pp. 57--67.

\bibitem{huang2017binsequence}
H.~Huang, A.~M. Youssef, and M.~Debbabi, ``{BinSequence}: Fast, accurate and
  scalable binary code reuse detection,'' in \emph{Proceedings of the {ACM}
  Symposium on Information, Computer and Communications Security}, 2017, pp.
  155--166.

\bibitem{hu2017binary}
Y.~Hu, Y.~Zhang, J.~Li, and D.~Gu, ``Binary code clone detection across
  architectures and compiling configurations,'' in \emph{Proceedings of the
  International Conference on Program Comprehension}.\hskip 1em plus 0.5em
  minus 0.4em\relax IEEE Press, 2017, pp. 88--98.

\bibitem{kim2017testing}
S.~Kim, M.~Faerevaag, M.~Jung, S.~Jung, D.~Oh, J.~Lee, and S.~K. Cha, ``Testing
  intermediate representations for binary analysis,'' in \emph{Proceedings of
  the IEEE/ACM International Conference on Automated Software Engineering},
  2017, pp. 353--364.

\bibitem{schwartz:usec:2013}
E.~J. Schwartz, J.~Lee, M.~Woo, and D.~Brumley, ``Native x86 decompilation
  using semantics-preserving structural analysis and iterative control-flow
  structuring,'' in \emph{Proceedings of the USENIX Security Symposium}, 2013,
  pp. 353--368.

\bibitem{yakdan:ndss:2015}
K.~Yakdan, S.~Eschweiler, E.~Gerhards-Padilla, and M.~Smith, ``No more gotos:
  Decompilation using pattern-independent control-flow structuring and
  semantics-preserving transformations,'' in \emph{Proceedings of the Network
  and Distributed System Security Symposium}, 2015.

\bibitem{real1996probabilistic}
R.~Real and J.~M. Vargas, ``The probabilistic basis of jaccard's index of
  similarity,'' \emph{Systematic biology}, vol.~45, no.~3, pp. 380--385, 1996.

\bibitem{bunke1997relation}
H.~Bunke, ``On a relation between graph edit distance and maximum common
  subgraph,'' \emph{Pattern Recognition Letters}, vol.~18, no.~8, pp. 689--694,
  1997.

\bibitem{bromley1994signature}
J.~Bromley, I.~Guyon, Y.~LeCun, E.~S{\"a}ckinger, and R.~Shah, ``Signature
  verification using a" siamese" time delay neural network,'' in \emph{Advances
  in neural information processing systems}, 1994, pp. 737--744.

\bibitem{geurts2006extremely}
P.~Geurts, D.~Ernst, and L.~Wehenkel, ``Extremely randomized trees,''
  \emph{Machine learning}, vol.~63, no.~1, pp. 3--42, 2006.

\bibitem{gao2008binhunt}
D.~Gao, M.~K. Reiter, and D.~Song, ``Binhunt: Automatically finding semantic
  differences in binary programs,'' in \emph{International Conference on
  Information and Communications Security}.\hskip 1em plus 0.5em minus
  0.4em\relax Springer, 2008, pp. 238--255.

\bibitem{dullien2005graph}
T.~Dullien and R.~Rolles, ``Graph-based comparison of executable objects
  (english version),'' \emph{SSTIC}, vol.~5, no.~1, p.~3, 2005.

\bibitem{flake2004structural}
H.~Flake, ``Structural comparison of executable objects,'' in
  \emph{International Conference on Detection of Intrusions and Malware, and
  Vulnerability Assessment}.\hskip 1em plus 0.5em minus 0.4em\relax Citeseer,
  2004, pp. 161--174.

\bibitem{bourquin2013binslayer}
M.~Bourquin, A.~King, and E.~Robbins, ``Binslayer: accurate comparison of
  binary executables,'' in \emph{Proceedings of the 2nd ACM SIGPLAN Program
  Protection and Reverse Engineering Workshop}.\hskip 1em plus 0.5em minus
  0.4em\relax ACM, 2013, p.~4.

\bibitem{ming2012ibinhunt}
J.~Ming, M.~Pan, and D.~Gao, ``ibinhunt: Binary hunting with inter-procedural
  control flow,'' in \emph{International Conference on Information Security and
  Cryptology}.\hskip 1em plus 0.5em minus 0.4em\relax Springer, 2012, pp.
  92--109.

\bibitem{jin2012binary}
W.~Jin, S.~Chaki, C.~Cohen, A.~Gurfinkel, J.~Havrilla, C.~Hines, and
  P.~Narasimhan, ``Binary function clustering using semantic hashes,'' in
  \emph{Machine Learning and Applications (ICMLA), 2012 11th International
  Conference on}, vol.~1.\hskip 1em plus 0.5em minus 0.4em\relax IEEE, 2012,
  pp. 386--391.

\bibitem{lakhotia2013fast}
A.~Lakhotia, M.~D. Preda, and R.~Giacobazzi, ``Fast location of similar code
  fragments using semantic'juice','' in \emph{Proceedings of the 2nd ACM
  SIGPLAN Program Protection and Reverse Engineering Workshop}.\hskip 1em plus
  0.5em minus 0.4em\relax ACM, 2013, p.~5.

\bibitem{alrabaee2015sigma}
S.~Alrabaee, P.~Shirani, L.~Wang, and M.~Debbabi, ``Sigma: A semantic
  integrated graph matching approach for identifying reused functions in binary
  code,'' \emph{Digital Investigation}, vol.~12, pp. S61--S71, 2015.

\bibitem{alrabaee2016bingold}
S.~Alrabaee, L.~Wang, and M.~Debbabi, ``{BinGold}: Towards robust binary
  analysis by extracting the semantics of binary code as semantic flow graphs
  (sfgs),'' \emph{Digital Investigation}, vol.~18, pp. S11--S22, 2016.

\bibitem{kim2019binary}
T.~Kim, Y.~R. Lee, B.~Kang, and E.~G. Im, ``Binary executable file similarity
  calculation using function matching,'' \emph{The Journal of Supercomputing},
  vol.~75, no.~2, pp. 607--622, 2019.

\bibitem{guo2020lightweight}
H.~Guo, S.~Huang, C.~Huang, M.~Zhang, Z.~Pan, F.~Shi, H.~Huang, D.~Hu, and
  X.~Wang, ``A lightweight cross-version binary code similarity detection based
  on similarity and correlation coefficient features,'' \emph{IEEE Access},
  vol.~8, pp. 120\,501--120\,512, 2020.

\bibitem{bindiff}
\BIBentryALTinterwordspacing
``Bindiff.'' [Online]. Available: \url{https://www.zynamics.com/bindiff.html}
\BIBentrySTDinterwordspacing

\bibitem{diaphora}
\BIBentryALTinterwordspacing
``Diaphora, a {F}ree and {O}pen {S}ource program diffing tool.'' [Online].
  Available: \url{http://diaphora.re/}
\BIBentrySTDinterwordspacing

\bibitem{oh2015darungrim}
J.~W. Oh, ``Darungrim: a patch analysis and binary diffing too,'' 2015.

\bibitem{david2014tracelet}
Y.~David and E.~Yahav, ``Tracelet-based code search in executables,'' in
  \emph{Proceedings of the ACM SIGPLAN Conference on Programming Language
  Design and Implementation}, 2014, pp. 349--360.

\bibitem{farhadi:2014}
M.~R. Farhadi, B.~C. Fung, P.~Charland, and M.~Debbabi, ``{BinClone}: Detecting
  code clones in malware,'' in \emph{Proceedings of the International
  Conference on Software Security and Reliability}, 2014, pp. 78--87.

\bibitem{david2016statistical}
Y.~David, N.~Partush, and E.~Yahav, ``Statistical similarity of binaries,'' in
  \emph{Proceedings of the ACM SIGPLAN Conference on Programming Language
  Design and Implementation}, 2016, pp. 266--280.

\bibitem{lageman2016bindnn}
N.~Lageman, E.~D. Kilmer, R.~J. Walls, and P.~D. McDaniel, ``{BinDNN}:
  Resilient function matching using deep learning,'' in \emph{International
  Conference on Security and Privacy in Communication Systems}.\hskip 1em plus
  0.5em minus 0.4em\relax Springer, 2016, pp. 517--537.

\bibitem{nouh2017binsign}
L.~Nouh, A.~Rahimian, D.~Mouheb, M.~Debbabi, and A.~Hanna, ``Binsign:
  fingerprinting binary functions to support automated analysis of code
  executables,'' in \emph{IFIP International Conference on ICT Systems Security
  and Privacy Protection}.\hskip 1em plus 0.5em minus 0.4em\relax Springer,
  2017, pp. 341--355.

\bibitem{david2017similarity}
Y.~David, N.~Partush, and E.~Yahav, ``Similarity of binaries through
  re-optimization,'' in \emph{Proceedings of the ACM SIGPLAN Conference on
  Programming Language Design and Implementation}, 2017, pp. 79--94.

\bibitem{ming2017binsim}
J.~Ming, D.~Xu, Y.~Jiang, and D.~Wu, ``{BinSim}: Trace-based semantic binary
  diffing via system call sliced segment equivalence checking,'' in
  \emph{Proceedings of the {USENIX} Security Symposium}, 2017, pp. 253--270.

\bibitem{kargen2017towards}
U.~Karg{\'e}n and N.~Shahmehri, ``Towards robust instruction-level trace
  alignment of binary code,'' in \emph{Proceedings of the IEEE/ACM
  International Conference on Automated Software Engineering}.\hskip 1em plus
  0.5em minus 0.4em\relax IEEE, 2017, pp. 342--352.

\bibitem{karamitas2018efficient}
C.~Karamitas and A.~Kehagias, ``Efficient features for function matching
  between binary executables,'' in \emph{Proceedings of the {IEEE}
  International Conference on Software Analysis, Evolution, and
  Reengineering}.\hskip 1em plus 0.5em minus 0.4em\relax IEEE, 2018, pp.
  335--345.

\bibitem{yuan2018new}
B.~Yuan, J.~Wang, Z.~Fang, and L.~Qi, ``A new software birthmark based on
  weight sequences of dynamic control flow graph for plagiarism detection,''
  \emph{The Computer Journal}, 2018.

\bibitem{hu2018binmatch}
Y.~Hu, Y.~Zhang, J.~Li, H.~Wang, B.~Li, and D.~Gu, ``Binmatch: A
  semantics-based hybrid approach on binary code clone analysis,'' in
  \emph{Software Maintenance and Evolution (ICSME), 2017 IEEE International
  Conference on}.\hskip 1em plus 0.5em minus 0.4em\relax IEEE, 2018.

\bibitem{shalev2018binary}
N.~Shalev and N.~Partush, ``Binary similarity detection using machine
  learning,'' in \emph{Proceedings of the 13th Workshop on Programming
  Languages and Analysis for Security}.\hskip 1em plus 0.5em minus 0.4em\relax
  ACM, 2018, pp. 42--47.

\bibitem{luo2019funcnet}
M.~Luo, C.~Yang, X.~Gong, and L.~Yu, ``{F}unc{N}et: A euclidean embedding
  approach for lightweight cross-platform binary recognition,'' in
  \emph{International Conference on Security and Privacy in Communication
  Systems}.\hskip 1em plus 0.5em minus 0.4em\relax Springer, 2016, pp.
  517--537.

\bibitem{jiang2020similarity}
J.~Jiang, G.~Li, M.~Yu, G.~Li, C.~Liu, Z.~Lv, B.~Lv, and W.~Huang, ``Similarity
  of binaries across optimization levels and obfuscation,'' in
  \emph{Proceedings of the European Symposium on Research in Computer
  Security}, 2020, pp. 295--315.

\bibitem{shirani2017binshape}
P.~Shirani, L.~Wang, and M.~Debbabi, ``{BinShape}: Scalable and robust binary
  library function identification using function shape,'' in
  \emph{International Conference on Detection of Intrusions and Malware, and
  Vulnerability Assessment}.\hskip 1em plus 0.5em minus 0.4em\relax Springer,
  2017, pp. 301--324.

\bibitem{chen2014achieving}
K.~Chen, P.~Liu, and Y.~Zhang, ``Achieving accuracy and scalability
  simultaneously in detecting application clones on android markets,'' in
  \emph{Proceedings of the 36th International Conference on Software
  Engineering}.\hskip 1em plus 0.5em minus 0.4em\relax ACM, 2014, pp. 175--186.

\bibitem{hu:ccs:2009}
X.~Hu, T.-c. Chiueh, and K.~G. Shin, ``Large-scale malware indexing using
  function-call graphs,'' in \emph{Proceedings of the ACM Conference on
  Computer and Communications Security}, 2009, pp. 611--620.

\bibitem{henry1981software}
S.~Henry and D.~Kafura, ``Software structure metrics based on information
  flow,'' \emph{IEEE transactions on Software Engineering}, no.~5, pp.
  510--518, 1981.

\bibitem{jang2012redebug}
J.~Jang, A.~Agrawal, and D.~Brumley, ``{ReDeBug}: Finding unpatched code clones
  in entire os distributions,'' in \emph{Proceedings of the {IEEE} Symposium on
  Security and Privacy}, 2012, pp. 48--62.

\bibitem{cha:nsdi:2010}
S.~K. Cha, I.~Moraru, J.~Jang, J.~Truelove, D.~Brumley, and D.~G. Andersen,
  ``{SplitScreen}: Enabling efficient, distributed malware detection,'' in
  \emph{Proceedings of the {USENIX} Symposium on Networked Systems Design and
  Implementation}, 2010, pp. 377--390.

\bibitem{khoo2013rendezvous}
W.~M. Khoo, A.~Mycroft, and R.~Anderson, ``Rendezvous: a search engine for
  binary code,'' in \emph{Proceedings of the 10th Working Conference on Mining
  Software Repositories}.\hskip 1em plus 0.5em minus 0.4em\relax IEEE Press,
  2013, pp. 329--338.

\bibitem{schkufza2013stochastic}
E.~Schkufza, R.~Sharma, and A.~Aiken, ``Stochastic superoptimization,'' in
  \emph{Proceedings of the International Conference on Architectural Support
  for Programming Languages and Operating Systems}, 2013, pp. 305--316.

\bibitem{ming2016deviation}
J.~Ming, F.~Zhang, D.~Wu, P.~Liu, and S.~Zhu, ``Deviation-based
  obfuscation-resilient program equivalence checking with application to
  software plagiarism detection,'' \emph{IEEE Transactions on Reliability},
  vol.~65, no.~4, pp. 1647--1664, 2016.

\bibitem{luo2017semantics}
L.~Luo, J.~Ming, D.~Wu, P.~Liu, and S.~Zhu, ``Semantics-based
  obfuscation-resilient binary code similarity comparison with applications to
  software and algorithm plagiarism detection,'' \emph{IEEE Transactions on
  Software Engineering}, no.~12, pp. 1157--1177, 2017.

\bibitem{forrest:1996}
S.~Forrest, S.~A. Hofmeyr, A.~Somayaji, and T.~A. Longstaff, ``A sense of self
  for {Unix} processes,'' in \emph{Proceedings of the IEEE Symposium on
  Security and Privacy}, 1996, pp. 120--128.

\bibitem{tian2020plagiarism}
Z.~Tian, Q.~Wang, C.~Gao, L.~Chen, and D.~Wu, ``Plagiarism detection of
  multi-threaded programs via siamese neural networks,'' \emph{IEEE Access},
  vol.~8, pp. 160\,802--160\,814, 2020.

\bibitem{manes:tse:2019}
V.~J.~M. Man{\`{e}}s, H.~Han, C.~Han, S.~K. Cha, M.~Egele, E.~J. Schwartz, and
  M.~Woo, ``The art, science, and engineering of fuzzing: A survey,''
  \emph{IEEE Transactions on Software Engineering}, 2019.

\bibitem{grobert2011automated}
F.~Gr{\"o}bert, C.~Willems, and T.~Holz, ``Automated identification of
  cryptographic primitives in binary programs,'' in \emph{International
  Workshop on Recent Advances in Intrusion Detection}.\hskip 1em plus 0.5em
  minus 0.4em\relax Springer, 2011, pp. 41--60.

\bibitem{horwitz:1988}
S.~Horwitz, T.~Reps, and D.~Binkley, ``Interprocedural slicing using dependence
  graphs,'' in \emph{Proceedings of the ACM SIGPLAN Conference on Programming
  Language Design and Implementation}, 1988, pp. 35--46.

\bibitem{ferrante:1987}
J.~Ferrante, K.~J. Ottenstein, and J.~D. Warren, ``The program dependence graph
  and its use in optimization,'' \emph{ACM Transactions on Programming
  Languages and Systems}, vol.~9, no.~3, pp. 319--349, 1987.

\bibitem{ng2002spectral}
A.~Y. Ng, M.~I. Jordan, and Y.~Weiss, ``On spectral clustering: Analysis and an
  algorithm,'' in \emph{Advances in neural information processing systems},
  2002, pp. 849--856.

\bibitem{yang2007evaluating}
J.~Yang, Y.-G. Jiang, A.~G. Hauptmann, and C.-W. Ngo, ``Evaluating
  bag-of-visual-words representations in scene classification,'' in
  \emph{Proceedings of the international workshop on Workshop on multimedia
  information retrieval}.\hskip 1em plus 0.5em minus 0.4em\relax ACM, 2007, pp.
  197--206.

\bibitem{arandjelovic2013all}
R.~Arandjelovic and A.~Zisserman, ``All about vlad,'' in \emph{Proceedings of
  the IEEE conference on Computer Vision and Pattern Recognition}, 2013, pp.
  1578--1585.

\bibitem{dai2016discriminative}
H.~Dai, B.~Dai, and L.~Song, ``Discriminative embeddings of latent variable
  models for structured data,'' in \emph{International Conference on Machine
  Learning}, 2016, pp. 2702--2711.

\bibitem{mikolov2013efficient}
T.~Mikolov, K.~Chen, G.~Corrado, and J.~Dean, ``Efficient estimation of word
  representations in vector space,'' \emph{arXiv preprint arXiv:1301.3781},
  2013.

\bibitem{kim2014convolutional}
Y.~Kim, ``Convolutional neural networks for sentence classification,''
  \emph{arXiv preprint arXiv:1408.5882}, 2014.

\bibitem{hochreiter1997long}
S.~Hochreiter and J.~Schmidhuber, ``Long short-term memory,'' \emph{Neural
  computation}, vol.~9, no.~8, pp. 1735--1780, 1997.

\bibitem{le2014distributed}
Q.~Le and T.~Mikolov, ``Distributed representations of sentences and
  documents,'' in \emph{International Conference on Machine Learning}, 2014,
  pp. 1188--1196.

\bibitem{ida}
\BIBentryALTinterwordspacing
Hex-Rays, ``{IDA} {P}ro.'' [Online]. Available:
  \url{https://www.hex-rays.com/products/ida/}
\BIBentrySTDinterwordspacing

\bibitem{andriesse2016depth}
D.~Andriesse, X.~Chen, V.~van~der Veen, A.~Slowinska, and H.~Bos, ``An in-depth
  analysis of disassembly on full-scale x86/x64 binaries,'' in
  \emph{Proceedings of the {USENIX} Security Symposium}, 2016, pp. 583--600.

\bibitem{jung:bar:2019}
M.~Jung, S.~Kim, H.~Han, J.~Choi, and S.~K. Cha, ``{B2R2}: Building an
  efficient front-end for binary analysis,'' in \emph{Proceedings of the NDSS
  Workshop on Binary Analysis Research}, 2019.

\bibitem{funseeker}
H.~Kim, J.~Lee, S.~Kim, S.~Jung, and S.~K. Cha, ``How'd security benefit
  reverse engineers? the implication of {Intel} {CET} on function
  identification,'' in \emph{Proceedings of the International Conference on
  Dependable Systems Networks}, 2022, pp. 559--566.

\bibitem{andriesse2017compiler}
D.~Andriesse, A.~Slowinska, and H.~Bos, ``Compiler-agnostic function detection
  in binaries,'' in \emph{Proceedings of the {IEEE} European Symposium on
  Security and Privacy}, 2017, pp. 177--189.

\bibitem{wang2017semantics}
S.~Wang, P.~Wang, and D.~Wu, ``Semantics-aware machine learning for function
  recognition in binary code,'' in \emph{Proceedings of the IEEE International
  Conference on Software Maintenance and Evolution}, 2017, pp. 388--398.

\bibitem{qiao2017function}
R.~Qiao and R.~Sekar, ``Function interface analysis: A principled approach for
  function recognition in cots binaries,'' in \emph{Proceedings of the Annual
  IEEE/IFIP International Conference on Dependable Systems and Networks}, 2017,
  pp. 201--212.

\bibitem{kinder:cav:2008}
J.~Kinder and H.~Veith, ``Jakstab: A static analysis platform for binaries,''
  in \emph{Proceedings of the International Conference on Computer Aided
  Verification}, 2008, pp. 423--427.

\bibitem{sec2016pie}
\BIBentryALTinterwordspacing
{SecurityTeam}, ``Pie,'' 2016. [Online]. Available:
  \url{https://wiki.ubuntu.com/SecurityTeam/PIE}
\BIBentrySTDinterwordspacing

\bibitem{gnupacks}
\BIBentryALTinterwordspacing
``{GNU} packages.'' [Online]. Available: \url{https://ftp.gnu.org/gnu/}
\BIBentrySTDinterwordspacing

\bibitem{junod2015obfuscator}
P.~Junod, J.~Rinaldini, J.~Wehrli, and J.~Michielin,
  ``Obfuscator-llvm--software protection for the masses,'' in \emph{Software
  Protection (SPRO), 2015 IEEE/ACM 1st International Workshop on}.\hskip 1em
  plus 0.5em minus 0.4em\relax IEEE, 2015, pp. 3--9.

\bibitem{madou2006loco}
M.~Madou, L.~Van~Put, and K.~De~Bosschere, ``Loco: An interactive code (de)
  obfuscation tool,'' in \emph{Proceedings of the 2006 ACM SIGPLAN symposium on
  Partial evaluation and semantics-based program manipulation}.\hskip 1em plus
  0.5em minus 0.4em\relax ACM, 2006, pp. 140--144.

\bibitem{vmprotect}
\BIBentryALTinterwordspacing
``{VMP}rotect.'' [Online]. Available: \url{http://vmpsoft.com}
\BIBentrySTDinterwordspacing

\bibitem{stunnix:obfus}
\BIBentryALTinterwordspacing
``{S}tunnix {C}/{C}++ {Obfuscator}.'' [Online]. Available:
  \url{http://stunnix.com/prod/cxxo/}
\BIBentrySTDinterwordspacing

\bibitem{semdesigns:obfus}
\BIBentryALTinterwordspacing
``{S}emantic {D}esigns: {S}ource {C}ode {O}bfuscators.'' [Online]. Available:
  \url{http://www.semdesigns.com/Products/Obfuscators/}
\BIBentrySTDinterwordspacing

\bibitem{collberg2015tigress}
C.~Collberg, ``The tigress c diversifier/obfuscator,'' \emph{Retrieved August},
  vol.~14, p. 2015, 2015.

\bibitem{crosstool-ng}
\BIBentryALTinterwordspacing
``{C}rosstool-{NG}.'' [Online]. Available:
  \url{https://github.com/crosstool-ng/crosstool-ng}
\BIBentrySTDinterwordspacing

\bibitem{mackenzie1996autoconf}
D.~MacKenzie, B.~Elliston, and A.~Demaille, ``Autoconf --- creating automatic
  configuration scripts,'' 1996.

\bibitem{Tange2011a}
\BIBentryALTinterwordspacing
O.~Tange, ``{GNU} parallel - the command-line power tool,'' \emph{;login: The
  USENIX Magazine}, vol.~36, no.~1, pp. 42--47, Feb 2011. [Online]. Available:
  \url{http://www.gnu.org/s/parallel}
\BIBentrySTDinterwordspacing

\bibitem{hagberg2008exploring}
A.~A. Hagberg, D.~A. Schult, and P.~J. Swart, ``Exploring network structure,
  dynamics, and function using {NetworkX},'' in \emph{Proceedings of the Python
  in Science Conference}, 2008, pp. 11--15.

\bibitem{intelmanual}
{Intel Corporation}, ``Intel{\textregistered} 64 and ia-32 architectures
  software developer's manual,''
  \url{https://software.intel.com/en-us/articles/intel-sdm}.

\bibitem{seal2001arm}
D.~Seal, \emph{ARM Architecture Reference Manual}.\hskip 1em plus 0.5em minus
  0.4em\relax Pearson Education, 2001.

\bibitem{mips2001mips}
{MIPS Technologies, Inc.}, ``Mips32 architecture for programmers volume ii: The
  mips32 instruction set,'' 2001.

\bibitem{capstone}
\BIBentryALTinterwordspacing
Capstone, ``The ultimate disassembler.'' [Online]. Available:
  \url{https://www.capstone-engine.org/}
\BIBentrySTDinterwordspacing

\bibitem{wiki:reldiff}
\BIBentryALTinterwordspacing
Wikipedia, ``Relative change and difference --- wikipedia{,} the free
  encyclopedia,'' 2018, [Online; accessed <today>]. [Online]. Available:
  \url{"https://en.wikipedia.org/w/index.php?title=Relative_change_and_difference&oldid=872867886"}
\BIBentrySTDinterwordspacing

\bibitem{guyon2003introduction}
I.~Guyon and A.~Elisseeff, ``An introduction to variable and feature
  selection,'' \emph{Journal of machine learning research}, vol.~3, no. Mar,
  pp. 1157--1182, 2003.

\bibitem{caruana94greedyattribute}
R.~Caruana and D.~Freitag, ``Greedy attribute selection,'' in \emph{Proceedings
  of the Eleventh International Conference on Machine Learning}.\hskip 1em plus
  0.5em minus 0.4em\relax Morgan Kaufmann, 1994, pp. 28--36.

\bibitem{pedregosa2011scikit}
F.~Pedregosa, G.~Varoquaux, A.~Gramfort, V.~Michel, B.~Thirion, O.~Grisel,
  M.~Blondel, P.~Prettenhofer, R.~Weiss, V.~Dubourg \emph{et~al.},
  ``Scikit-learn: Machine learning in python,'' \emph{Journal of machine
  learning research}, vol.~12, no. Oct, pp. 2825--2830, 2011.

\bibitem{scipy}
\BIBentryALTinterwordspacing
E.~Jones, T.~Oliphant, P.~Peterson \emph{et~al.}, ``{SciPy}: Open source
  scientific tools for {Python},'' 2001--. [Online]. Available:
  \url{http://www.scipy.org/}
\BIBentrySTDinterwordspacing

\bibitem{walt2011numpy}
S.~v.~d. Walt, S.~C. Colbert, and G.~Varoquaux, ``The {NumPy} array: A
  structure for efficient numerical computation,'' \emph{Computing in Science
  \& Engineering}, vol.~13, no.~2, pp. 22--30, 2011.

\bibitem{gccopt}
\BIBentryALTinterwordspacing
``Using the {GNU} compiler collection ({GCC}): Optimize options.'' [Online].
  Available: \url{https://gcc.gnu.org/onlinedocs/gcc/Optimize-Options.html}
\BIBentrySTDinterwordspacing

\bibitem{clangopt}
\BIBentryALTinterwordspacing
``Clang - the clang c, c++, and objective-c compiler.'' [Online]. Available:
  \url{https://clang.llvm.org/docs/CommandGuide/clang.html}
\BIBentrySTDinterwordspacing

\bibitem{laszlo2009obfuscating}
T.~L{\'a}szl{\'o} and {\'A}.~Kiss, ``Obfuscating c++ programs via control flow
  flattening,'' \emph{Annales Universitatis Scientarum Budapestinensis de
  Rolando E{\"o}tv{\"o}s Nominatae, Sectio Computatorica}, vol.~30, pp. 3--19,
  2009.

\bibitem{themida}
\BIBentryALTinterwordspacing
``Themida: Advanced windows software protection system.'' [Online]. Available:
  \url{https://www.oreans.com/themida.php}
\BIBentrySTDinterwordspacing

\bibitem{kim:2020:firmae}
M.~Kim, D.~Kim, E.~Kim, S.~Kim, Y.~Jang, and Y.~Kim, ``{FirmAE}: Towards
  large-scale emulation of iot firmware for dynamic analysis,'' in \emph{Annual
  Computer Security Applications Conference (ACSAC)}, Online, Dec. 2020.

\bibitem{chua2017neural}
Z.~L. Chua, S.~Shen, P.~Saxena, and Z.~Liang, ``Neural nets can learn function
  type signatures from binaries,'' in \emph{Proceedings of the USENIX Security
  Symposium}, 2017, pp. 99--116.

\bibitem{van2016tough}
V.~van~der Veen, E.~G{\"o}ktas, M.~Contag, A.~Pawoloski, X.~Chen, S.~Rawat,
  H.~Bos, T.~Holz, E.~Athanasopoulos, and C.~Giuffrida, ``A tough call:
  Mitigating advanced code-reuse attacks at the binary level,'' in
  \emph{Proceedings of the {IEEE} Symposium on Security and Privacy}, 2016, pp.
  934--953.

\bibitem{hiebert1999exuberant}
D.~Hiebert, ``Exuberant {Ctags},'' 1999.

\bibitem{lee2011tie}
J.~Lee, T.~Avgerinos, and D.~Brumley, ``{TIE}: Principled reverse engineering
  of types in binary programs,'' in \emph{Proceedings of the Network and
  Distributed System Security Symposium}, 2011.

\bibitem{elwazeer2013scalable}
K.~ElWazeer, K.~Anand, A.~Kotha, M.~Smithson, and R.~Barua, ``Scalable variable
  and data type detection in a binary rewriter,'' \emph{ACM SIGPLAN Notices},
  vol.~48, no.~6, pp. 51--60, 2013.

\bibitem{he2018debin}
J.~He, P.~Ivanov, P.~Tsankov, V.~Raychev, and M.~Vechev, ``Debin: Predicting
  debug information in stripped binaries,'' in \emph{Proceedings of the 2018
  ACM SIGSAC Conference on Computer and Communications Security}.\hskip 1em
  plus 0.5em minus 0.4em\relax ACM, 2018, pp. 1667--1680.

\bibitem{artuso2019nomine}
F.~Artuso, G.~A. Di~Luna, L.~Massarelli, and L.~Querzoni, ``In nomine function:
  Naming functions in stripped binaries with neural networks,'' \emph{arXiv},
  pp. arXiv--1912, 2019.

\bibitem{rosenblum2011recovering}
N.~Rosenblum, B.~P. Miller, and X.~Zhu, ``Recovering the toolchain provenance
  of binary code,'' in \emph{Proceedings of the International Symposium on
  Software Testing and Analysis}, 2011, pp. 100--110.

\bibitem{tol2020fastspec}
M.~C. Tol, K.~Yurtseven, B.~Gulmezoglu, and B.~Sunar, ``{F}ast{S}pec: Scalable
  generation and detection of spectre gadgets using neural embeddings,''
  \emph{arXiv preprint arXiv:2006.14147}, 2020.

\bibitem{canali2012quantitative}
D.~Canali, A.~Lanzi, D.~Balzarotti, C.~Kruegel, M.~Christodorescu, and
  E.~Kirda, ``A quantitative study of accuracy in system call-based malware
  detection,'' in \emph{Proceedings of the 2012 International Symposium on
  Software Testing and Analysis}.\hskip 1em plus 0.5em minus 0.4em\relax ACM,
  2012, pp. 122--132.

\bibitem{babic2011malware}
D.~Babi{\'c}, D.~Reynaud, and D.~Song, ``Malware analysis with tree automata
  inference,'' in \emph{International Conference on Computer Aided
  Verification}.\hskip 1em plus 0.5em minus 0.4em\relax Springer, 2011, pp.
  116--131.

\bibitem{xiao2016matching}
Y.~Xiao, S.~Cao, Z.~Cao, F.~Wang, F.~Lin, J.~Wu, and H.~Bi, ``Matching similar
  functions in different versions of a malware,'' in \emph{2016 IEEE
  Trustcom/BigDataSE/ISPA}.\hskip 1em plus 0.5em minus 0.4em\relax IEEE, 2016,
  pp. 252--259.

\bibitem{ming2015memoized}
J.~Ming, D.~Xu, and D.~Wu, ``Memoized semantics-based binary diffing with
  application to malware lineage inference,'' in \emph{IFIP International
  Information Security and Privacy Conference}.\hskip 1em plus 0.5em minus
  0.4em\relax Springer, 2015, pp. 416--430.

\bibitem{alrabaee2018fossil}
S.~Alrabaee, P.~Shirani, L.~Wang, and M.~Debbabi, ``Fossil: a resilient and
  efficient system for identifying foss functions in malware binaries,''
  \emph{ACM Transactions on Privacy and Security}, vol.~21, no.~2, pp. 1--34,
  2018.

\bibitem{calvet2012aligot}
J.~Calvet, J.~M. Fernandez, and J.-Y. Marion, ``Aligot: cryptographic function
  identification in obfuscated binary programs,'' in \emph{Proceedings of the
  2012 ACM conference on Computer and communications security}.\hskip 1em plus
  0.5em minus 0.4em\relax ACM, 2012, pp. 169--182.

\bibitem{xu2017cryptographic}
D.~Xu, J.~Ming, and D.~Wu, ``Cryptographic function detection in obfuscated
  binaries via bit-precise symbolic loop mapping,'' in \emph{Proceedings of the
  {IEEE} Symposium on Security and Privacy}, 2017, pp. 921--937.

\bibitem{qiu2015library}
J.~Qiu, X.~Su, and P.~Ma, ``Library functions identification in binary code by
  using graph isomorphism testings,'' in \emph{Proceedings of the {IEEE}
  International Conference on Software Analysis, Evolution, and
  Reengineering}.\hskip 1em plus 0.5em minus 0.4em\relax IEEE, 2015, pp.
  261--270.

\bibitem{jia2020neural}
L.~Jia, A.~Zhou, P.~Jia, L.~Liu, Y.~Wang, and L.~Liu, ``A neural network-based
  approach for cryptographic function detection in malware,'' \emph{IEEE
  Access}, vol.~8, pp. 23\,506--23\,521, 2020.

\bibitem{alrabaee2020cpa}
S.~Alrabaee, M.~Debbabi, and L.~Wang, ``Cpa: Accurate cross-platform binary
  authorship characterization using lda,'' \emph{IEEE Transactions on
  Information Forensics and Security}, vol.~15, pp. 3051--3066, 2020.

\bibitem{xu2017spain}
Z.~Xu, B.~Chen, M.~Chandramohan, Y.~Liu, and F.~Song, ``Spain: security patch
  analysis for binaries towards understanding the pain and pills,'' in
  \emph{Proceedings of the 39th International Conference on Software
  Engineering}.\hskip 1em plus 0.5em minus 0.4em\relax IEEE Press, 2017, pp.
  462--472.

\bibitem{hu2019automatically}
Y.~Hu, Y.~Zhang, and D.~Gu, ``Automatically patching vulnerabilities of binary
  programs via code transfer from correct versions,'' \emph{IEEE Access},
  vol.~7, pp. 28\,170--28\,184, 2019.

\bibitem{zhao2020patchscope}
L.~Zhao, Y.~Zhu, J.~Ming, Y.~Zhang, H.~Zhang, and H.~Yin, ``{P}atch{S}cope:
  Memory object centric patch diffing,'' in \emph{Proceedings of the ACM
  Conference on Computer and Communications Security}, 2020.

\bibitem{haq2019survey}
I.~U. Haq and J.~Caballero, ``A survey of binary code similarity,'' \emph{arXiv
  preprint arXiv:1909.11424}, 2019.

\bibitem{qasem2021automatic}
A.~Qasem, P.~Shirani, M.~Debbabi, L.~Wang, B.~Lebel, and B.~L. Agba,
  ``Automatic vulnerability detection in embedded devices and firmware: survey
  and layered taxonomies,'' \emph{ACM Computing Surveys (CSUR)}, vol.~54,
  no.~2, pp. 1--42, 2021.

\bibitem{kamiya2002ccfinder}
T.~Kamiya, S.~Kusumoto, and K.~Inoue, ``{CCFinder}: a multilinguistic
  token-based code clone detection system for large scale source code,''
  \emph{IEEE Transactions on Software Engineering}, vol.~28, no.~7, pp.
  654--670, 2002.

\bibitem{schleimer:2003}
S.~Schleimer, D.~S. Wilkerson, and A.~Aiken, ``Winnowing: Local algorithms for
  document fingerprinting,'' in \emph{Proceedings of the ACM SIGMOD
  International Conference on Management of Data}, 2003, pp. 76--85.

\bibitem{li2004cp}
Z.~Li, S.~Lu, S.~Myagmar, and Y.~Zhou, ``{CP-Miner}: A tool for finding
  copy-paste and related bugs in operating system code,'' in \emph{OSdi},
  vol.~4, no.~19, 2004, pp. 289--302.

\bibitem{jiang2007deckard}
L.~Jiang, G.~Misherghi, Z.~Su, and S.~Glondu, ``Deckard: Scalable and accurate
  tree-based detection of code clones,'' in \emph{Proceedings of the
  International Conference on Software Engineering}, 2007, pp. 96--105.

\bibitem{dam2017automatic}
H.~K. Dam, T.~Tran, T.~Pham, S.~W. Ng, J.~Grundy, and A.~Ghose, ``Automatic
  feature learning for vulnerability prediction,'' \emph{arXiv preprint
  arXiv:1708.02368}, 2017.

\bibitem{lahiri2012symdiff}
S.~K. Lahiri, C.~Hawblitzel, M.~Kawaguchi, and H.~Reb{\^e}lo, ``{SymDiff}: A
  language-agnostic semantic diff tool for imperative programs,'' in
  \emph{Proceedings of the International Conference on Computer Aided
  Verification}, 2012, pp. 712--717.

\bibitem{kim:oakland:2017}
S.~Kim, S.~Woo, H.~Lee, and H.~Oh, ``{VUDDY}: A scalable approach for
  vulnerable code clone discovery,'' in \emph{Proceedings of the {IEEE}
  Symposium on Security and Privacy}, 2017, pp. 595--614.

\bibitem{wang2016automatically}
S.~Wang, T.~Liu, and L.~Tan, ``Automatically learning semantic features for
  defect prediction,'' in \emph{Proceedings of the International Conference on
  Software Engineering}, 2016, pp. 297--308.

\bibitem{li2016vulpecker}
Z.~Li, D.~Zou, S.~Xu, H.~Jin, H.~Qi, and J.~Hu, ``{VulPecker}: an automated
  vulnerability detection system based on code similarity analysis,'' in
  \emph{Proceedings of the Annual Conference on Computer Security
  Applications}, 2016, pp. 201--213.

\bibitem{miyani2017binpro}
D.~Miyani, Z.~Huang, and D.~Lie, ``{BinPro}: A tool for binary source code
  provenance,'' \emph{arXiv preprint arXiv:1711.00830}, 2017.

\bibitem{rahimian2012resource}
A.~Rahimian, P.~Charland, S.~Preda, and M.~Debbabi, ``{RESource}: A framework
  for online matching of assembly with open source code,'' in
  \emph{International Symposium on Foundations and Practice of Security}, 2012,
  pp. 211--226.

\bibitem{hemel2011finding}
A.~Hemel, K.~T. Kalleberg, R.~Vermaas, and E.~Dolstra, ``Finding software
  license violations through binary code clone detection,'' in
  \emph{Proceedings of the 8th Working Conference on Mining Software
  Repositories}, 2011, pp. 63--72.

\bibitem{ji2021buggraph}
Y.~Ji, L.~Cui, and H.~H. Huang, ``{B}ug{G}raph: Differentiating source-binary
  code similarity with graph triplet-loss network,'' in \emph{Proceedings of
  the 2021 ACM Asia Conference on Computer and Communications Security}, 2021,
  pp. 702--715.

\end{thebibliography}

\normalsize
\clearpage
\onecolumn

\begin{appendices}

\section{Additional Results}
\label{app:opti-level}


\begin{table}[!h]
  \scriptsize
  \centering
  \setlength\tabcolsep{0.04cm}
  \def\arraystretch{0.50}

  \begin{minipage}[t]{.48\linewidth}
  \caption{Number of functions and basic blocks in the \normaldataset
  dataset for each compiler option.}
  \label{tab:numfuncsbbs}
  \begin{threeparttable}
\begin{tabular}{@{}cccrrrrrrrrrr@{}}
\toprule
\multicolumn{3}{c}{Options} & \multicolumn{1}{c}{} & \multicolumn{4}{c}{\# of Functions}   & \multicolumn{1}{c}{} & \multicolumn{4}{c}{\# of Basic Blocks}  \\ \cmidrule(r){1-3} \cmidrule(lr){5-8} \cmidrule(l){10-13}

Comp.\tnote{$\ast$}  & Arch  & Bit  & \multicolumn{1}{c}{} & \multicolumn{1}{c}{O0} & \multicolumn{1}{c}{O1} & \multicolumn{1}{c}{O2} & \multicolumn{1}{c}{O3} & \multicolumn{1}{c}{} & \multicolumn{1}{c}{O0} & \multicolumn{1}{c}{O1} & \multicolumn{1}{c}{O2} & \multicolumn{1}{c}{O3} \\ \midrule
 \clang   & arm   & 32  &  & 35,610   & 35,638   & 26,325\tnote{$\ddagger$}   & 26,210   &  & 921,181\tnote{$\dagger$}  & 422,807  & 490,029  & 512,890  \\
 \gcc   & arm   & 32  &  & 35,798   & 28,817\tnote{$\ddagger$}   & 28,270   & 26,755   &  & 561,575  & 451,965  & 450,850  & 523,404  \\ \midrule
 \clang   & arm   & 64  &  & 35,612   & 35,637   & 26,106   & 26,005   &  & 897,892  & 466,108  & 555,294  & 584,098  \\
 \gcc   & arm   & 64  &  & 35,790   & 28,701   & 28,155   & 26,605   &  & 566,364  & 490,574  & 488,070  & 576,724  \\ \midrule
 \clang   & mips  & 32  &  & 35,723   & 35,741   & 26,221   & 26,130   &  & 933,529  & 470,923  & 547,755  & 583,302  \\
 \gcc   & mips  & 32  &  & 35,898   & 28,832   & 28,311   & 26,778   &  & 580,432  & 552,430  & 548,533  & 634,923  \\ \midrule
 \clang   & mips  & 64  &  & 35,679   & 35,701   & 26,222   & 26,095   &  & 921,484  & 460,491  & 534,209  & 569,074  \\
 \gcc   & mips  & 64  &  & 35,775   & 28,732   & 28,218   & 26,638   &  & 560,243  & 547,273  & 533,655  & 618,634  \\ \midrule
 \clang   & mipseb  & 32  &  & 35,721   & 35,741   & 26,217   & 26,136   &  & 933,654  & 470,795  & 547,752  & 583,189  \\
 \gcc   & mipseb  & 32  &  & 35,895   & 28,831   & 28,305   & 26,775   &  & 580,544  & 554,791  & 549,768  & 635,195  \\ \midrule
 \clang   & mipseb  & 64  &  & 35,676   & 35,700   & 26,220   & 26,093   &  & 922,077  & 460,531  & 534,528  & 569,063  \\
 \gcc   & mipseb  & 64  &  & 35,772   & 28,728   & 28,208   & 26,635   &  & 560,230  & 547,265  & 533,678  & 618,663  \\ \midrule
 \clang   & x86   & 32  &  & 35,466   & 35,484   & 25,974   & 25,878   &  & 692,755  & 479,383  & 575,554  & 609,008  \\
 \gcc   & x86   & 32  &  & 35,602   & 28,543   & 28,476   & 27,074   &  & 562,037  & 501,925  & 503,681  & 580,059  \\ \midrule
 \clang   & x86   & 64  &  & 34,127   & 34,202   & 25,593   & 25,482   &  & 640,058  & 444,199  & 551,809  & 581,622  \\
 \gcc   & x86   & 64  &  & 35,837   & 28,803   & 28,749   & 27,308   &  & 567,578  & 499,713  & 503,899  & 592,185  \\ \bottomrule
\end{tabular}

\begin{tablenotes}
\item [$\ast$]
  We show the numbers for \gcc v8.2.0 and \clang v7.0 for clear comparison.

\item [$\dagger$]
  \clang inserts dummy basic blocks for the \tc{O0} option on ARM and MIPS.

\item [$\ddagger$]
  \gcc starts function inlining from \tc{O1}, but \clang does from \tc{O2}.


\end{tablenotes}
\end{threeparttable}
  \end{minipage}%
  \hfill
  \begin{minipage}[t]{.49\linewidth}
    \vspace{.2em}
    \normalsize

\autoref{tab:numfuncsbbs} presents the number of functions and basic blocks in
the \normaldataset dataset.
For comparison, we show the numbers for the latest versions of \gcc and \clang, which are
v8.2.0 and v7.0, respectively.
For both compilers, the number of functions significantly decreases for higher
optimization levels due to function inlining.
Meanwhile, the number of basic blocks does not decrease, as basic blocks can
survive in caller functions although function inlining is applied.
%
%
The number even increases as the optimization level increases from \tc{O2} to
\tc{O3} for both compilers.
By analyzing the cases, we confirmed that one possible reason is loop unrolling,
which unwinds the loops and generates multiple copies of basic blocks.
Consequently, the number of basic blocks for \tc{O3} reaches higher than that
for \tc{O2}.

In this paper, we have described multiple optimization issues that significantly
affect the resulting binary code and presemantic features. However, we believe
that there exist remaining issues for detailed optimization techniques. Therefore, we
conclude by encouraging further studies to investigate the implications of
detailed options at each optimization level across different compilers.

  \end{minipage}%

\end{table}


\begin{table*}[!h]
  \caption{Number of binaries, original functions, and filtered (final) functions in \sys.}
  \label{tab:fulldataset}
  \tiny
  \centering
  \setlength\tabcolsep{0.05cm}
  \def\arraystretch{0.50}
  \begin{threeparttable}

    \begin{tabular}{@{}lcrrrrlrrrlrrrlrrrlrrrlrrrl@{}}
\cmidrule(r){1-26}
\multicolumn{1}{c}{\textbf{}} & \multicolumn{1}{c}{\textbf{}}     & \multicolumn{1}{c}{\textbf{}} & \multicolumn{3}{c}{\textbf{\normaldataset}}      & \multicolumn{1}{c}{\textbf{}} & \multicolumn{3}{c}{\textbf{\sizeoptdataset}}     & \multicolumn{1}{c}{\textbf{}} & \multicolumn{3}{c}{\textbf{\piedataset}\tnote{$\ast$}}& \multicolumn{1}{c}{\textbf{}} & \multicolumn{3}{c}{\textbf{\noinlinedataset}}    & \multicolumn{1}{c}{\textbf{}} & \multicolumn{3}{c}{\textbf{\ltodataset}\tnote{$\ast$}}& \multicolumn{1}{c}{\textbf{}} & \multicolumn{3}{c}{\textbf{\obfuscationdataset}}       &  \\ \cmidrule(lr){4-6} \cmidrule(lr){8-10} \cmidrule(lr){12-14} \cmidrule(lr){16-18} \cmidrule(lr){20-22} \cmidrule(lr){24-26}
   & \multicolumn{1}{l}{}     & \multicolumn{1}{l}{} & \multicolumn{1}{l}{}        & \multicolumn{2}{c}{\textbf{\# of Functions}} &    & \multicolumn{1}{l}{}        & \multicolumn{2}{c}{\textbf{\# of Functions}} &    & \multicolumn{1}{l}{}        & \multicolumn{2}{c}{\textbf{\# of Functions}} &    & \multicolumn{1}{l}{}        & \multicolumn{2}{c}{\textbf{\# of Functions}} &    & \multicolumn{1}{l}{}        & \multicolumn{2}{c}{\textbf{\# of Functions}} &    & \multicolumn{1}{l}{}        & \multicolumn{2}{c}{\textbf{\# of Functions}} &  \\ \cmidrule(lr){5-6} \cmidrule(lr){9-10} \cmidrule(lr){13-14} \cmidrule(lr){17-18} \cmidrule(lr){21-22} \cmidrule(lr){25-26}
\multicolumn{1}{c}{\textbf{\begin{tabular}[c]{@{}c@{}}Package\\ Name\end{tabular}}} & \multicolumn{1}{c}{\textbf{Ver.}} & \multicolumn{1}{c}{\textbf{}} & \multicolumn{1}{c}{\textbf{\begin{tabular}[c]{@{}c@{}}\# of\\ Bins\end{tabular}}} & \multicolumn{1}{c}{\textbf{Orig.}} & \multicolumn{1}{c}{\textbf{Final}} & \multicolumn{1}{c}{\textbf{}} & \multicolumn{1}{c}{\textbf{\begin{tabular}[c]{@{}c@{}}\# of\\ Bins\end{tabular}}} & \multicolumn{1}{c}{\textbf{Orig.}} & \multicolumn{1}{c}{\textbf{Final}} & \multicolumn{1}{c}{\textbf{}} & \multicolumn{1}{c}{\textbf{\begin{tabular}[c]{@{}c@{}}\# of\\ Bins\end{tabular}}} & \multicolumn{1}{c}{\textbf{Orig.}} & \multicolumn{1}{c}{\textbf{Final}} & \multicolumn{1}{c}{\textbf{}} & \multicolumn{1}{c}{\textbf{\begin{tabular}[c]{@{}c@{}}\# of\\ Bins\end{tabular}}} & \multicolumn{1}{c}{\textbf{Orig.}} & \multicolumn{1}{c}{\textbf{Final}} & \multicolumn{1}{c}{\textbf{}} & \multicolumn{1}{c}{\textbf{\begin{tabular}[c]{@{}c@{}}\# of\\ Bins\end{tabular}}} & \multicolumn{1}{c}{\textbf{Orig.}} & \multicolumn{1}{c}{\textbf{Final}} & \multicolumn{1}{c}{\textbf{}} & \multicolumn{1}{c}{\textbf{\begin{tabular}[c]{@{}c@{}}\# of\\ Bins\end{tabular}}} & \multicolumn{1}{c}{\textbf{Orig.}} & \multicolumn{1}{c}{\textbf{Final}} &  \\ \cmidrule(r){1-26}

a2ps        & 4.14   &    & 576       & 364K    & 251K    &    & 144       & 88K     & 60K     &    & 576       & 364K    & 251K    &    & 576       & 402K    & 287K    &    & 576       & 244K    & 141K    &    & 256       & 166K    & 116K    &  \\
binutils    & 2.30   &    & 4K        & 7,752K  & 1,032K  &    & 1K        & 1,847K  & 240K    &    & 4K        & 7,748K  & 1,033K  &    & 4K        & 9,113K  & 1,266K  &    & 4K        & 5,388K  & 1,123K  &    & 2K        & 3,615K  & 486K    &  \\
bool        & 0.2.2  &    & 288       & 39K     & 15K     &    & 72        & 10K     & 3K      &    & 288       & 40K     & 15K     &    & 288       & 42K     & 16K     &    & 288       & 35K     & 11K     &    & 128       & 18K     & 7K      &  \\
ccd2cue     & 0.5    &    & 288       & 44K     & 8K      &    & 72        & 11K     & 2K      &    & 288       & 44K     & 8K      &    & 288       & 45K     & 8K      &    & 288       & 39K     & 5K      &    & 128       & 20K     & 4K      &  \\
cflow       & 1.5    &    & 288       & 151K    & 85K     &    & 72        & 36K     & 20K     &    & 288       & 151K    & 85K     &    & 288       & 173K    & 106K    &    & 288       & 109K    & 51K     &    & 128       & 70K     & 40K     &  \\
coreutils   & 8.29   &    & 30K       & 10,396K & 507K    &    & 8K        & 2,535K  & 105K    &    & $\cdot$ & $\cdot$     & $\cdot$     &    & 30K       & 11,592K & 733K    &    & 2K        & 381K    & 23K     &    & 13K       & 4,744K  & 251K    &  \\
cpio        & 2.12   &    & 576       & 303K    & 102K    &    & 144       & 76K     & 25K     &    & 576       & 303K    & 102K    &    & 576       & 341K    & 127K    &    & 576       & 228K    & 68K     &    & 256       & 138K    & 48K     &  \\
cppi        & 1.18   &    & 288       & 85K     & 32K     &    & 72        & 22K     & 8K      &    & 288       & 85K     & 32K     &    & 288       & 91K     & 37K     &    & 288       & 56K     & 12K     &    & 128       & 39K     & 15K     &  \\
dap& 3.10   &    & 1K        & 105K    & 26K     &    & 288       & 26K     & 6K      &    & 1K        & 108K    & 26K     &    & 1K        & 105K    & 26K     &    & 1K        & 94K     & 15K     &    & 512       & 46K     & 11K     &  \\
datamash    & 1.3    &    & 288       & 153K    & 80K     &    & 72        & 37K     & 18K     &    & 288       & 153K    & 80K     &    & 288       & 174K    & 99K     &    & $\cdot$ & $\cdot$     & $\cdot$     &    & 128       & 71K     & 38K     &  \\
direvent    & 5.1    &    & 288       & 222K    & 120K    &    & 72        & 54K     & 29K     &    & 288       & 222K    & 120K    &    & 288       & 239K    & 137K    &    & 288       & 156K    & 70K     &    & 128       & 101K    & 55K     &  \\
enscript    & 1.6.6  &    & 864       & 225K    & 59K     &    & 216       & 56K     & 14K     &    & 864       & 225K    & 59K     &    & 864       & 235K    & 66K     &    & 864       & 195K    & 46K     &    & 384       & 101K    & 27K     &  \\
findutils   & 4.6.0  &    & 2K        & 814K    & 210K    &    & 432       & 198K    & 48K     &    & 2K        & 815K    & 210K    &    & 2K        & 925K    & 269K    &    & 2K        & 603K    & 189K    &    & 768       & 373K    & 100K    &  \\
gawk        & 4.2.1  &    & 288       & 406K    & 252K    &    & 72        & 99K     & 60K     &    & 288       & 403K    & 252K    &    & 288       & 488K    & 332K    &    & 288       & 354K    & 218K    &    & 128       & 195K    & 123K    &  \\
gcal        & 4.1    &    & 1K        & 341K    & 155K    &    & 288       & 85K     & 38K     &    & 1K        & 343K    & 155K    &    & 1K        & 351K    & 160K    &    & 1K        & 325K    & 146K    &    & 512       & 151K    & 69K     &  \\
gdbm        & 1.15   &    & 1K        & 317K    & 98K     &    & 288       & 78K     & 23K     &    & 1K        & 318K    & 98K     &    & 1K        & 331K    & 110K    &    & $\cdot$ & $\cdot$     & $\cdot$     &    & 512       & 144K    & 45K     &  \\
glpk        & 4.65   &    & 576       & 596K    & 399K    &    & 144       & 145K    & 96K     &    & 576       & 596K    & 399K    &    & 576       & 643K    & 445K    &    & $\cdot$ & $\cdot$     & $\cdot$     &    & 256       & 244K    & 184K    &  \\
gmp& 6.1.2  &    & 288       & 273K    & 198K    &    & 72        & 67K     & 48K     &    & 288       & 273K    & 198K    &    & 288       & 291K    & 215K    &    & $\cdot$ & $\cdot$     & $\cdot$     &    & 128       & 123K    & 90K     &  \\
gnu-pw-mgr  & 2.3.1  &    & 576       & 289K    & 81K     &    & 144       & 66K     & 16K     &    & 576       & 291K    & 81K     &    & 576       & 358K    & 119K    &    & 576       & 201K    & 50K     &    & 256       & 139K    & 42K     &  \\
gnudos      & 1.11.4 &    & 576       & 188K    & 82K     &    & 144       & 47K     & 20K     &    & 576       & 188K    & 82K     &    & 576       & 189K    & 82K     &    & $\cdot$ & $\cdot$     & $\cdot$     &    & 256       & 83K     & 36K     &  \\
grep        & 3.1    &    & 288       & 237K    & 133K    &    & 72        & 57K     & 31K     &    & $\cdot$ & $\cdot$     & $\cdot$     &    & 288       & 286K    & 180K    &    & $\cdot$ & $\cdot$     & $\cdot$     &    & 128       & 110K    & 64K     &  \\
gsasl       & 1.8.0  &    & 288       & 125K    & 84K     &    & 72        & 31K     & 20K     &    & 288       & 125K    & 84K     &    & 288       & 131K    & 90K     &    & $\cdot$ & $\cdot$     & $\cdot$     &    & 128       & 56K     & 38K     &  \\
gsl& 2.5    &    & 1K        & 2,043K  & 1,694K  &    & 288       & 500K    & 412K    &    & 1K        & 2,044K  & 1,694K  &    & 1K        & 2,210K  & 1,851K  &    & $\cdot$ & $\cdot$     & $\cdot$     &    & 512       & 921K    & 770K    &  \\
gss& 1.0.3  &    & 576       & 82K     & 28K     &    & 144       & 20K     & 7K      &    & 576       & 82K     & 28K     &    & 576       & 86K     & 32K     &    & $\cdot$ & $\cdot$     & $\cdot$     &    & 256       & 36K     & 13K     &  \\
gzip        & 1.9    &    & 288       & 112K    & 38K     &    & 72        & 28K     & 9K      &    & $\cdot$ & $\cdot$     & $\cdot$     &    & 288       & 122K    & 47K     &    & $\cdot$ & $\cdot$     & $\cdot$     &    & 128       & 50K     & 18K     &  \\
hello       & 2.10   &    & 288       & 65K     & 20K     &    & 72        & 17K     & 6K      &    & 288       & 65K     & 20K     &    & 288       & 67K     & 22K     &    & 288       & 43K     & 6K      &    & 128       & 29K     & 9K      &  \\
inetutils   & 1.9.4  &    & 5K        & 2,083K  & 267K    &    & 1K        & 507K    & 65K     &    & 5K        & 2,086K  & 267K    &    & 5K        & 2,271K  & 309K    &    & 5K        & 1,719K  & 260K    &    & 2K        & 946K    & 122K    &  \\
libiconv    & 1.15   &    & 864       & 164K    & 89K     &    & 216       & 40K     & 21K     &    & 864       & 164K    & 89K     &    & 864       & 175K    & 101K    &    & $\cdot$ & $\cdot$     & $\cdot$     &    & 384       & 75K     & 42K     &  \\
libidn2     & 2.0.5  &    & 288       & 71K     & 23K     &    & 72        & 17K     & 5K      &    & 288       & 71K     & 23K     &    & 288       & 78K     & 31K     &    & $\cdot$ & $\cdot$     & $\cdot$     &    & 128       & 33K     & 12K     &  \\
libmicrohttpd        & 0.9.59 &    & 288       & 109K    & 46K     &    & 72        & 27K     & 11K     &    & 288       & 109K    & 46K     &    & 288       & 115K    & 52K     &    & $\cdot$ & $\cdot$     & $\cdot$     &    & 128       & 50K     & 22K     &  \\
libosip2    & 5.0.0  &    & 576       & 303K    & 188K    &    & 144       & 76K     & 46K     &    & 576       & 303K    & 188K    &    & 576       & 312K    & 196K    &    & $\cdot$ & $\cdot$     & $\cdot$     &    & 256       & 135K    & 85K     &  \\
libtasn1    & 4.13   &    & 1K        & 158K    & 35K     &    & 288       & 39K     & 8K      &    & 1K        & 160K    & 35K     &    & 1K        & 167K    & 42K     &    & $\cdot$ & $\cdot$     & $\cdot$     &    & 512       & 69K     & 16K     &  \\
libtool     & 2.4.6  &    & 288       & 66K     & 28K     &    & 72        & 16K     & 7K      &    & 288       & 66K     & 28K     &    & 288       & 69K     & 30K     &    & $\cdot$ & $\cdot$     & $\cdot$     &    & 128       & 30K     & 13K     &  \\
libunistring& 0.9.10 &    & 288       & 260K    & 180K    &    & 72        & 61K     & 41K     &    & 288       & 260K    & 180K    &    & 288       & 297K    & 215K    &    & $\cdot$ & $\cdot$     & $\cdot$     &    & 128       & 179K    & 86K     &  \\
lightning   & 2.1.2  &    & 288       & 131K    & 101K    &    & 72        & 29K     & 22K     &    & 288       & 131K    & 101K    &    & 288       & 175K    & 144K    &    & $\cdot$ & $\cdot$     & $\cdot$     &    & 128       & 85K     & 48K     &  \\
macchanger  & 1.6.0  &    & 288       & 37K     & 7K      &    & 72        & 9K      & 2K      &    & 288       & 37K     & 7K      &    & 288       & 38K     & 8K      &    & 288       & 32K     & 4K      &    & 128       & 16K     & 3K      &  \\
nettle      & 3.4    &    & 1K        & 92K     & 11K     &    & 288       & 23K     & 2K      &    & 1K        & 95K     & 11K     &    & 1K        & 95K     & 13K     &    & $\cdot$ & $\cdot$     & $\cdot$     &    & 512       & 41K     & 5K      &  \\
patch       & 2.7.6  &    & 288       & 224K    & 110K    &    & 72        & 54K     & 25K     &    & 288       & 224K    & 110K    &    & 288       & 261K    & 147K    &    & 288       & 168K    & 71K     &    & 128       & 104K    & 53K     &  \\
plotutils   & 2.6    &    & 864       & 168K    & 36K     &    & 216       & 42K     & 9K      &    & 864       & 170K    & 36K     &    & 864       & 171K    & 37K     &    & 864       & 145K    & 24K     &    & 384       & 75K     & 16K     &  \\
readline    & 7.0    &    & 576       & 353K    & 168K    &    & 144       & 88K     & 42K     &    & $\cdot$ & $\cdot$     & $\cdot$     &    & 576       & 375K    & 186K    &    & $\cdot$ & $\cdot$     & $\cdot$     &    & 256       & 158K    & 77K     &  \\
recutils    & 1.7    &    & 3K        & 1,776K  & 241K    &    & 720       & 446K    & 58K     &    & 3K        & 1,777K  & 241K    &    & 3K        & 1,984K  & 277K    &    & $\cdot$ & $\cdot$     & $\cdot$     &    & 1K        & 812K    & 111K    &  \\
sed& 4.5    &    & 288       & 188K    & 96K     &    & 72        & 46K     & 22K     &    & $\cdot$ & $\cdot$     & $\cdot$     &    & 288       & 225K    & 131K    &    & $\cdot$ & $\cdot$     & $\cdot$     &    & 128       & 88K     & 46K     &  \\
sharutils   & 4.15.2 &    & 1K        & 581K    & 98K     &    & 288       & 136K    & 21K     &    & 1K        & 583K    & 97K     &    & 1K        & 710K    & 136K    &    & 1K        & 404K    & 57K     &    & 512       & 278K    & 49K     &  \\
spell       & 1.1    &    & 288       & 35K     & 3K      &    & 72        & 9K      & 864     &    & 288       & 36K     & 3K      &    & 288       & 36K     & 4K      &    & 288       & 33K     & 3K      &    & 128       & 16K     & 2K      &  \\
tar& 1.30   &    & 576       & 556K    & 300K    &    & 144       & 135K    & 70K     &    & 576       & 555K    & 300K    &    & 576       & 661K    & 388K    &    & 576       & 439K    & 235K    &    & 256       & 257K    & 142K    &  \\
texinfo     & 6.5    &    & 288       & 113K    & 47K     &    & 72        & 28K     & 11K     &    & 288       & 114K    & 47K     &    & 288       & 131K    & 63K     &    & 288       & 93K     & 35K     &    & 128       & 52K     & 22K     &  \\
time        & 1.9    &    & 288       & 37K     & 6K      &    & 72        & 9K      & 1K      &    & 288       & 38K     & 6K      &    & 288       & 39K     & 8K      &    & 288       & 32K     & 3K      &    & 128       & 17K     & 3K      &  \\
units       & 2.16   &    & 288       & 102K    & 37K     &    & 72        & 26K     & 9K      &    & 288       & 102K    & 37K     &    & 288       & 103K    & 37K     &    & 288       & 86K     & 26K     &    & 128       & 46K     & 16K     &  \\
wdiff       & 1.2.2  &    & 288       & 62K     & 12K     &    & 72        & 15K     & 3K      &    & 288       & 62K     & 12K     &    & 288       & 65K     & 15K     &    & 288       & 50K     & 8K      &    & 128       & 28K     & 6K      &  \\
which       & 2.21   &    & 288       & 41K     & 8K      &    & 72        & 10K     & 2K      &    & 288       & 42K     & 8K      &    & 288       & 43K     & 9K      &    & 288       & 37K     & 5K      &    & 128       & 19K     & 3K      &  \\
xorriso     & 1.4.8  &    & 288       & 921K    & 784K    &    & 72        & 226K    & 190K    &    & 288       & 920K    & 783K    &    & 288       & 990K    & 851K    &    & 288       & 590K    & 473K    &    & 128       & 420K    & 358K    &  \\ \cmidrule(r){1-26}
\multicolumn{2}{c}{\textbf{Total}}     &    & 68K       & 34,356K & 8,708K  &    & 17K       & 8,350K  & 2,061K  &    & 36K       & 23,091K & 7,766K  &    & 68K       & 38,617K & 10,291K &    & 25K       & 12,280K & 3,375K  &    & 30K       & 15,809K & 4,055K  &  \\ \cmidrule(r){1-26}
\end{tabular}

\begin{tablenotes}


\item [$\cdot$]
  The dot symbol ($\cdot$) denotes that these packages were unable to be compiled
due to architecture-specific code or incompatible dependencies.

\end{tablenotes}

\end{threeparttable}
\end{table*}

\end{appendices}

\clearpage
\twocolumn

\begin{IEEEbiography}[{\includegraphics[width=1in,height=1.25in,clip,keepaspectratio]{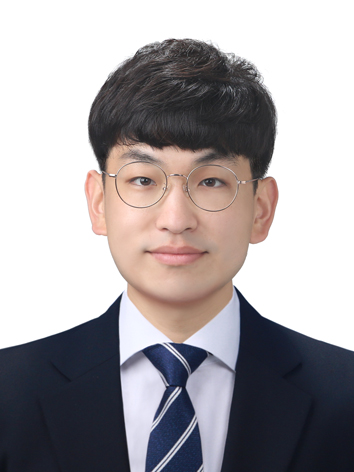}}]%
{Dongkwan Kim} is a freelancer security researcher. He received a Ph.D. from the School of Electrical Engineering at Korea Advanced Institute of Science and Technology (KAIST). His research interests include securing software, embedded \& cyber-physical systems, and cellular infrastructures. He competed in various hacking contests, such as DEFCON, Codegate, and Whitehat Contest. 
\end{IEEEbiography}

\begin{IEEEbiography}[{\includegraphics[width=1in,height=1.25in,clip,keepaspectratio]{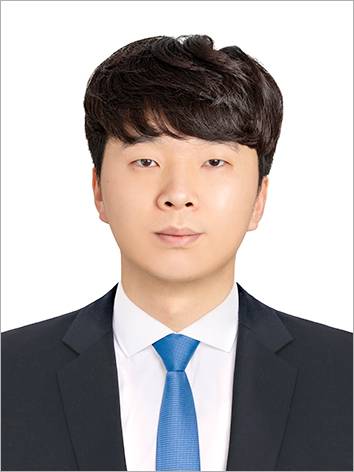}}]%
{Eunsoo Kim}
is a security researcher at Samsung Research. He received a Ph.D. from the Graduate School of Information Security at KAIST. His research interests include finding vulnerabilities in various software and embedded systems.
\end{IEEEbiography}

\begin{IEEEbiography}[{\includegraphics[width=1in,height=1.25in,clip,keepaspectratio]{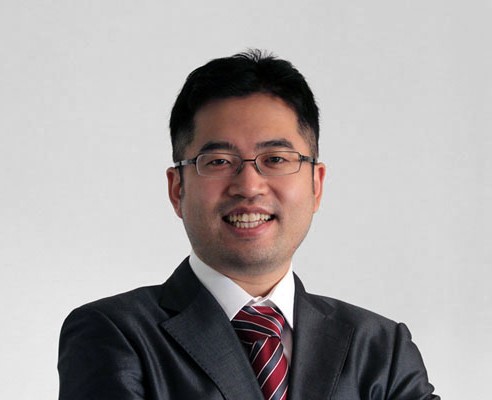}}]%
{Sang Kil Cha}
is an associate professor of Computer Science at KAIST. He completed his Ph.D. in the Electrical \& Computer Engineering department of Carnegie Mellon University. His current research interests revolve mainly around software security, software engineering, and program analysis. He received an ACM distinguished paper award in 2014. He is currently supervising GoN and KaisHack, which are, respectively, undergraduate and graduate hacking team at KAIST.
\end{IEEEbiography}

\begin{IEEEbiography}[{\includegraphics[width=1in,height=1.25in,clip,keepaspectratio]{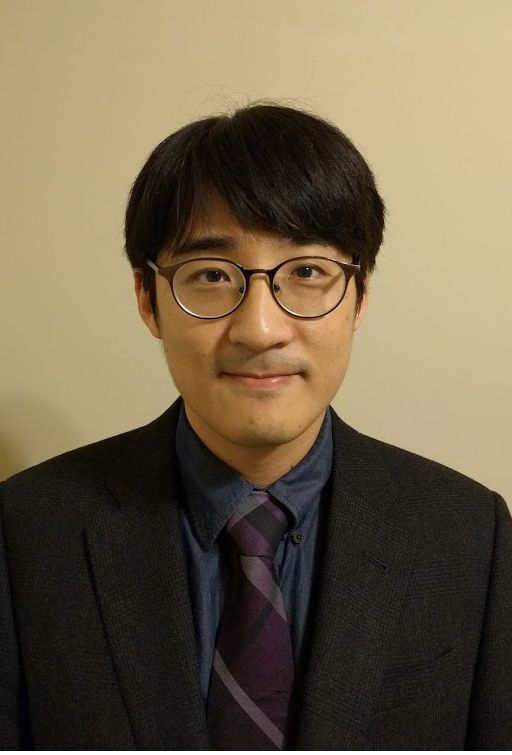}}]%
{Sooel Son}
is an associate professor of School of Computing at KAIST. He received his Ph.D. in the department of computer science at the University of Texas at Austin. He is working on various topics regarding web security and privacy.
\end{IEEEbiography}

\begin{IEEEbiography}[{\includegraphics[width=1in,height=1.25in,clip,keepaspectratio]{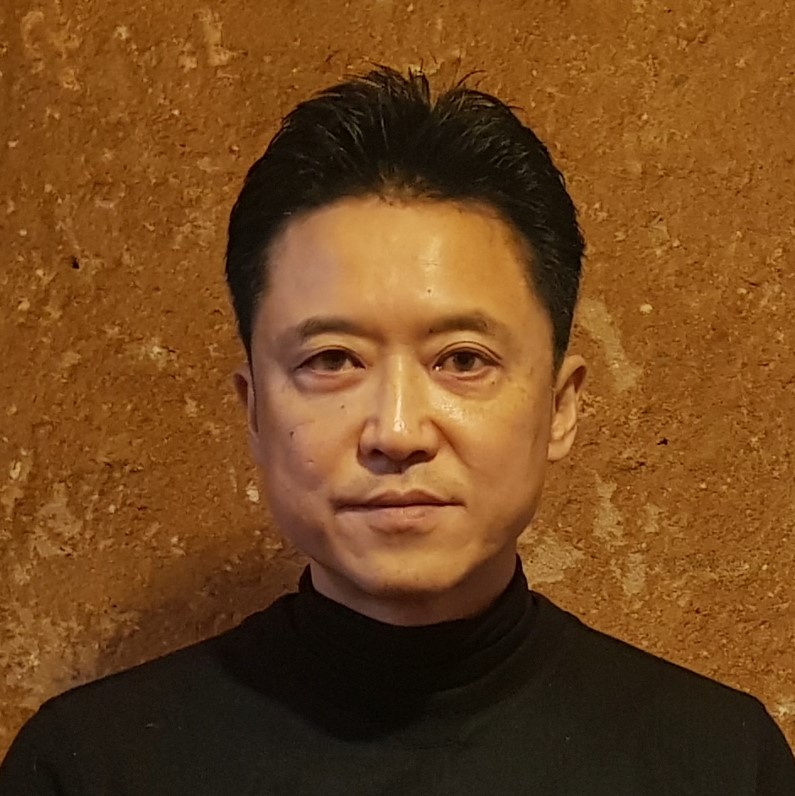}}]%
{Yongdae Kim}
is a Professor in the Department of Electrical Engineering, and an affiliate professor in the Graduate School of Information Security, KAIST. He received a Ph.D. degree from the computer science department at the University of Southern California. Between 2002 and 2012, he was a professor in the Department of Computer Science and Engineering at the University of Minnesota - Twin Cities. Before coming to the US, he worked 6 years in ETRI for securing Korean cyber-infrastructure. He served as a KAIST Chair Professor between 2013 and 2016, and received NSF career award on storage security and McKnight Land-Grant Professorship Award from University of Minnesota in 2005. His main research includes novel attacks and analysis methodologies for emerging technologies, such as 4G/5G cellular networks, drone/self-driving cars, and blockchain.
\end{IEEEbiography}



\end{document}